\documentclass[aps, prb, longbibliography, nofootinbib, showpacs, amsfonts, amsmath, amssymb, twocolumn, floatfix, superscriptaddress, 10pt]{revtex4-2}
\usepackage[final]{graphicx}
\usepackage{xcolor}
\usepackage[colorlinks=true, linkcolor=blue, citecolor=blue, urlcolor=blue]{hyperref}
\usepackage{verbatim}
\usepackage{csquotes}
\usepackage{braket}
\usepackage{mathtools}
\usepackage[normalem]{ulem}

\newcommand{\vct}[1]{\mathbf{#1}}

\DeclareMathOperator{\sgn}{sgn}
\renewcommand\Re{\operatorname{Re}}
\renewcommand\Im{\operatorname{Im}}
\newcommand\Tr{\operatorname{Tr}}
\allowdisplaybreaks

\DeclareSymbolFont{bbgreek}{U}{bbold}{m}{n}
\DeclareMathSymbol{\bbmu}{\mathbb}{bbgreek}{'26}
\DeclareMathSymbol{\bbeps}{\mathbb}{bbgreek}{'17}

\begin{document}

% HEADER_begin ---------------------------------------------

\title{Time-dependent radiative heat flux after the beginning of thermal radiation}

\author{Kiryl Asheichyk}
\email[]{asheichyk@bsu.by}
\affiliation{Department of Theoretical Physics and Astrophysics, Belarusian State University, 5 Babruiskaya Street, 220006 Minsk, Belarus}

\begin{abstract} 
We develop a theoretical formalism for time-dependent radiative heat flux from one object to another in the case where the former starts radiating at a certain time. The time dependence is demonstrated for the heat flux between two isolated nanoparticles. After one particle starts radiating, the emitted energy first reaches the other one with a delay according to electromagnetic retardation, and afterwards the flux exhibits oscillatory exponential relaxation to its stationary value. For the room- or higher-temperature radiation, the oscillation period and relaxation time are determined by the resonance frequency and damping rate of the particle polarizability, respectively, being equal to dozens of femtoseconds and one picosecond for silicon carbide particles. At cryogenic temperatures, the relaxation time depends on the thermal wavelength.
\end{abstract}

\maketitle

% HEADER_end -----------------------------------------------

% MAIN_PART_begin ------------------------------------------

\section{Introduction}
\label{sec:Introduction}
Radiative heat transfer at micro- and nanoscale has been studied intensively in the current century, with promising ideas for a variety of applications~\cite{Chen2005, Basu2009, Novotny2012, Park2013, Bimonte2017, Cuevas2018, Zhang2020, Modest2021, Song2021, Biehs2021, Vazquez-Lozano2024}. However, most of the studies consider stationary conditions, where the heat transfer does not depend on time. Note that one can still study temporal correlations of the stationary transfer via computing the corresponding second order correlation functions~\cite{Biehs2018}, describing the heat transfer fluctuations, which are the cornerstone for the Green-Kubo relations~\cite{Golyk2013_1, Herz2019}. These functions were also computed for electromagnetic energy density~\cite{Herz2023}.

One important and by now well-understood nonstationary problem concerns system thermalization, i.e., relaxation towards global thermal equilibrium of the system of objects with different initial temperatures. This process has been studied for various systems using the heat balance equation, where the heat fluxes between the objects depend on time via the time-dependent temperatures~\cite{Tschikin2012, Messina2013, Dyakov2014, Nikbakht2015, Yannopapas2013, Kruger2012}. A more recently studied nonstationary scenarios involve time modulation of the objects' dielectric functions~\cite{Buddhiraju2020, Vazquez-Lozano2023, Yu2023, Yu2024_1, Liberal2024, Tang2024} or temperatures~\cite{Latella2018, Ordonez-Miranda2019_1, Ordonez-Miranda2019_2, Messina2020, Li2024}. It was shown that this modulation can fundamentally affect heat radiation and transfer: For example, the heat emission of an object can be coherent between different frequencies~\cite{Vazquez-Lozano2023, Yu2023}, and a colder object in the proximity of a hotter one can experience an active cooling~\cite{Yu2024_1}.

In practice, the strength of thermal radiation may change abruptly, leading to another type of the nonstationarity. This can happen because of a sudden increase of the radiation intensity of an object, either via abruptly rising its temperature or changing its dielectric properties (for example, from a weakly radiating good conductor to a strongly radiating dielectric), or because of electromagnetic (EM) shielding of the radiated field until a certain time. The heat fluxes following such changes become time dependent, and their fundamental understanding may be important for a variety of applications at nanoscale, including thermal logic~\cite{Ben-Abdallah2015} and nanomedicine (see, e.g., Ref.~\cite{Pustovalov2024} and references therein).  In this paper, we aim to provide a theoretical description for such scenarios by considering that the radiation exists only starting from a certain time. This nonstationary condition mimics an abrupt radiation increase and leads to a transient regime for the radiated EM field until a steady state is reached.

Specifically, in Sec.~\ref{sec:Flux_general}, we derive a formula for heat flux from one object (object~$ 1 $) to another (object~$ 2 $) in the case where the former starts its radiation at time $ t = 0 $. The two objects can be of arbitrary shape, size, and material, and they can be surrounded by other arbitrary objects. Our derivation employs fluctuational electrodynamics in time domain, under the assumption of local equilibrium inside each object. The derived flux contains the permittivity and Green's tensors in time domain, thus rigorously accounting for the time dependence, including EM retardation. In Sec.~\ref{sec:HT_PP}, we give the flux in the limit where objects $ 1 $ and $ 2 $ reduce to small spherical particles, a practically important scenario being feasible for numerical computations. We demonstrate the developed formalism in Sec.~\ref{sec:HT_example} for two small particles in isolation, finding the time dependence of the flux for different temperatures and interparticle distances.

\section{Maxwell equations and Green's functions in time domain}
\label{sec:MEGF}
We consider a system of objects at different temperatures in vacuum (see Fig.~\ref{fig:System})\footnote{One can also take into account environment and its radiation~\cite{Kruger2012}, which, however, plays no role in our quantities of interest.}. Following fluctuational electrodynamics~\cite{Rytov1958, Rytov1989}, finite temperatures lead to fluctuating current $ \vct{j}(\vct{r},t) $ inside the objects, which is responsible for thermal radiation. The resulting EM field obeys Maxwell equations~\cite{Jackson1998},
\begin{subequations}
\begin{equation}
\nabla \times \vct{E} = -\frac{1}{c}\frac{\partial\vct{B}}{\partial t},
\label{eq:ME1}
\end{equation}
\begin{equation}
\nabla \times \vct{H} = \frac{4\pi}{c}\vct{j} + \frac{1}{c}\frac{\partial\vct{D}}{\partial t},
\label{eq:ME2}
\end{equation}
\end{subequations}
with $ c $ being the speed of light in vacuum. We restrict to nonmagnetic objects (magnetic permeability $ \mu = 1 $) with spatially local dielectric response, such that the constitutive relations read~\cite{Jackson1998,Rahi2009}\footnote{Unless otherwise stated, the absence of integration limits implies integration over all space, time, or frequencies.}
\begin{subequations}
\begin{equation}
\vct{D}(\vct{r},t) = \int\! dt' \bbeps(\vct{r},t')\vct{E}(\vct{r},t-t'),
\label{eq:ConstE}
\end{equation}
\begin{equation}
\vct{B}(\vct{r},t) = \vct{H}(\vct{r},t),
\label{eq:ConstH}
\end{equation}
\end{subequations}
where $ \bbeps $ is the permittivity tensor.

Fields $ \vct{E} $ and $ \vct{H} $ are related to current $ \vct{j} $ via electric and magnetic dyadic Green's functions (GFs) $ \mathbb{G}^{\textrm{E}} $ and $ \mathbb{G}^{\textrm{H}} $ (tensors) of the system~\cite{Felsen1994},
\begin{subequations}
\begin{equation}
\vct{E}(\vct{r},t) = \vct{E}_0(\vct{r},t) + \int\! d^3r' \int\! dt' \mathbb{G}^{\textrm{E}}(\vct{r},\vct{r}';t-t')\vct{j}(\vct{r}',t'),
\label{eq:GEdef}
\end{equation}
\begin{equation}
\vct{H}(\vct{r},t) = \vct{H}_0(\vct{r},t) + \int\! d^3r' \int\! dt' \mathbb{G}^{\textrm{H}}(\vct{r},\vct{r}';t-t')\vct{j}(\vct{r}',t'),
\label{eq:GHdef}
\end{equation}
\end{subequations}
where $ \vct{E}_0 $ and $ \vct{H}_0 $ are the homogeneous solutions, i.e., with no current present. For our problem, these solutions play no role, and we hence formally set them to zero. Depending on the EM field of interest, the integration in Eqs.~\eqref{eq:GEdef} and~\eqref{eq:GHdef} can run over specific region in space (e.g., a certain object) and a certain time interval.

\section{Formula for heat flux}
\label{sec:Flux_general}
\subsection{Problem of interest}
\label{subsec:Problem}
Let us consider the field emitted by a certain object (object~$ 1 $), such that the spatial integration in Eqs.~\eqref{eq:GEdef} and~\eqref{eq:GHdef} runs over the volume $ V_1 $ of this object. In a stationary scenario, the object is assumed to radiate forever (or starting far away in the past), such that the lower and upper time integration limits are set to $ -\infty $ and $ \infty $, respectively\footnote{Physically, the upper limit of Eqs.~\eqref{eq:GEdef} and~\eqref{eq:GHdef} cannot be larger than $ t $ because of the causality principle. However, it can be formally extended to $ \infty $ owing to the causality of the GFs.}. This allows to work entirely in frequency domain when computing physical observables~\cite{Bimonte2017, Kruger2011, Messina2011, Kruger2012, Rodriguez2012, Muller2017, Messina2013, Asheichyk2017, Soo2018} (see also Appendix~\ref{app:sec:FluxSt}).

\begin{figure}[!t]
\begin{center}
\includegraphics[width=1.0\linewidth]{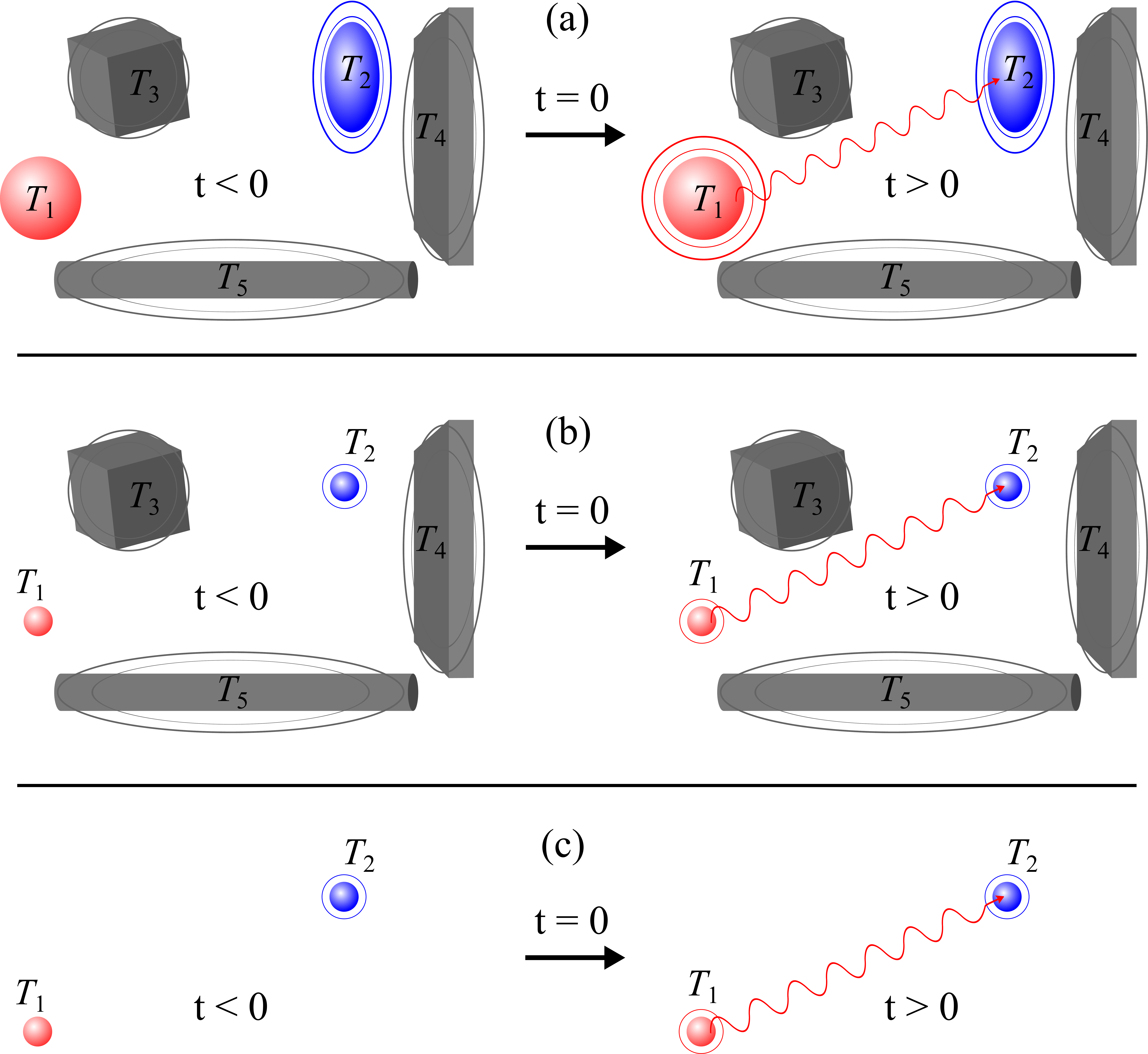}
\end{center}
\caption{\label{fig:System}Radiative heat flux after started thermal radiation. For time $ t < 0 $, object $ 1 $ (red) does not radiate energy, whereas other objects may radiate. At $ t = 0 $, object $ 1 $ starts radiating, which leads to a time-dependent heat flux to object $ 2 $ (blue) for $ t > 0 $, which we aim to study. $ T_i \geq 0 $ is the temperature of object $ i $. Different subfigures depict different systems (by their complexity): (a) general system, where all objects are arbitrary (considered in Sec.~\ref{sec:Flux_general}); (b) objects $ 1 $ and $ 2 $ are small particles (Sec.~\ref{sec:HT_PP}); (c) objects $ 1 $ and $ 2 $ are small particles, and there no other objects present (Sec.~\ref{sec:HT_example}).}
\end{figure}

In this paper, we consider that no radiation of object $ 1 $ is present for time $ t < 0 $, such that the time integration starts at time zero:
\begin{subequations}
\begin{equation}
\vct{E}(\vct{r},t) = \int_{V_1}\! d^3r' \int_0^{\infty}\! dt' \mathbb{G}^{\textrm{E}}(\vct{r},\vct{r}';t-t')\vct{j}(\vct{r}',t'),
\label{eq:E}
\end{equation}
\begin{equation}
\vct{H}(\vct{r},t) =  \int_{V_1}\! d^3r' \int_0^{\infty}\! dt' \mathbb{G}^{\textrm{H}}(\vct{r},\vct{r}';t-t')\vct{j}(\vct{r}',t').
\label{eq:H}
\end{equation}
\end{subequations}
In other words, EM field which can be sourced by object $ 1 $ is zero everywhere for $ t < 0 $, and only at positive times the thermal current inside the object can lead to finite radiation. While introducing this initial condition, we assume that current $ \vct{j} $ in Eqs.~\eqref{eq:E} and~\eqref{eq:H} is in a stationary state and its autocorrelation function is given by the equilibrium fluctuation-dissipation theorem (FDT) in Eq.~\eqref{eq:C_current}. In contrast, fields $ \vct{E} $ and $ \vct{H} $ in Eqs.~\eqref{eq:E} and~\eqref{eq:H}, and the corresponding observables (EM energies, energy fluxes, or forces) follow nonstationary (transient) dynamics until the system relaxes to a stationary state.  

We aim to study this transient regime for heat flux $ \Phi_1^{(2)} $ from object $ 1 $ to another object (object $ 2 $), in the presence of other objects (see Fig.~\ref{fig:System}). This flux is given by the divergence of the Poynting vector~\cite{Jackson1998}
\begin{equation}
\vct{S} = \frac{c}{4\pi}\vct{E}\times\vct{H},
\label{eq:S_def}
\end{equation}
with $ \vct{E} $ and $ \vct{H} $ given by Eqs.~\eqref{eq:E} and~\eqref{eq:H}, respectively, integrated over the volume $ V_2 $ of object $ 2 $,
\begin{equation}
\Phi_1^{(2)}(t) = -\int_{V_2}\!d^3r\langle\nabla\cdot\vct{S}(\vct{r},t)\rangle,
\label{eq:Flux_def}
\end{equation}
where $ \langle \cdots \rangle $ indicates the thermal average~\cite{Rytov1958, Rytov1989, Bimonte2017} and the minus sign stands for the inward direction with respect to object $ 2 $.

\subsection{Poynting's theorem}
\label{subsec:PoyntingsTheorem}
To proceed, we use the vector identity~\cite{Jackson1998}
\begin{equation}
\nabla\cdot\left(\vct{E}\times\vct{H}\right) = \left[\vct{H}\cdot\left(\nabla\times\vct{E}\right)-\vct{E}\cdot\left(\nabla\times\vct{H}\right)\right],
\label{eq:nablaS}
\end{equation}
and employ Eqs.~\eqref{eq:ME1}--\eqref{eq:ConstH} to obtain
\begin{align}
\notag - &\nabla\cdot\vct{S}(\vct{r},t) = \frac{1}{8\pi}\frac{\partial}{\partial t}\vct{H}^2(\vct{r},t)\\
& + \frac{1}{4\pi}\int\! dt'\vct{E}(\vct{r},t)\bbeps(\vct{r},t')\frac{\partial\vct{E}(\vct{r},t-t')}{\partial t} + \vct{E}(\vct{r},t)\cdot\vct{j}(\vct{r},t).
\label{eq:PoyntingsTheorem}
\end{align}
Equation~\eqref{eq:PoyntingsTheorem} is a version of the Poynting's theorem, representing conservation of energy~\cite{Jackson1998}: The energy entering some volume (given by $ -\nabla\cdot\vct{S} $) is stored and dissipated within the volume (the first two terms on the right-hand side), accounting for the presence of explicit Ohmic loss (the last term). 

Importantly, in a nonstationary scenario, not all the energy emitted by an object and entering another one is necessarily absorbed by the latter, such that heat flux $ \Phi_1^{(2)} $ is, in general, not equal to heat transfer $ H_1^{(2)} $. It is not evident from Eq.~\eqref{eq:PoyntingsTheorem} what part of the energy is dissipated, i.e., turns into heat, and it hence appears challenging to extract the heat transfer from the heat flux for a general case. However, this becomes clear when considering concrete examples~\cite{Jackson1998, Loudon1970, Ruppin2002, Cui2004, Luan2009, Nunes2011, Shivanand2012}, as we demonstrate in Sec.~\ref{sec:HT_example}.

\subsection{Current correlator in time domain}
\label{subsec:C_current}
We substitute Eq.~\eqref{eq:PoyntingsTheorem} into Eq.~\eqref{eq:Flux_def} to get
\begin{align}
\notag \Phi_1^{(2)}(t) & = \frac{1}{8\pi}\frac{\partial}{\partial t}\int_{V_2}\! d^3r\langle\vct{H}^2(\vct{r},t)\rangle\\
\notag & +\frac{1}{4\pi}\int_{V_2}\! d^3r\int\! dt'\left\langle\vct{E}(\vct{r},t)\bbeps(\vct{r},t')\frac{\partial\vct{E}(\vct{r},t-t')}{\partial t}\right\rangle\\
& +\int_{V_2}\! d^3r\langle \vct{E}(\vct{r},t)\cdot\vct{j}(\vct{r},t)\rangle.
\label{eq:FluxPT}
\end{align}
Note that $ \langle\vct{H}^2(\vct{r},t)\rangle $ is the equal-time autocorrelation function, and the first (magnetic) term of Eq.~\eqref{eq:FluxPT} hence vanishes in a stationary case.

Substituting Eqs.~\eqref{eq:E} and~\eqref{eq:H} into Eq.~\eqref{eq:FluxPT}, we obtain
\begin{widetext}
\begin{align}
\notag \Phi_1^{(2)}(t) = \ & \frac{1}{8\pi}\frac{\partial}{\partial t}\int_{V_1}\! d^3r_1\int_{V_1}\! d^3r_1'\int_{V_2}\! d^3r_2\int_0^{\infty}\! dt' \int_0^{\infty}\! dt'' \Tr\left\{\langle\vct{j}(\vct{r}_1,t')\otimes\vct{j}(\vct{r}_1',t'')\rangle  \mathbb{G}^{\textrm{H}T}(\vct{r}_2,\vct{r}_1';t-t'') \mathbb{G}^{\textrm{H}}(\vct{r}_2,\vct{r}_1;t-t')\right\}\\
\notag & +\frac{1}{4\pi}\int_{V_1}\! d^3r_1\int_{V_1}\! d^3r_1'\int_{V_2}\! d^3r_2\int_0^{\infty}\! dt' \int_0^{\infty}\! dt'' \int\! d\tilde{t}\\
\notag & \ \ \ \ \ \ \ \ \times \Tr\left\{\langle\vct{j}(\vct{r}_1,t')\otimes\vct{j}(\vct{r}_1',t'')\rangle  \frac{\partial\mathbb{G}^{\textrm{E}T}(\vct{r}_2,\vct{r}_1';t-\tilde{t}-t'')}{\partial t}\bbeps^T(\vct{r}_2,\tilde{t})\mathbb{G}^{\textrm{E}}(\vct{r}_2,\vct{r}_1;t-t')\right\}\\
& +\int_{V_1}\! d^3r_1\int_{V_2}\! d^3r_2\int_0^{\infty}\! dt' \Tr\left\{\langle\vct{j}(\vct{r}_1,t')\otimes\vct{j}(\vct{r}_2,t)\rangle\mathbb{G}^{\textrm{E}}(\vct{r}_2,\vct{r}_1;t-t')\right\},
\label{eq:FluxEH}
\end{align}
\end{widetext}
where $ \otimes $ and the superscript $ T $ denote the outer product and matrix transpose (not including arguments permutation), respectively, and $ \Tr $ indicates the trace over matrix indices. Equation~\eqref{eq:FluxEH} is valid for reciprocal or nonreciprocal objects~\cite{Ben-Abdallah2016, Zhu2016, Zhu2018, Herz2019, Gelbwaser-Klimovsky2021, Yang2022}.

From now on, we restrict to reciprocal objects, such that $ \bbeps^T = \bbeps $, $ \mathbb{G}^{\textrm{E}T}(\vct{r},\vct{r}') = \mathbb{G}^{\textrm{E}}(\vct{r}',\vct{r}) $, and $ \mathbb{G}^{\textrm{H}T}(\vct{r},\vct{r}') = \mathbb{G}^{\textrm{H}}(\vct{r}',\vct{r}) $~\cite{Felsen1994, Tai1994, Eckhardt1984}. The time-domain current correlator in Eq.~\eqref{eq:FluxEH} can be obtained from the inverse Fourier transform of the well-known frequency-domain correlator, given by the fluctuation-dissipation theorem (FDT)~\cite{Rytov1989, Bimonte2017, Zhu2018, Yang2022, Eckhardt1984, Joulain2005},
\begin{align}
\notag &\langle \vct{j}(\vct{r},\omega)\otimes\vct{j}^*(\vct{r}',\omega')\rangle_{\textrm{full}}\\
& =\frac{\hbar\omega^2}{2}\coth\left(\frac{\omega}{2\omega_T}\right)\Im[\bbeps(\vct{r},\omega)]\delta^{(3)}(\vct{r}-\vct{r}')\delta(\omega-\omega'),
\label{eq:FDTomega_full}
\end{align}
where $ \bbeps $ is the dielectric response in frequency domain of the medium and $ \omega_T = k_{\textrm{B}}T/\hbar $ is the thermal frequency, with $ T $ being the temperature of the medium ($ \hbar $ and $ k_{\textrm{B}} $ are Planck's and Boltzmann's constants, respectively).

Equation~\eqref{eq:FDTomega_full} contains zero-point (temperature-independent) fluctuations,
\begin{align}
\notag &\lim_{T\to 0}\langle \vct{j}(\vct{r},\omega)\otimes\vct{j}^*(\vct{r}',\omega')\rangle_{\textrm{full}}\\
& =\frac{\hbar\omega^2}{2}\sgn(\omega)\Im[\bbeps(\vct{r},\omega)]\delta^{(3)}(\vct{r}-\vct{r}')\delta(\omega-\omega'),
\label{eq:FDTomega_zp}
\end{align}
which play no role in the considered problem~\cite{Bimonte2017, Kruger2011, Kruger2012}. Subtracting Eq.~\eqref{eq:FDTomega_zp} from Eq.~\eqref{eq:FDTomega_full}, we get the thermal part, which is responsible for heat fluxes (and thus used in our computations),
\begin{align}
\notag &\langle \vct{j}(\vct{r},\omega)\otimes\vct{j}^*(\vct{r}',\omega')\rangle\\
\notag & = \langle \vct{j}(\vct{r},\omega)\otimes\vct{j}^*(\vct{r}',\omega')\rangle_{\textrm{full}} - \lim_{T\to 0}\langle \vct{j}(\vct{r},\omega)\otimes\vct{j}^*(\vct{r}',\omega')\rangle_{\textrm{full}}\\
& =\frac{\hbar\omega^2\sgn(\omega)}{e^{\frac{|\omega|}{\omega_T}}-1}\Im[\bbeps(\vct{r},\omega)]\delta^{(3)}(\vct{r}-\vct{r}')\delta(\omega-\omega').
\label{eq:FDTomega}
\end{align}
Note that the sign function in Eq.~\eqref{eq:FDTomega_zp}, $ \sgn(\omega) $, distinguishes between $ T \to 0 $ limit of the full correlator for positive and negative frequencies~\cite{Kruger2012}. As a consequence, the thermal correlator~\eqref{eq:FDTomega} also contains $ \sgn(\omega) $, as well as $ |\omega| $; for $ T \to 0 $, it vanishes, independently of the sign of $ \omega $.

We assume that the currents of different objects are uncorrelated, such that the last term of Eq.~\eqref{eq:FluxEH} vanishes. This means that explicit Ohmic loss within object $ 2 $ does not contribute to $ \Phi_1^{(2)}(t) $, because $ \vct{E} $ (sourced by object $ 1 $) and $ \vct{j} $ (originated in object $ 2 $) are uncorrelated. However, this loss is, in general, finite when considering the self-emission of an object, where the object is both emitter and receiver~\cite{Kruger2012}.

Performing the inverse Fourier transform\footnote{For a function $ f(t) $, we define its Fourier transform as $ f (\omega) = \int dt f(t)e^{i\omega t} $, and the inverse Fourier transform as $ f(t) = \frac{1}{2\pi}\int d\omega f(\omega)e^{-i\omega t} $.} of Eq.~\eqref{eq:FDTomega}, and using the fact that $ \Im[\bbeps(\vct{r},\omega)] $ is an odd function of $ \omega $ [owing to the reality of $ \bbeps(\vct{r},t) $], we obtain
\begin{align}
\notag &\langle \vct{j}(\vct{r},t)\otimes\vct{j}(\vct{r}',t')\rangle\\
& =\delta(\vct{r}-\vct{r}')\frac{\hbar}{2\pi^2}\int_0^{\infty}\! d\omega \frac{\omega^2}{e^{\frac{\omega}{\omega_T}}-1}\Im[\bbeps(\vct{r},\omega)]\cos[\omega(t-t')].
\label{eq:C_current}
\end{align}

\subsection{Formula for time-dependent heat flux, and its stationary analog}
\label{subsec:Formula}
Substituting Eq.~\eqref{eq:C_current} into Eq.~\eqref{eq:FluxEH}, taking into account that the temperature of object $ 1 $ is $ T_1 $, we obtain
\begin{widetext}
\begin{align}
\notag \Phi_1^{(2)}(t) = \ & \frac{\hbar}{16\pi^3}\frac{\partial}{\partial t}\int_0^{\infty}\! d\omega \frac{\omega^2}{e^{\frac{\omega}{\omega_{T_1}}}-1}\int_0^{\infty}\! dt' \int_0^{\infty}\! dt''\cos[\omega(t'-t'')]\int_{V_1}\! d^3r_1\int_{V_2}\! d^3r_2\\
\notag & \ \ \ \ \ \ \ \ \ \ \ \times \Tr\left\{\Im[\bbeps(\vct{r}_1,\omega)]\mathbb{G}^{\textrm{H}}(\vct{r}_1,\vct{r}_2;t-t') \mathbb{G}^{\textrm{H}T}(\vct{r}_1,\vct{r}_2;t-t'')\right\}\\
\notag & +\frac{\hbar}{8\pi^3}\int_0^{\infty}\! d\omega \frac{\omega^2}{e^{\frac{\omega}{\omega_{T_1}}}-1}\int_0^{\infty}\! dt' \int_0^{\infty}\! dt''\cos[\omega(t'-t'')]\int\! d\tilde{t}\int_{V_1}\! d^3r_1\int_{V_2}\! d^3r_2\\
& \ \ \ \ \ \ \ \ \ \ \ \times \Tr\left\{\Im[\bbeps(\vct{r}_1,\omega)]\frac{\partial\mathbb{G}^{\textrm{E}}(\vct{r}_1,\vct{r}_2;t-\tilde{t}-t')}{\partial t}\bbeps(\vct{r}_2,\tilde{t})\mathbb{G}^{\textrm{E}T}(\vct{r}_1,\vct{r}_2;t-t'')\right\}.
\label{eq:Flux}
\end{align}
\end{widetext}
Formula~\eqref{eq:Flux} is the main result of this paper. It gives the time-dependent radiative heat flux from object $ 1 $ to object $ 2 $ in the presence of other objects if object $ 1 $ starts radiating at time $ t = 0 $ (see Fig.~\ref{fig:System}). In this nonequilibrium nonstationary state, local equilibrium for each object is assumed, such that the thermal current fluctuations inside each object obey FDT~\eqref{eq:FDTomega} at the object's temperature. All objects are assumed to be nonmagnetic, with a local and reciprocal dielectric response, but otherwise can be of arbitrary material [including inhomogeneous and (or) unisotropic] and arbitrary shape and size. We remind that the GFs in Eq.~\eqref{eq:Flux} are the GFs of the full system, i.e., of all objects, including objects $ 1 $ and $ 2 $ (see Fig.~\ref{fig:System}). Note that $ \Phi_1^{(2)}(t) $ depends on the temperature of only object $ 1 $, whereas other temperatures may be relevant for other fluxes in the system (which are not studied in this paper).

In the corresponding stationary scenario, the current existed far away in the past, such that the stationary flux, denoted as $ \Phi_{1\textrm{st}}^{(2)} $, is given by Eq.~\eqref{eq:Flux} with $ 0 $ replaced by $ -\infty $ in the lower time integration limits. Because of this replacement and the stationarity of the current correlator~\eqref{eq:C_current}, $ t' $, $ t'' $, and $ \tilde{t} $ in Eq.~\eqref{eq:Flux} can be integrated out in a way that $ t $ cancels out and only original $ \omega $ integral remains (see Appendix~\ref{app:subsec:FluxSt_from_tdep}),
\begin{align}
\notag & \Phi_{1\textrm{st}}^{(2)} = H_{1\textrm{st}}^{(2)} = \frac{2\hbar}{\pi c^4}\int_0^{\infty}\! d\omega \frac{\omega^5}{e^{\frac{\omega}{\omega_{T_1}}}-1}\int_{V_1}\! d^3r_1\int_{V_2}\! d^3r_2\\
& \times\Tr\left\{\Im[\bbeps(\vct{r}_1,\omega)]\mathbb{G}^{\textrm{E}}(\vct{r}_1,\vct{r}_2;\omega)\Im[\bbeps(\vct{r}_2,\omega)]\mathbb{G}^{\textrm{E}\dagger}(\vct{r}_1,\vct{r}_2;\omega)\right\},
\label{eq:FluxSt}
\end{align}
where $ \bbeps $ and $ \mathbb{G}^{\textrm{E}} $ are in frequency domain, indicating that Eq.~\eqref{eq:FluxSt} can also be derived starting from frequency-dependent Maxwell equations (see Appendix~\ref{app:subsec:FluxSt_omega} and Ref.~\cite{Soo2018}); $ \dagger $ stands for the conjugate transpose (not including arguments permutation). Since the flux spectrum in Eq.~\eqref{eq:FluxSt} is proportional to $ \Im[\bbeps(\vct{r}_2,\omega)] $, all the energy entering object $ 2 $ is absorbed, such that the heat flux $ \Phi_{1\textrm{st}}^{(2)} $ equals the heat transfer $ H_{1\textrm{st}}^{(2)} $, as indicated. Equation~\eqref{eq:FluxSt} was derived in Refs.~\cite{Soo2018, Kruger2024}, and it can also be cast into the formulas in terms of the scattering operators~\cite{Bimonte2017, Kruger2011, Messina2011, Kruger2012, Muller2017, Soo2018}. In contrast to the nonstationary flux~\eqref{eq:Flux}, which can be either positive, zero, or negative (see Sec.~\ref{sec:HT_example}), $ \Phi_{1\textrm{st}}^{(2)} $ in Eq.~\eqref{eq:FluxSt} is manifestly nonnegative\footnote{\label{fn:positive_matrix}For a passive material, $ \Im[\bbeps] $ is a positive-semidefinite matrix~\cite{Kruger2012}, and therefore, $ \mathbb{G}^{\textrm{E}}\Im[\bbeps]\mathbb{G}^{\textrm{E}\dagger} $ is also positive semidefinite~\cite{Horn2012}. The trace of a product of two positive-semidefinite matrices is nonnegative~\cite{Coope1994}.}. Also, the stationary flux is symmetric upon interchanging objects $ 1 $ and $ 2 $ (i.e., $ \Phi_{1\textrm{st}}^{(2)} =  \Phi_{2\textrm{st}}^{(1)} $ if $ T_1 = T_2 $), whereas the nonstationary one is, in general, not. The magnetic term of Eq.~\eqref{eq:Flux} (containing $ \mathbb{G}^{\textrm{H}} $) does not contribute to $ \Phi_{1\textrm{st}}^{(2)} $ [see Eq.~\eqref{eq:FluxPT} and the discussion below it]. 

For large $ t $, $ \Phi_1^{(2)} $ in Eq.~\eqref{eq:Flux} must converge to a time-independent value, which, from physical grounds, is expected to be the same as $ \Phi_{1\textrm{st}}^{(2)} $ in Eq.~\eqref{eq:FluxSt},
\begin{equation}
\underset{t\to\infty}{\lim}\Phi_1^{(2)}(t) = \Phi_{1\textrm{st}}^{(2)}.
\label{eq:LimStFluxSt}
\end{equation} 
However, mathematically, the limit $ \lim_{t\to\infty}\Phi_1^{(2)}(t) $ [$ t $ approaches infinity in Eq.~\eqref{eq:Flux}, i.e., in the GFs and their derivatives] is taken differently than the limit corresponding to $ \Phi_{1\textrm{st}}^{(2)} $ (described above). In Sec.~\ref{sec:HT_example}, we show that the nonstationary flux computed for a specific example satisfies Eq.~\eqref{eq:LimStFluxSt}.

\section{Point particle limit}
\label{sec:HT_PP}
For arbitrary objects $ 1 $ and $ 2 $, computation of the stationary heat transfer in Eq.~\eqref{eq:FluxSt} is a technically difficult problem even if no other objects [the gray ones in Fig.~\ref{fig:System}(a)] are present; the existing computations are mainly restricted to two objects, with each object being either a plate~\cite{Bimonte2017, Kruger2011, Messina2011, Muller2017, Dyakov2014, Polder1971, Bimonte2009, Otey2011, McCauley2012, Guerout2012, Golyk2013_2, Nefzaoui2013}, sphere~\cite{Bimonte2017, Kruger2011, Kruger2012, Rodriguez2012, Zhu2016, Zhu2018, Otey2011, McCauley2012, Golyk2013_2, Narayanaswamy2008, Sasihithlu2011, Rodriguez2013}, or cylinder~\cite{Bimonte2017, Rodriguez2012, McCauley2012, Rodriguez2013, Barura2022, Xiao2023}. Computing the nonstationary heat flux in formula~\eqref{eq:Flux} is a way more complicated, which may limit the use of the formula in practice.

A much more simple yet practically relevant and extensively studied~\cite{Biehs2021} situation is when objects $ 1 $ and $ 2 $ are small particles [see Fig.~\ref{fig:System}(b)], more precisely, they are spherical nonmagnetic particles, which are smaller than any other length scale related to them\footnote{The radius of each particle is assumed to be small compared to the thermal wavelength $ \lambda_{T_1} = \hbar c/(k_{\textrm{B}}T_1) $, the particle's skin depth, and the distances between the particle and other objects.}. In this point particle limit, Eq.~\eqref{eq:FluxSt} simplifies to~\cite{Messina2013, Asheichyk2017, Biehs2021, Ben-Abdallah2011, Dong2017}
\begin{align}
\notag & \Phi_{1\textrm{st}}^{(2)} = H_{1\textrm{st}}^{(2)} = \frac{32\pi\hbar}{c^4} \int_0^\infty\! d\omega \frac{\omega^5}{e^{\frac{\omega}{\omega_{T_1}}}-1}\\
& \times \Im[\alpha_1(\omega)]\Im[\alpha_2(\omega)]\Tr\left\{\mathbb{G}_{\overline{12}}^{\textrm{E}}(\vct{r}_1,\vct{r}_2;\omega)\mathbb{G}_{\overline{12}}^{\textrm{E}\dagger}(\vct{r}_1,\vct{r}_2,\omega)\right\},
\label{eq:FluxStPP}
\end{align}
where
\begin{equation}
\alpha_i (\omega)= \frac{\varepsilon_i(\omega)-1}{\varepsilon_i(\omega)+2}R_i^3
\label{eq:polarizability}
\end{equation}
is the polarizability of particle $ i $ (with $ \varepsilon_i $ and $ R_i $ being the scalar permittivity and radius of the particle, respectively), and $ \mathbb{G}_{\overline{12}}^{\textrm{E}}(\vct{r}_1,\vct{r}_2;\omega) $ (where $ \vct{r}_i $ is the position of particle $ i $) is the GF of the system not including the two particles. The latter fact greatly simplifies the computations: For example, to compute the heat transfer between two point particles in the presence of another object, one needs the GF of only that single object (see, e.g., Refs.~\cite{Saaskilahti2014, Dong2018, Messina2018, Asheichyk2018, Zhang2019_1, Zhang2019_2, He2019, Fang2022},~\cite{Asheichyk2017}, and~\cite{Asheichyk2022, Asheichyk2023, Wang2024, Asheichyk2024} for the heat transfer in the presence of a plate, sphere, and cylinder, respectively). Note that the flux in Eq.~\eqref{eq:FluxStPP} [as well as in Eq.~\eqref{eq:FluxPP}] is proportional to $ V_1V_2 $, where $ V_1 $ and $ V_2 $ are the volumes of the particles.

Similarly to the stationary case, we can apply the point particle limit for the considered nonstationary scenario [see Fig.~\ref{fig:System}(b)]: The nonstationary flux in Eq.~\eqref{eq:Flux} simplifies to
\begin{widetext}
\begin{align}
\notag & \Phi_1^{(2)}(t) = \frac{\partial}{\partial t}V_2\frac{\hbar}{(2\pi)^2}\int_0^{\infty}\! d\omega \frac{\omega^2}{e^{\frac{\omega}{\omega_{T_1}}}-1}\Im[\alpha_1(\omega)]\int_0^{\infty}\! dt' \int_0^{\infty}\! dt''\cos[\omega(t'-t'')]\Tr\left\{\mathbb{G}_{\overline{12}}^{\textrm{H}}(\vct{r}_1,\vct{r}_2;t-t')\mathbb{G}_{\overline{12}}^{\textrm{H}T}(\vct{r}_1,\vct{r}_2;t-t'')\right\}\\
\notag & \ \ \ \ \ \ \ \ \ + \frac{\partial}{\partial t}V_2\frac{\hbar}{(2\pi)^2}\int_0^{\infty}\! d\omega \frac{\omega^2}{e^{\frac{\omega}{\omega_{T_1}}}-1}\Im[\alpha_1(\omega)]\int_0^{\infty}\! dt' \int_0^{\infty}\! dt''\cos[\omega(t'-t'')]\Tr\left\{\mathbb{G}_{\overline{12}}^{\textrm{E}}(\vct{r}_1,\vct{r}_2;t-t')\mathbb{G}_{\overline{12}}^{\textrm{E}T}(\vct{r}_1,\vct{r}_2;t-t'')\right\}\\
& +\frac{2\hbar}{\pi}\int_0^{\infty}\! d\omega \frac{\omega^2}{e^{\frac{\omega}{\omega_{T_1}}}-1}\Im[\alpha_1(\omega)]\int_0^{\infty}\! dt' \int_0^{\infty}\! dt''\cos[\omega(t'-t'')]\int\! d\tilde{t}\alpha_2(\tilde{t}) \Tr\left\{\frac{\partial\mathbb{G}_{\overline{12}}^{\textrm{E}}(\vct{r}_1,\vct{r}_2;t-\tilde{t}-t')}{\partial t}\mathbb{G}_{\overline{12}}^{\textrm{E}T}(\vct{r}_1,\vct{r}_2;t-t'')\right\},
\label{eq:FluxPP}
\end{align}
\end{widetext}
which is derived in Appendix~\ref{app:sec:FluxPP}. Note that Eq.~\eqref{eq:FluxPP} contains the polarizabilty of particle $ 2 $ in time domain, $ \alpha_2(\tilde{t}) $, i.e., the inverse Fourier transform of $ \alpha_2(\omega) $ in Eq.~\eqref{eq:polarizability}. In the corresponding stationary scenario, one can show, following the steps in Appendix~\ref{app:sec:FluxSt}, that Eq.~\eqref{eq:FluxPP} reduces to Eq.~\eqref{eq:FluxStPP}.

Interestingly, the first (magnetic) term of Eq.~\eqref{eq:FluxPP} contains no properties of particle $ 2 $ except its position and volume, i.e., in the point particle limit, the magnetic part of the energy radiated by particle $ 1 $ does not interact with the second particle, thus making no contribution to the dissipative part of the flux. Therefore, we can identify this term with the change of the magnetic energy produced by particle $ 1 $ within the volume occupied by particle $ 2 $, i.e., $ \frac{\partial}{\partial t}V_2u_{1\textrm{H}}^{(2)}(t) $, with the corresponding energy density
\begin{align}
\notag & u_{1\textrm{H}}^{(2)}(t) = \frac{\hbar}{(2\pi)^2}\int_0^{\infty}\! d\omega \frac{\omega^2}{e^{\frac{\omega}{\omega_{T_1}}}-1}\Im[\alpha_1(\omega)]\int_0^{\infty}\! dt' \int_0^{\infty}\! dt''\\
& \times\cos[\omega(t'-t'')]\Tr\left\{\mathbb{G}_{\overline{12}}^{\textrm{H}}(\vct{r}_1,\vct{r}_2;t-t')\mathbb{G}_{\overline{12}}^{\textrm{H}T}(\vct{r}_1,\vct{r}_2;t-t'')\right\}.
\label{eq:uH}
\end{align}
For large $ t $, similar to the flux in Eq.~\eqref{eq:LimStFluxSt}, $ u_{1\textrm{H}}^{(2)}(t) $ is expected to converge to the stationary energy density (see Appendix~\ref{app:sec:uHst} for the derivation),
\begin{align}
\notag u_{1\textrm{Hst}}^{(2)} = \ & \frac{4\hbar}{c^2}\int_0^{\infty}\! d\omega \frac{\omega^2}{e^{\frac{\omega}{\omega_{T_1}}}-1}\Im[\alpha_1(\omega)]\\
& \times \Tr\left\{\mathbb{G}_{\overline{12}}^{\textrm{H}}(\vct{r}_1,\vct{r}_2;\omega)\mathbb{G}_{\overline{12}}^{\textrm{H}\dagger}(\vct{r}_1,\vct{r}_2;\omega)\right\},
\label{eq:uHst}
\end{align}
such that its change (time derivative) is zero, thereby not contributing to the stationary heat flux, in agreement with Eq.~\eqref{eq:FluxStPP}. Since it is this change, and not the energy itself, which enters the heat flux, not much attention was paid to $ u_{1\textrm{Hst}}^{(2)} $ in the literature investigating the stationary heat transfer. Note that $ u_{1\textrm{Hst}}^{(2)} $ in Eq.~\eqref{eq:uHst} is manifestly nonnegative (see Appendix~\ref{app:sec:uHst}). It is also worth noticing that, according to Eqs.~\eqref{eq:FluxPT},~\eqref{eq:FluxPP}, and~\eqref{eq:uH}, $ \langle \vct{H}^2(\vct{r}_2,t) \rangle = 8\pi u_{1\textrm{H}}^{(2)}(t) $, i.e., in the point particle limit, the magnetic field amplitude is, up to prefactor $ 8\pi $, identical to the magnetic energy density. 

A similar identification of the energy density also applies for the second term of Eq.~\eqref{eq:FluxPP}, with the difference that this term does not represent the full electric energy but only the part which does not interact with particle $ 2 $, whereas the other part (which depends on the response of particle $ 2 $) is hidden in the third term of Eq.~\eqref{eq:FluxPP} [see Eq.~\eqref{eq:UdotPPVac} for the total energy change for two particles in vacuum]. Similar to Eqs.~\eqref{eq:uH} and~\eqref{eq:uHst}, we identify the energy density
\begin{align}
\notag & u_{1\textrm{E}0}^{(2)}(t) = \frac{\hbar}{(2\pi)^2}\int_0^{\infty}\! d\omega \frac{\omega^2}{e^{\frac{\omega}{\omega_{T_1}}}-1}\Im[\alpha_1(\omega)]\int_0^{\infty}\! dt' \int_0^{\infty}\! dt''\\
& \times\cos[\omega(t'-t'')]\Tr\left\{\mathbb{G}_{\overline{12}}^{\textrm{E}}(\vct{r}_1,\vct{r}_2;t-t')\mathbb{G}_{\overline{12}}^{\textrm{E}T}(\vct{r}_1,\vct{r}_2;t-t'')\right\}
\label{eq:uE0}
\end{align}
and its stationary analog (see Appendix~\ref{app:sec:uHst})
\begin{align}
\notag u_{1\textrm{E}0\textrm{st}}^{(2)} = \ & \frac{4\hbar}{c^4}\int_0^{\infty}\! d\omega \frac{\omega^4}{e^{\frac{\omega}{\omega_{T_1}}}-1}\Im[\alpha_1(\omega)]\\
& \times \Tr\left\{\mathbb{G}_{\overline{12}}^{\textrm{E}}(\vct{r}_1,\vct{r}_2;\omega)\mathbb{G}_{\overline{12}}^{\textrm{E}\dagger}(\vct{r}_1,\vct{r}_2;\omega)\right\},
\label{eq:uE0st}
\end{align}
where the subscript $ 0 $ indicates the aforementioned noninteracting part. $ u_{1\textrm{E}0\textrm{st}}^{(2)} $ in Eq.~\eqref{eq:uE0st} is manifestly nonnegative. Similar to the magnetic field, we notice that $ \langle \vct{E}^2(\vct{r}_2,t) \rangle = 8\pi u_{1\textrm{E}0}^{(2)}(t) $.

The third term of Eq.~\eqref{eq:FluxPP}, the interaction term, vanishes if $ \alpha_2 = 0 $ (i.e., no particle $ 2 $ is present).

\section{Example: nanoparticles in isolation}
\label{sec:HT_example}
\subsection{Free-space Green's functions and their traces}
\label{subsec:G0}
In this section, we study the nonstationary heat flux for two particles in vacuum, i.e., with no other objects present (indicated by the subscript \enquote{$ \textrm{vac} $} below), as depicted in Fig.~\ref{fig:System}(c). In this case, $ \mathbb{G}_{\overline{12}}^{\textrm{E(H)}} = \mathbb{G}_0^{\textrm{E(H)}} $, where $ \mathbb{G}_0^{\textrm{E(H)}} $ is the free-space electric (magnetic) GF, i.e., the GF of vacuum. These GFs are best known in frequency space~\cite{Asheichyk2017, Tai1994, Dong2017, Tsang2004, VanBladel1961, Yaghjian1980, Weiglhofer1989, Weiglhofer1993},
\begin{subequations}
\begin{align}
\notag & \mathbb{G}^{\textrm{E}}_0(\vct{r}, \vct{r}'; \omega) = -\frac{1}{3k^2}\mathcal{I}\delta^{(3)}(\vct{r}-\vct{r}')\\
\notag & \ \ \ \ \ \ \ \ \ \ \ \ \ \ \ \ \ \ + \frac{e^{ikd}}{4\pi k^2d^5}\big[d^2(-1+ikd+k^2d^2)\mathcal{I}\\
& \ \ \ \ \ \ \ \ \ \ \ \ \ \ \ \ \ \ + (3-3ikd-k^2d^2)(\vct{r}-\vct{r}')\otimes(\vct{r}-\vct{r}')\big],
\label{eq:G0Eomega}\\
& \mathbb{G}^{\textrm{H}}_0(\vct{r}, \vct{r}'; \omega) = \frac{e^{ikd}}{4\pi d^3}\left(-1+ikd\right)(\vct{r}-\vct{r'})\times\mathcal{I},
\label{eq:G0Homega}
\end{align}
\end{subequations}
where $ k = \omega/c $, $ d $ is the distance between the points $ \vct{r} $ and $ \vct{r}' $, $ \mathcal{I} $ is the $ 3 \times 3 $ identity matrix, and $ (\vct{r}-\vct{r'})\times\mathcal{I} $ means~\cite{Felsen1994}
\begin{equation}
(\vct{r}-\vct{r'})\times\mathcal{I}=
\begin{pmatrix}
0 & -(z-z') & (y-y')\\
(z-z') & 0 & -(x-x')\\
-(y-y') & (x-x') & 0
\end{pmatrix}.
\label{eq:rxI}
\end{equation}

The time-domain GFs can be obtained from Eqs~\eqref{eq:G0Eomega} and~\eqref{eq:G0Homega} by performing the inverse Fourier transform according to Eqs.~\eqref{eq:GEtviaGEomega} and~\eqref{eq:GHtviaGHomega}\footnote{For the electric GF, the integrand in Eq.~\eqref{eq:GEtviaGEomega} contains a pole at $ \omega = 0 $, which has to be shifted to the lower complex plain in order to obtain the retarded GF.}. We get~\cite{Felsen1994, Nevels2004}
\begin{subequations}
\begin{align}
\notag & \mathbb{G}^{\textrm{E}}_0(\vct{r},\vct{r}';t-t') = -\frac{4\pi}{3}\mathcal{I}\delta^{(3)}(\vct{r}-\vct{r}')\theta(t-t')\\
\notag & + \frac{1}{d^5}\left[-d^2\mathcal{I} + 3(\vct{r}-\vct{r}')\otimes(\vct{r}-\vct{r}')\right]\theta\left(t-t'-\frac{d}{c}\right)\\
\notag & + \frac{1}{cd^4}\left[-d^2\mathcal{I} + 3(\vct{r}-\vct{r}')\otimes(\vct{r}-\vct{r}')\right]\delta\left(t-t'-\frac{d}{c}\right)\\
& + \frac{1}{c^2d^3}\left[-d^2\mathcal{I} + (\vct{r}-\vct{r}')\otimes(\vct{r}-\vct{r}')\right]\delta'\left(t-t'-\frac{d}{c}\right),
\label{eq:G0Et}\\
\notag & \mathbb{G}^{\textrm{H}}_0(\vct{r},\vct{r}';t-t') = -\frac{1}{cd^3}\delta\left(t-t'-\frac{d}{c}\right)(\vct{r}-\vct{r'})\times\mathcal{I}\\
& - \frac{1}{c^2d^2}\delta'\left(t-t'-\frac{d}{c}\right)(\vct{r}-\vct{r'})\times\mathcal{I},
\label{eq:G0Ht}
\end{align}
\end{subequations}
where $ \theta $ is the Heaviside step function and $ \delta' $ means the derivative of delta function with respect to its full argument. The causality principle is manifest in Eqs.~\eqref{eq:G0Et} and~\eqref{eq:G0Ht}: The GFs are zero if the difference between the observation and source times is smaller than the propagation time $ d/c $; the first term of $ \mathbb{G}^{\textrm{E}}_0 $ corresponds to the field in the source region, and hence to the instantaneous propagation. The magnetic GF and the last two terms of the electric GF, being proportional to the delta function and its derivative, lead to the fields separated in time from the source by strictly $ d/c $ (e.g., the pulse of the current creates the pulse of the field after time $ d/c $). In contrast, the second (near-field) term of the electric GF leads to the field for all times elapsed from the source time plus $ d/c $. This indicates a fundamental difference in the wave dynamics between the near- and far-field regimes.

Using the GFs in Eqs.~\eqref{eq:G0Eomega} and \eqref{eq:G0Homega}, we compute the traces required for the stationary observables,
\begin{subequations}
\begin{align}
\notag & \Tr\left\{\mathbb{G}_0^{\textrm{E}}(\vct{r}_1,\vct{r}_2;\omega)\mathbb{G}_0^{\textrm{E}\dagger}(\vct{r}_1,\vct{r}_2,\omega)\right\}\\
& \ \ \ \ \ \ \ \ \ \ \ \ \ \ \ \ \ \ \ \ \ \ =\frac{1}{8\pi^2 d^2}\left[1+\frac{1}{k^2d^2}+\frac{3}{k^4d^4}\right],
\label{eq:TrGEGEdaggeromega}\\
& \Tr\left\{\mathbb{G}_0^{\textrm{H}}(\vct{r}_1,\vct{r}_2;\omega)\mathbb{G}_0^{\textrm{H}\dagger}(\vct{r}_1,\vct{r}_2,\omega)\right\} = \frac{k^2}{8\pi^2 d^2}\left[1+\frac{1}{k^2d^2}\right],
\label{eq:TrGHGHdaggeromega}
\end{align}
\end{subequations}
where $ d = |\vct{r}_1 - \vct{r}_2| > 0 $ is the distance between the particles. Note the absence of the near-field term in Eq.~\eqref{eq:TrGHGHdaggeromega}. The traces for the nonstationary observables are computed using Eqs.~\eqref{eq:G0Et} and~\eqref{eq:G0Ht},
\begin{subequations}
\begin{align}
\notag & \Tr\left\{\mathbb{G}_0^{\textrm{E}}(\vct{r}_1,\vct{r}_2;t-t')\mathbb{G}_0^{\textrm{E}T}(\vct{r}_1,\vct{r}_2;t-t'')\right\}\\
\notag & =\frac{2}{d^2}\Bigg\{\frac{1}{c^4}\delta'(\tau-t')\delta'(\tau-t'')\\
\notag & \ \ \ \ \ \ \ \ \ +\frac{1}{c^3d}\left[\delta(\tau-t')\delta'(\tau-t'')+\delta'(\tau-t')\delta(\tau-t'')\right]\\
\notag & \ \ \ \ \ \ \ \ \ +\frac{1}{c^2d^2}\big[\theta(\tau-t')\delta'(\tau-t'')+3\delta(\tau-t')\delta(\tau-t'')\\
\notag & \ \ \ \ \ \ \ \ \ \ \ \ \ \ \ \ \ \ \ \ +\delta'(\tau-t')\theta(\tau-t'')\big]\\
\notag & \ \ \ \ \ \ \ \ \ +\frac{3}{cd^3}\left[\theta(\tau-t')\delta(\tau-t'')+\delta(\tau-t')\theta(\tau-t'')\right]\\
& \ \ \ \ \ \ \ \ \ +\frac{3}{d^4}\theta(\tau-t')\theta(\tau-t'')\Bigg\},
\label{eq:TrGEGETt}\\
\notag & \Tr\left\{\frac{\partial\mathbb{G}_0^{\textrm{E}}(\vct{r}_1,\vct{r}_2;t-\tilde{t}-t')}{\partial t}\mathbb{G}_0^{\textrm{E}T}(\vct{r}_1,\vct{r}_2;t-t'')\right\}\\
\notag & =\frac{2}{d^2}\Bigg\{\frac{1}{c^4}\delta''(\tau-\tilde{t}-t')\delta'(\tau-t'')\\
\notag & +\frac{1}{c^3d}\left[\delta'(\tau-\tilde{t}-t')\delta'(\tau-t'')+\delta''(\tau-\tilde{t}-t')\delta(\tau-t'')\right]\\
\notag & +\frac{1}{c^2d^2}\big[\delta(\tau-\tilde{t}-t')\delta'(\tau-t'')+3\delta'(\tau-\tilde{t}-t')\delta(\tau-t'')\\
\notag & \ \ \ \ \ \ \ \ \ \ \ +\delta''(\tau-\tilde{t}-t')\theta(\tau-t'')\big]\\
\notag & +\frac{3}{cd^3}\left[\delta(\tau-\tilde{t}-t')\delta(\tau-t'')+\delta'(\tau-\tilde{t}-t')\theta(\tau-t'')\right]\\
& +\frac{3}{d^4}\delta(\tau-\tilde{t}-t')\theta(\tau-t'')\Bigg\},
\label{eq:TrGEprimeGETt}\\
\notag & \Tr\left\{\mathbb{G}_0^{\textrm{H}}(\vct{r}_1,\vct{r}_2;t-t')\mathbb{G}_0^{\textrm{H}T}(\vct{r}_1,\vct{r}_2;t-t'')\right\}\\
\notag & =\frac{2}{d^2}\Bigg\{\frac{1}{c^4}\delta'(\tau-t')\delta'(\tau-t'')\\
\notag & \ \ \ \ \ \ \ \ \ +\frac{1}{c^3d}\left[\delta(\tau-t')\delta'(\tau-t'')+\delta'(\tau-t')\delta(\tau-t'')\right]\\
& \ \ \ \ \ \ \ \ \ +\frac{1}{c^2d^2}\delta(\tau-t')\delta(\tau-t'')\Bigg\},
\label{eq:TrGHGHTt}
\end{align}
\end{subequations}
where $ \tau \equiv t - d/c $. The time-dependent observables are thus functions of $ \tau $. Compared to the frequency-dependent traces, Eqs.~\eqref{eq:TrGEGETt}--\eqref{eq:TrGHGHTt} contain new power laws: $ d^{-3} $ and $ d^{-5} $ in Eq.~\eqref{eq:TrGEGETt} and~\eqref{eq:TrGEprimeGETt}, and $ d^{-3} $ in Eq.~\eqref{eq:TrGHGHTt}.

\subsection{Stationary flux and energy densities}
\label{subsec:FluxSt}
The stationary heat flux (transfer) is obtained by substituting trace~\eqref{eq:TrGEGEdaggeromega} into formula~\eqref{eq:FluxStPP},
\begin{align}
\notag & \Phi_{1\textrm{st,vac}}^{(2)} = H_{1\textrm{st,vac}}^{(2)} = \frac{4\hbar}{\pi c^4} \int_0^\infty\! d\omega \frac{\omega^5}{e^{\frac{\omega}{\omega_{T_1}}}-1}\\
& \times \Im[\alpha_1(\omega)]\Im[\alpha_2(\omega)]\left[\frac{1}{d^2}+\frac{c^2}{\omega^2d^4}+\frac{3c^4}{\omega^4d^6}\right],
\label{eq:FluxStPPvac}
\end{align}
in agreement with the literature \{see, e.g., Eq.~(62) in Ref.~\cite{Volokitin2001}, Eq.~(9) in Ref.~\cite{Chapuis2008}, Eq.~(137) in Ref.~\cite{Kruger2012}, and Eq.~(23) in Ref.~\cite{Asheichyk2017}\}.

Substituting traces~\eqref{eq:TrGHGHdaggeromega} and~\eqref{eq:TrGEGEdaggeromega} into Eqs.~\eqref{eq:uHst} and~\eqref{eq:uE0st}, respectively, we get for the stationary energy densities
\begin{subequations}
\begin{align}
\notag u_{1\textrm{Hst,vac}}^{(2)} = \ & \frac{\hbar}{2\pi^2 c^4}\int_0^{\infty}\! d\omega \frac{\omega^4}{e^{\frac{\omega}{\omega_{T_1}}}-1}\Im[\alpha_1(\omega)]\\
& \times \left[\frac{1}{d^2}+\frac{c^2}{\omega^2d^4}\right],
\label{eq:uHstvac}\\
\notag u_{1\textrm{E}0\textrm{st,vac}}^{(2)} = \ & \frac{\hbar}{2\pi^2 c^4}\int_0^{\infty}\! d\omega \frac{\omega^4}{e^{\frac{\omega}{\omega_{T_1}}}-1}\Im[\alpha_1(\omega)]\\
& \times \left[\frac{1}{d^2}+\frac{c^2}{\omega^2d^4}+\frac{3c^4}{\omega^4d^6}\right],
\label{eq:uE0stvac}
\end{align}
\end{subequations}
where $ u_{1\textrm{E}0\textrm{st,vac}}^{(2)} $ is reminiscent of $ H_{1\textrm{st,vac}}^{(2)} $ in Eq.~\eqref{eq:FluxStPPvac}. Indeed, $ H_{1\textrm{st,vac}}^{(2)} $  differs from $ V_2u_{1\textrm{E}0\textrm{st,vac}}^{(2)} $ by $ 6\omega \Im[(\varepsilon_2-1)/(\varepsilon_2+2)] $ in the spectrum, indicating that the heat transfer is directly determined by how well the second particle is able to absorb the electric energy radiated by the first one.

\subsection{Nonstationary energy densities}
\label{subsec:uNonstat}
In order to compute the nonstationary energy densities in Eqs.~\eqref{eq:uH} and~\eqref{eq:uE0}, we have to perform the time integrals of the traces in Eqs.~\eqref{eq:TrGHGHTt} and~\eqref{eq:TrGEGETt}, respectively, multiplied by $ \cos[\omega(t'-t'')] $. We find that no energy is present at $ \vct{r}_2 $ for $ \tau \equiv t - d/c < 0 $, while for $ \tau > 0 $, we obtain
\begin{subequations}
\begin{align}
& u_{1\textrm{H,vac}}^{(2)} = u_{1\textrm{Hst,vac}}^{(2)},
\label{eq:uHvac}\\
\notag & u_{1\textrm{E}0\textrm{,vac}}^{(2)}(\tau) = \frac{\hbar}{2\pi^2 c^4}\int_0^{\infty}\! d\omega \frac{\omega^4}{e^{\frac{\omega}{\omega_{T_1}}}-1}\Im[\alpha_1(\omega)]\\
\notag & \times \Bigg\{\frac{1}{d^2}+\frac{c^2}{\omega^2d^4}\left[1+2\cos(\omega\tau)\right]+\frac{6c^3}{\omega^3d^5}\sin(\omega\tau)\\
& \ \ \ \ \ \ \ \ \ \ +\frac{6c^4}{\omega^4d^6}\left[1-\cos(\omega\tau)\right]\Bigg\},
\label{eq:uE0vac}
\end{align}
\end{subequations}
with $ u_{1\textrm{Hst,vac}}^{(2)} $ given by Eq.~\eqref{eq:uHstvac}. The magnetic energy density thus immediately becomes stationary once the waves reached the second particle. Therefore, $ \frac{\partial}{\partial t}V_2u_{1\textrm{H}}^{(2)} = 0 $ for all times (apart from a possible delta-function pulse at $ t = d/c $), such that the first term in Eq.~\eqref{eq:FluxPP} does not contribute to the nonstationary flux.

The time dependence of the electric energy density is a way more complicated, featuring an oscillatory behavior\footnote{We note that the near-field term of Eq.~\eqref{eq:uE0vac} converges not to its stationary value in Eq.~\eqref{eq:uE0stvac} but to twice this value as $ \tau \to \infty $, i.e., our requirement in Eq.~\eqref{eq:LimStFluxSt} does not hold for $ u_{1\textrm{E}0\textrm{,vac}}^{(2)}(t) $. For now, we were not able to find a physical interpretation or mathematical reason for this issue, which, however, does not appear in $ \Phi_{1\textrm{,vac}}^{(2)}(t) $, our main quantity of interest, because the latter includes not $ u_{1\textrm{E}0\textrm{,vac}}^{(2)}(t) $ but its time derivative.}. Compared to the stationary density in Eq.~\eqref{eq:uE0stvac}, an additional power law $ d^{-5} $ appears in Eq.~\eqref{eq:uE0vac}. While the term with this power law and the near-field term, $ \propto d^{-6} $, grow from zero once the waves reached particle $ 2 $, the other two terms, $ \propto d^{-2} $ and $ \propto d^{-4} $, jump to finite values, with the $ d^{-2} $ term reaching the steady state immediately. Performing the time derivative of Eq.~\eqref{eq:uE0vac}, we obtain the change of the electric energy within $ V_2 $ [which is the second term of Eq.~\eqref{eq:FluxPP}] for $ \tau > 0 $ [note that $ \frac{\partial}{\partial t}V_2u_{1\textrm{E}0\textrm{,vac}}^{(2)}(t) = \frac{\partial}{\partial \tau}V_2u_{1\textrm{E}0\textrm{,vac}}^{(2)}(\tau) $],
\begin{align}
\notag & \frac{\partial}{\partial \tau}V_2u_{1\textrm{E}0\textrm{,vac}}^{(2)}(\tau) = \frac{\hbar V_2}{\pi^2 c^2}\int_0^{\infty}\! d\omega \frac{\omega^3}{e^{\frac{\omega}{\omega_{T_1}}}-1}\Im[\alpha_1(\omega)]\\
& \times \left[-\frac{1}{d^4}\sin(\omega\tau)+\frac{3c}{\omega d^5}\cos(\omega\tau)+\frac{3c^2}{\omega^2 d^6}\sin(\omega\tau)\right].
\label{eq:ddtuE0vac}
\end{align}
For $ \tau < 0 $, $ \frac{\partial}{\partial \tau}V_2u_{1\textrm{E}0\textrm{,vac}}^{(2)}(\tau) = 0 $, and a possible delta-function pulse appears at $ \tau = 0 $. It can be shown numerically that Eq.~\eqref{eq:ddtuE0vac} approaches zero for large $ \tau $ (see Sec.~\ref{sec:HT_example}), and it hence does not contribute to the stationary flux, as expected.

\subsection{Drude-Lorentz model for polarizability}
\label{subsec:DL}
To proceed with the last term of Eq.~\eqref{eq:FluxPP}, we have to specify the polarizability of particle $ 2 $. We consider a common model for the dielectric function, the Drude-Lorentz one~\cite{Bohren2004},
\begin{equation}
\varepsilon_2(\omega) = \varepsilon_{\infty} - \frac{\omega_{\textrm{p}}^2}{\omega^2-\omega_0^2+i\gamma\omega},
\label{eq:epsilonDL}
\end{equation}
where $ \omega_{\textrm{p}} $ is the plasma frequency, $ \omega_0 $ is the resonance frequency, $ \gamma $ is the damping rate, and $ \varepsilon_{\infty} $ is the high-frequency dielectric constant. The resulting polarizability, Eq.~\eqref{eq:polarizability}, can be written as
\begin{equation}
\alpha_2(\omega) = \alpha_{\infty}\frac{\left[\omega-\left(\widetilde{\beta}-i\frac{\gamma}{2}\right)\right]\left[\omega-\left(-\widetilde{\beta}-i\frac{\gamma}{2}\right)\right]}{\left[\omega-\left(\beta-i\frac{\gamma}{2}\right)\right]\left[\omega-\left(-\beta-i\frac{\gamma}{2}\right)\right]},
\label{eq:alphaDL}
\end{equation}
where $ \widetilde{\beta} = \sqrt{\omega_0^2+\omega_{\textrm{p}}^2/(\varepsilon_{\infty}-1)-\gamma^2/4} $ and $ \beta = \sqrt{\omega_0^2+\omega_{\textrm{p}}^2/(\varepsilon_{\infty}+2)-\gamma^2/4} $, both assumed to be real and positive; $ \alpha_{\infty} = (\varepsilon_{\infty}-1)/(\varepsilon_{\infty}+2)R_2^3 $. The first two terms inside the square root of $ \beta $ are identified with the resonance frequency of $ \alpha_2 $,
\begin{equation}
\omega_{0\alpha} = \sqrt{\omega_0^2+\frac{\omega_{\textrm{p}}^2}{(\varepsilon_{\infty}+2)}}.
\label{eq:omega0alpha}
\end{equation}
For the heat flux computation, it is useful to identify the real and imaginary parts of $ \alpha_2 $,
\begin{subequations}
\begin{align}
& \Re[\alpha_2(\omega)] = \alpha_{\infty} - \frac{3\omega_{\textrm{p}}^2R_2^3}{(\varepsilon_{\infty}+2)^2}\frac{\omega^2-\omega_{0\alpha}^2}{\left(\omega^2-\omega_{0\alpha}^2\right)^2+\gamma^2\omega^2},
\label{eq:alphaDLRe}\\
& \Im[\alpha_2(\omega)] = \frac{3\omega_{\textrm{p}}^2R_2^3}{(\varepsilon_{\infty}+2)^2}\frac{\gamma\omega}{\left(\omega^2-\omega_{0\alpha}^2\right)^2+\gamma^2\omega^2}.
\label{eq:alphaDLIm}
\end{align}
\end{subequations}

The time-dependent polarizability is given by the inverse Fourier transform of $ \alpha_2(\omega) $ in Eq.~\eqref{eq:alphaDL}, which can be evaluated using the residue theorem (similar to the time-dependent dielectric function in Ref.~\cite{Jackson1998}). Assuming $ \gamma > 0 $, we obtain
\begin{equation}
\alpha_2(t) = \alpha_{\infty}\delta(t) + \frac{3\omega_{\textrm{p}}^2R_2^3}{(\varepsilon_{\infty}+2)^2\beta}e^{-\frac{\gamma}{2}t}\sin(\beta t)\theta(t),
\label{eq:alphat}
\end{equation}
which is manifestly causal\footnote{For a lossless material ($ \gamma = 0 $), the polarizability is not causal and cannot be obtained as $ \gamma \to 0 $ limit of Eq.~\eqref{eq:alphat}~\cite{Jackson1998, Novotny2012}: It is given by Eq.~\eqref{eq:alphat}, but with the theta function replaced by the sign function and the exponent replaced by $ 1 $.} and reminiscent of Eq.~(7.110) in Ref.~\cite{Jackson1998}. The first term in Eq.~\eqref{eq:alphat} corresponds to the instantaneous response to the incident field, whereas the second term describes the relaxation after the interaction with the field.

\subsection{Nonstationary flux}
\label{subsec:FluxNonstat}
We have now all the quantities to compute the nonstationary heat flux in Eq.~\eqref{eq:FluxPP}. As discussed in Sec.~\ref{subsec:uNonstat}, its first term is zero, whereas the second one is given by Eq.~\eqref{eq:ddtuE0vac}. As for the third term, it is straightforward to show [using Eq.~\eqref{eq:derivtive_identity}] that the delta-function part of the polarizability in Eq.~\eqref{eq:alphat} gives the same contribution as Eq.~\eqref{eq:ddtuE0vac} multiplied by $ 3(\varepsilon_{\infty}-1)/(\varepsilon_{\infty}+2) $. 

The second term of polarizability~\eqref{eq:alphat} requires a more tedious evaluation of the time integrals in the third term of Eq.~\eqref{eq:FluxPP}. As detailed in Appendix~\ref{app:sec:FluxPPVacInt}, this evaluation leads to expressions proportional to the real or imaginary part of $ \alpha_2(\omega) $. Using the results of Appendix~\ref{app:sec:FluxPPVacInt}, we give here the final result for the total heat flux, which can be decomposed into two parts,
\begin{equation}
\Phi_{1\textrm{,vac}}^{(2)}(\tau) = \dot{U}_{1\textrm{E,vac}}^{(2)}(\tau) + H_{1\textrm{,vac}}^{(2)}(\tau),
\label{eq:FluxPPVac}
\end{equation}
where $ \dot{U}_{1\textrm{E,vac}}^{(2)}(\tau) $ is the change of the electric energy which is transported to particle $ 2 $ but not dissipated inside it, and $ H_{1\textrm{,vac}}^{(2)}(\tau) $ is the dissipative heat transfer rate, both finite only for $ \tau \equiv t - d/c > 0 $. For $ \tau < 0 $, $ \Phi_{1\textrm{,vac}}^{(2)}(\tau) = \dot{U}_{1\textrm{E,vac}}^{(2)}(\tau) = H_{1\textrm{,vac}}^{(2)}(\tau) = 0 $, i.e., the field first reaches particle $ 2 $ only after time $ t = d/c $ elapsed, in agreement with the retardation of EM field. 

For the energy change, we have ($ \tau >0 $)
\begin{widetext}
\begin{align}
\notag & \dot{U}_{1\textrm{E,vac}}^{(2)}(\tau) = \frac{\partial}{\partial \tau}V_2u_{1\textrm{E}0\textrm{,vac}}^{(2)}(\tau) +4\pi\alpha_{\infty}\frac{\partial}{\partial \tau}u_{1\textrm{E}0\textrm{,vac}}^{(2)}(\tau) + \frac{4\hbar}{\pi c^2}\int_0^{\infty}\! d\omega \frac{\omega^3}{e^{\frac{\omega}{\omega_{T_1}}}-1}\Im[\alpha_1(\omega)]\left(\Re[\alpha_2(\omega)]-\alpha_{\infty}\right)\\
\notag & \times\left[-\frac{1}{d^4}\sin(\omega\tau)+\frac{3c}{\omega d^5}\cos(\omega\tau)+\frac{3c^2}{\omega^2 d^6}\sin(\omega\tau)\right] +\frac{4\hbar}{\pi c^4}e^{-\frac{\gamma}{2}\tau}\int_0^{\infty}\! d\omega \frac{\omega^5}{e^{\frac{\omega}{\omega_{T_1}}}-1}\Im[\alpha_1(\omega)]\left(\Re[\alpha_2(\omega)]-\alpha_{\infty}\right)\\
\notag & \times\Bigg\{\frac{\omega_{0\alpha}^2}{\omega^2d^2}\left[-\cos(\beta\tau)\sin(\omega\tau)+\frac{\omega}{\beta}\sin(\beta\tau)\cos(\omega\tau)\right] - \frac{c}{\omega^3d^3}\left(\omega^2-\omega_{0\alpha}^2\right)\cos(\beta\tau)\cos(\omega\tau)\\
\notag & \ \ \ \ \ \ +\frac{c^2}{\omega^2 d^4}\left[-\frac{2\omega^2-\omega_{0\alpha}^2}{\omega^2}\cos(\beta\tau)\sin(\omega\tau)+\frac{\omega}{\beta}\sin(\beta\tau)\left(\frac{\omega_{0\alpha}^2}{\omega^2}-\frac{\omega^2-2\omega_{0\alpha}^2}{\omega^2}\cos(\omega\tau)\right)\right]\\
& \ \ \ \ \ \ -\frac{3c^3}{\omega^3d^5}\left[\cos(\beta\tau)+\frac{\omega^2-\omega_{0\alpha}^2}{\beta\omega}\sin(\beta\tau)\sin(\omega\tau)\right] +\frac{3c^4}{\omega^4d^6}\left[-\cos(\beta\tau)\sin(\omega\tau)-\frac{\omega}{\beta}\sin(\beta\tau)\left[1-\cos(\omega\tau)\right]\right]\!\!\Bigg\},
\label{eq:UdotPPVac}
\end{align}
\end{widetext}
where the first term, given by Eq.~\eqref{eq:ddtuE0vac}, contains no properties of particle $ 2 $ except its position and volume, thus remaining finite in the absence of the particle and corresponding to the energy change inside this empty volume. The rest depends on the particle dielectric properties and vanishes if the particle is absent, but it remains if the damping rate $ \gamma $ of the particle material approaches zero, indicating a nondissipative nature of $ \dot{U}_{1\textrm{E,vac}}^{(2)}(\tau) $. The second term of Eq.~\eqref{eq:UdotPPVac}, $ \propto \alpha_{\infty} $, is present even for a nondispersive material [i.e., $ \varepsilon_2 $ in Eq.~\eqref{eq:epsilonDL} has no frequency dependence, or equivalently, $ \omega_{\textrm{p}} = 0 $], while the third and the last terms require dispersion.
All terms of Eq.~\eqref{eq:UdotPPVac} vanish as $ t \to \infty $ (equivalently, $ \tau \to \infty $), but different terms approach zero differently. The spectra of the first three ones oscillate forever, and only the integration over frequency leads to a decay (whose form is studied numerically in Sec.~\ref{subsec:FluxPPSiC}), while the spectrum of the last term oscillates and decays exponentially with the relaxation time $ 2/\gamma $.

The heat transfer reads as ($ \tau >0 $)
\begin{widetext}
\begin{align}
\notag & H_{1\textrm{,vac}}^{(2)}(\tau) = H_{1\textrm{st,vac}}^{(2)} + \frac{4\hbar}{\pi c^2}\int_0^{\infty}\! d\omega \frac{\omega^3}{e^{\frac{\omega}{\omega_{T_1}}}-1}\Im[\alpha_1(\omega)]\Im[\alpha_2(\omega)]\!\left[\frac{1}{d^4}\cos(\omega\tau)+\frac{3c}{\omega d^5}\sin(\omega\tau)-\frac{3c^2}{\omega^2 d^6}\cos(\omega\tau)\right]\\
\notag & + \frac{4\hbar}{\pi c^4}e^{-\frac{\gamma}{2}\tau}\int_0^{\infty}\! d\omega \frac{\omega^5}{e^{\frac{\omega}{\omega_{T_1}}}-1}\Im[\alpha_1(\omega)]\Im[\alpha_2(\omega)]\Bigg\{\frac{1}{d^2}\Bigg[-\cos(\beta\tau)\cos(\omega\tau) -\frac{\omega_{0\alpha}^2\left(3\omega^2-\omega_{0\alpha}^2\right)}{2\beta\omega^3}\sin(\beta\tau)\sin(\omega\tau)\\
\notag & -\frac{\gamma}{\omega}\left(\cos(\beta\tau)\sin(\omega\tau)-\frac{\omega}{2\beta}\sin(\beta\tau)\cos(\omega\tau)\right)+\frac{\gamma^2}{2\beta\omega}\sin(\beta\tau)\sin(\omega\tau)\Bigg] +\frac{c}{\omega d^3}\Bigg[-\frac{\left(\omega^2-\omega_{0\alpha}^2\right)^2}{2\beta\omega^3}\sin(\beta\tau)\cos(\omega\tau)\\
\notag & +\frac{\gamma}{\omega}\cos(\beta\tau)\cos(\omega\tau) -\frac{\gamma^2}{2\beta\omega}\sin(\beta\tau)\cos(\omega\tau)\Bigg] +\frac{c^2}{\omega^2d^4}\Bigg[-\cos(\beta\tau)\left[1+\cos(\omega\tau)\right] -\frac{2\omega^4-\omega_{0\alpha}^2\omega^2+\omega_{0\alpha}^4}{2\beta\omega^3}\sin(\beta\tau)\sin(\omega\tau)\\
\notag & +\frac{\gamma}{\omega}\left(\cos(\beta\tau)\sin(\omega\tau) + \frac{\omega}{2\beta}\sin(\beta\tau)\left[1+3\cos(\omega\tau)\right]\right) -\frac{\gamma^2}{2\beta\omega}\sin(\beta\tau)\sin(\omega\tau)\Bigg] +\frac{3c^3}{\omega^3d^5}\Bigg[-\frac{\omega^2+\omega_{0\alpha}^2}{2\beta\omega}\sin(\beta\tau)\\
& +\frac{\gamma}{\beta}\sin(\beta\tau)\sin(\omega\tau)\Bigg] +\frac{3c^4}{\omega^4d^6}\Bigg[\cos(\beta\tau)\left[1-\cos(\omega\tau)\right]-\frac{\omega^2+\omega_{0\alpha}^2}{2\beta\omega}\sin(\beta\tau)\sin(\omega\tau) + \frac{\gamma}{2\beta}\sin(\beta\tau)\left[1-\cos(\omega\tau)\right]\Bigg]\Bigg\},
\label{eq:HPPVac}
\end{align}
\end{widetext}
where the first term is the stationary heat transfer given by Eq.~\eqref{eq:FluxStPPvac}; the other terms vanish as $ \tau \to \infty $ (in different ways, as discussed in the previous paragraph). All the terms go to zero as $ \gamma \to 0 $, indicating a dissipative nature of $ H_{1\textrm{,vac}}^{(2)}(\tau) $. In contrast to the stationary heat transfer in Eq.~\eqref{eq:FluxStPPvac}, which contains $ d^{-2} $, $ d^{-4} $, and $ d^{-6} $ power laws, Eqs.~\eqref{eq:UdotPPVac} and~\eqref{eq:HPPVac} also have terms $ \propto d^{-3} $ and $ \propto d^{-5} $.

Note that the dielectric response of particle $ 1 $ in Eqs.~\eqref{eq:UdotPPVac} and~\eqref{eq:HPPVac} does not necessarily have to be described by the Drude-Lorentz model discussed in Sec.~\ref{subsec:DL}, i.e., $ \Im[\alpha_1(\omega)] $ can differ from $ \Im[\alpha_2(\omega)] $ in Eq.~\eqref{eq:alphaDLIm}. However, in the numerical example below, we consider the two polarizabilities to be equal.

\subsection{Numerical example for silicon carbide particles}
\label{subsec:FluxPPSiC}
\subsubsection{Parameters}
\label{subsubsec:Parameters}
Let us finally study the nonstationary heat flux for two silicon carbide (SiC) particles in vacuum, where we numerically compute the remaining integrals over frequency in Eqs.~\eqref{eq:UdotPPVac} and~\eqref{eq:HPPVac}. The permittivity of SiC is given by Eq.~\eqref{eq:epsilonDL} [$ \varepsilon_1 = \varepsilon_2 $], with $ \varepsilon_\infty=6.7 $, $ \omega_0=1.49\times10^{14} \ \textrm{rad} \ \textrm{s}^{-1} $, $ \omega_{\textrm{p}}=2.71\times10^{14} \ \textrm{rad} \ \textrm{s}^{-1} $, and $ \gamma=8.93\times10^{11} \ \textrm{rad} \ \textrm{s}^{-1} $~\cite{Spitzer1959}. The corresponding polarizability is given by Eq.~\eqref{eq:polarizability}, or equivalently by Eq.~\eqref{eq:alphaDL} [$ \alpha_1 = \alpha_2 $], with the resonance frequency $ \omega_{0\alpha}=1.75\times10^{14} \ \textrm{rad} \ \textrm{s}^{-1} $. The real and imaginary parts of $ \alpha_i(\omega) $ are expressed in Eqs.~\eqref{eq:alphaDLRe} and~\eqref{eq:alphaDLIm}, respectively. Since $ \gamma \ll \omega_{0\alpha} $, $ \beta \approx \omega_{0\alpha} $.

We consider two temperatures of particle $ 1 $, $ T_1 = 300 \ \textrm{K} $ and $ T_1 = 30 \ \textrm{K} $, with the corresponding thermal frequencies $ \omega_{T_1} = 3.93\times10^{13} \ \textrm{rad} \ \textrm{s}^{-1} $ and $ \omega_{T_1} = 3.93\times10^{12} \ \textrm{rad} \ \textrm{s}^{-1} $, and thermal wavelengths $ \lambda_{T_1} = 7.63 \ \mu\textrm{m} $ and $ \lambda_{T_1} = 76.3 \ \mu\textrm{m} $, respectively.

\subsubsection{Stationary heat transfer}
\label{subsubsec:StatH}
\begin{figure}[!b]
\begin{center}
\includegraphics[width=1.0\linewidth]{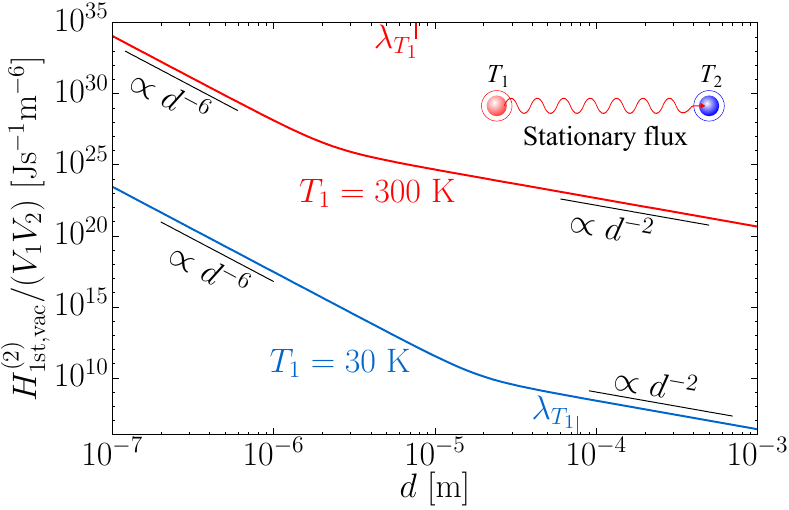}
\end{center}
\caption{\label{fig:Stat}Stationary radiative heat transfer from SiC particle $ 1 $ to SiC particle $ 2 $ in vacuum, as a function of the interparticle distance, normalized by the particle volumes. The results are given for two different temperatures $ T_1 $ of particle $ 1 $ (the corresponding thermal wavelengths $ \lambda_{T_1} $ are indicated on the horizontal axes).}
\end{figure}

\begin{figure*}[!t]
\begin{center}
\includegraphics[width=1.0\linewidth]{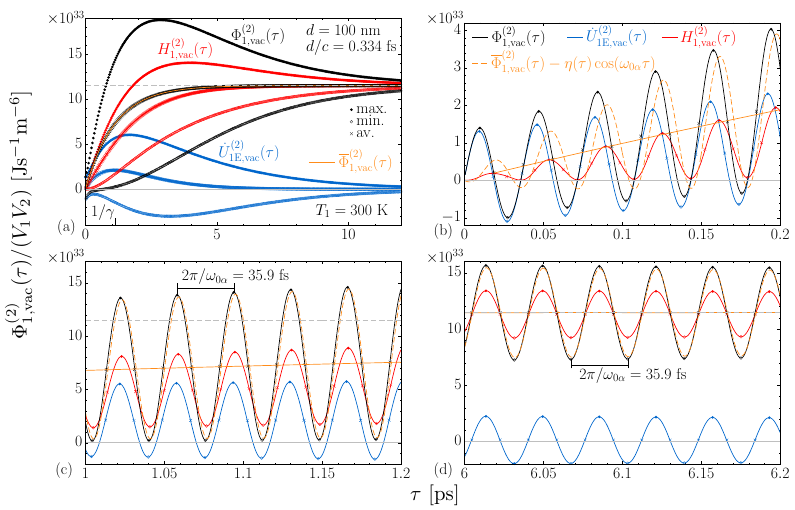}
\end{center}
\caption{\label{fig:T300nf}Time dependence of the normalized radiative heat flux from SiC particle $ 1 $ to SiC particle $ 2 $ after the former starts its radiation at time $ t = 0 $ with temperature $ T_1 = 300 \ \textrm{K} $ (the corresponding thermal wavelength is $ \lambda_{T_1} = 7.63 \ \mu\textrm{m} $). The results are given as functions of $ \tau \equiv t - d/c > 0 $, where $ d = 100 \ \textrm{nm} $ is the interparticle distance and $ c $ is the speed of light in vacuum. Here, $ d \ll \lambda_{T_1} $, meaning that the near-field flux is dominant. The full flux, nondissipative energy change, and heat transfer are given by the black, blue, and red symbols and lines, respectively. While these quantities oscillate with a small period [as shown in (b), (c), and (d)], in (a), we plot only their maxima, minima, and averages to demonstrate the behavior on a larger time scale. The gray dashed line is the stationary heat flux (transfer), while the gray solid line at a value of $ 0 $ is included as a guide to the eye.}
\end{figure*}

We first revisit the stationary heat flux (transfer) given by Eq.~\eqref{eq:FluxStPPvac}. Since it is proportional to $ V_1V_2 $, we compute the flux normalized by the particle volumes (the same applies for the nonstationary flux below). The exponent in the heat transfer spectrum cuts off frequencies (wavelengths) being large (small) compared to the thermal frequency (wavelength). At $ T_1 = 300 \ \textrm{K} $ (room temperature), the resonance frequency $ \omega_{0\alpha} $ lies within the range of relevant frequencies, such that the spectrum is strongly peaked at $ \omega_{0\alpha} $, with the corresponding dominant wavelength being close to $ \lambda_{T_1} $. When the temperature decreases down to $ 30 \ \textrm{K} $, the dominant frequencies (wavelengths), being concentrated around $ \omega_{T_1} $ ($ \lambda_{T_1} $), are much smaller (larger) than $ \omega_{0\alpha} $ (the corresponding wavelength).

Figure~\ref{fig:Stat} shows the flux as a function of the interparticle distance $ d $ for the two temperatures specified above. One can distinguish near-field ($ H_{1\textrm{st,vac}}^{(2)} \propto d^{-6} $) and far-field ($ H_{1\textrm{st,vac}}^{(2)} \propto d^{-2} $) regimes. The former is present for $ d \ll \lambda_{T_1} $, thereby lasting longer in $ d $ for $ T_1 = 30 \ \textrm{K} $ than for $ T_1 = 300 \ \textrm{K} $, whereas the latter appears for $ d > \lambda_{T_1} $, and hence starts at a larger distance for the smaller temperature. For all distances, the heat transfer for $ T_1 = 300 \ \textrm{K} $ is more than $ 10 $ orders of magnitude larger than that for $ T_1 = 30 \ \textrm{K} $, with the ratio increasing as $ d $ increases.

\subsubsection{Nonstationary heat flux at room temperature}
\label{subsubsec:NonstatT300}
\begin{figure*}[!t]
\begin{center}
\includegraphics[width=1.0\linewidth]{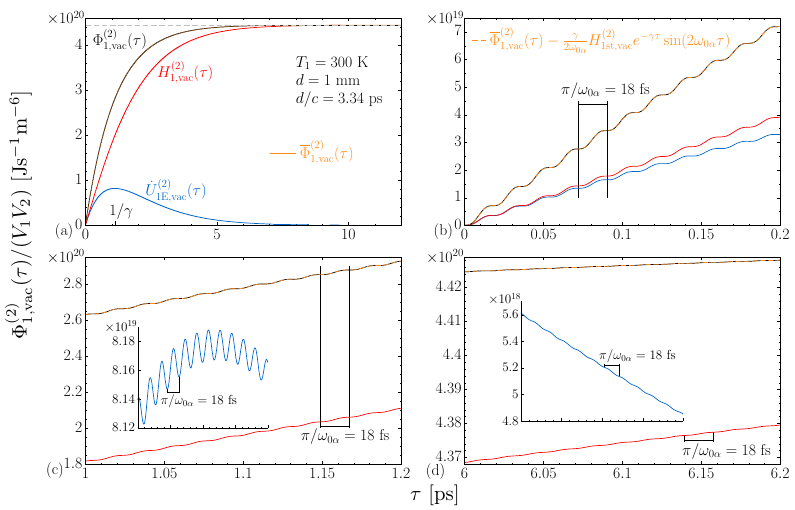}
\end{center}
\caption{\label{fig:T300ff}The same as Fig.~\ref{fig:T300nf}, but for $ d = 1 \ \textrm{mm} $ (far field). In contrast to the near-field flux in Fig.~\ref{fig:T300nf}, here, the oscillatory (or step-like) behavior [(b), (c), and (d)] is barely significant on the scale of the flux relaxation [(a)].}
\end{figure*}

Consider now the nonstationary scenario, where particle $ 1 $ starts its radiation at $ t = 0 $ with temperature $ T_1 = 300 \ \textrm{K} $. First, we separate the particles by the near-field distance $ d = 100 \ \textrm{nm} $. As discussed in Sec.~\ref{subsec:FluxNonstat}, partice $ 2 $ starts receiving the energy after time $ d/c = 0.334 \ \textrm{fs} $ elapsed. Figure~\ref{fig:T300nf} shows the time dependence of the heat flux after this time, i.e., as a function of $ \tau \equiv t - d/c > 0 $. Both the nondissipative energy change and heat transfer grow from positive values and then oscillate with the frequency given by the dominant frequency of the flux spectrum $ \omega_{0\alpha} $. For small $ \tau $, as shown in Fig.~\ref{fig:T300nf}(b), there is a phase shift [approximately $ \pi/(2\omega_{0\alpha}) $] between $ \dot{U}_{1\textrm{E,vac}}^{(2)}(\tau) $ or $ \Phi_{1\textrm{,vac}}^{(2)}(\tau) $ and $ H_{1\textrm{,vac}}^{(2)}(\tau) $, but it diminishes as $ \tau $ grows, because the two former quantities oscillate with a larger average\footnote{In contrast to $ H_{1\textrm{,vac}}^{(2)}(\tau) $ (which is periodic with the period $ 2\pi/\omega_{0\alpha} $ for all $ \tau $), no strict periodicity is observed for $ \dot{U}_{1\textrm{E,vac}}^{(2)}(\tau) $ and $ \Phi_{1\textrm{,vac}}^{(2)}(\tau) $ when $ \tau < 1/\gamma $ (the distances between the maxima are significantly larger than or equal to $ 2\pi/\omega_{0\alpha} $).} period than the latter one for small $ \tau $. When $ \tau $ becomes comparable or larger than $ 1/\gamma = 1.12 \ \textrm{ps} $ [Figs.~\ref{fig:T300nf}(c) and (d)], all the quantities oscillate almost in phase, with the the period $ 2\pi/\omega_{0\alpha} = 35.9 \ \textrm{fs} $ and the maxima appearing at $ \tau \approx \pi(2n-1)/\omega_{0\alpha} $ ($ n \in \mathbb{N} $).

To demonstrate the behavior of the flux on a large time scale, we plot the maxima and minima of the oscillations in Fig.~\ref{fig:T300nf}(a). In addition, we plot the oscillation averages, whose positions and values are estimated as $ (\tau_{\textrm{max}} + \tau_{\textrm{min}})/2 $ and $ [\Phi_{1\textrm{,vac}}^{(2)}(\tau_{\textrm{max}}) + \Phi_{1\textrm{,vac}}^{(2)}(\tau_{\textrm{min}})]/2 $, respectively, for a maximum at $ \tau_{\textrm{max}} $ and the subsequent (preceding) minimum at $ \tau_{\textrm{min}} $ (the same applies for the energy change and heat transfer). The maxima of $ \dot{U}_{1\textrm{E,vac}}^{(2)}(\tau) $ are positive and form a curve with a global maximum at $ \tau \approx 1.67 \ \textrm{ps} $, with its value being about twice smaller than the stationary flux. The curve for the averages shows a similar behavior, but with a smaller amplitude and the global maximum shifted to a smaller $ \tau $. The minima of $ \dot{U}_{1\textrm{E,vac}}^{(2)}(\tau) $ are negative and feature the global maximum and minimum. All the curves converge to zero for $ \tau \gg 1/\gamma $.

In contrast to $ \dot{U}_{1\textrm{E,vac}}^{(2)}(\tau) $, both maxima and minima of $ H_{1\textrm{,vac}}^{(2)}(\tau) $ are positive, i.e., the heat transfer is positive for any $ \tau $. The curve for the maxima features a global maximum at $ \tau \approx 4 \ \textrm{ps} $, with a value of $ 1.22H_{1\textrm{st,vac}}^{(2)} $, and approaches $ H_{1\textrm{st,vac}}^{(2)} $ for $ \tau \gg 1/\gamma$, whereas the curves for the minima and averages have no extrema and approach $ H_{1\textrm{st,vac}}^{(2)} $ from below.

The full flux, being the sum of the energy change and the heat transfer, has larger maxima, forming a curve studied in detail in Appendix~\ref{app:sec:FluxPPVacFormula}. The minima are negative for $ \tau \lesssim 1/\gamma $ and positive otherwise. Remarkably, the averages of $ \Phi_{1\textrm{,vac}}^{(2)}(\tau) $ form a curve which is accurately described by a simple exponential relaxation,
\begin{equation}
\overline{\Phi}_{1\textrm{,vac}}^{(2)}(\tau) = H_{1\textrm{st,vac}}^{(2)}\left(1-e^{-\gamma\tau}\right),
\label{eq:FluxAv}
\end{equation}
plotted as a solid orange line in Fig.~\ref{fig:T300nf}. This remains valid for any distance, i.e., only $ H_{1\textrm{st,vac}}^{(2)} $ changes with changing $ d $ [for the far-field $ d = 1 \ \textrm{mm} $, we demonstrate the accuracy of Eq.~\eqref{eq:FluxAv} in Fig.~\ref{fig:T300ff}(a)]. Note that, although looking similar, the average curve for $ H_{1\textrm{,vac}}^{(2)}(\tau) $ cannot be accurately described by a single exponent.

Taking formula~\eqref{eq:FluxAv} as a basis, we can approximate the near-field flux for any positive $ \tau $ by the following expression (see Appendix~\ref{app:sec:FluxPPVacFormula}):
\begin{equation}
\underset{d \ll \lambda_{T_1}}{\Phi_{1\textrm{,vac}}^{(2)}(\tau)} \approx \overline{\Phi}_{1\textrm{,vac}}^{(2)}(\tau) - \eta(\tau)\cos(\omega_{0\alpha}\tau),
\label{eq:FluxAnNF}
\end{equation}
where $ \eta(\tau) $, the amplitude of the oscillations, is given by Eq.~\eqref{eq:eta}. While being inaccurate for small $ \tau $ [see Fig.~\ref{fig:T300nf}(b)], formula~\eqref{eq:FluxAnNF} can be considered as a qualitatively good approximation for $ \tau > 1/\gamma $ [see Figs.~\ref{fig:T300nf}(c) and~\ref{fig:T300nf}(d)]. 

We note that an oscillatory time dependence with the correlation time of approximately $ 1/\gamma $ was also observed for the stationary autocorrelation function of the Poynting vector between two semi-infinite SiC plates~\cite{Biehs2018}.

When the interparticle distance is large compared to the thermal wavelength (far-field regime), the heat flux still oscillates in time, but with a much smaller relative oscillation amplitude, as shown in Fig.~\ref{fig:T300ff}. The oscillations form a step-like behavior, with a period $ \pi/\omega_{0\alpha} $, which is twice smaller than the period of the near-field oscillations; clear minima and maxima are observed only for $ \dot{U}_{1\textrm{E,vac}}^{(2)}(\tau) $ near its global maximum, as shown in the inset of Fig.~\ref{fig:T300ff}(c).

On a large time scale [Fig.~\ref{fig:T300ff}(a)], the oscillations are almost invisible. The aforementioned global maximum of $ \dot{U}_{1\textrm{E,vac}}^{(2)}(\tau) $ appears at $ \tau \approx 1/\gamma $; afterwards, $ \dot{U}_{1\textrm{E,vac}}^{(2)}(\tau) $ converges to zero (being positive for the plotted range of $ \tau $, but also taking negative values at larger times). $ H_{1\textrm{,vac}}^{(2)}(\tau) $ and $ \Phi_{1\textrm{,vac}}^{(2)}(\tau) $ relax monotonically (on average) to the stationary value. Figure~\ref{fig:T300ff}(a) confirms that the relaxation of $ \Phi_{1\textrm{,vac}}^{(2)}(\tau) $ is accurately described by Eq.~\eqref{eq:FluxAv}, which, because of the insignificance of the oscillations, can be considered as a good approximation on any time scale. Yet we found a formula describing these far-field oscillations,
\begin{equation}
\underset{d \gg \lambda_{T_1}}{\Phi_{1\textrm{,vac}}^{(2)}(\tau)} \approx \overline{\Phi}_{1\textrm{,vac}}^{(2)}(\tau) - \frac{\gamma}{2\omega_{0\alpha}}H_{1\textrm{st,vac}}^{(2)}\sin(2\omega_{0\alpha}\tau),
\label{eq:FluxAnFF}
\end{equation}
which is much more precise than its near-field analog in Eq.~\eqref{eq:FluxAnNF}: The relative error between Eq.~\eqref{eq:FluxAnFF} and the numerical results does not exceed $0.2\%$ for $ \tau > \pi/(2\omega_{0\alpha}) = 9 \ \textrm{fs} $ [see also the comparison in Figs.~\ref{fig:T300ff}(b)--~\ref{fig:T300ff}(d)]. Equation~\eqref{eq:FluxAnFF} is not applicable for extremely small $ \tau $, as it approaches zero in this case, whereas the exact flux does not.

Indeed, for $ \tau \to 0^+ $ (here, $ \tau $ is considered arbitrary small, but still positive), we find (for any $ T_1 $)
\begin{subequations}
\begin{align}
\notag & \lim_{\tau \to 0^+}\dot{U}_{1\textrm{E,vac}}^{(2)}(\tau) = -\frac{4\hbar}{\pi c^3d^3}\int_0^{\infty}\! d\omega \frac{\omega^2}{e^{\frac{\omega}{\omega_{T_1}}}-1}\Im[\alpha_1(\omega)]\\
\notag & \ \ \ \ \ \ \ \ \ \ \ \ \ \ \ \ \ \ \ \ \ \times \left(\Re[\alpha_2(\omega)]-\alpha_{\infty}\right)\left(\omega^2-\omega_{0\alpha}^2\right)\\
& + 3\left(V_2+4\pi\alpha_{\infty}\right)\frac{\hbar}{\pi^2cd^5}\int_0^{\infty}\! d\omega \frac{\omega^2}{e^{\frac{\omega}{\omega_{T_1}}}-1}\Im[\alpha_1(\omega)],
\label{eq:UdotPPVac_tau0}\\
\notag & \lim_{\tau \to 0^+}H_{1\textrm{,vac}}^{(2)}(\tau) = \frac{4\hbar\gamma}{\pi c^3d^3}\int_0^{\infty}\! d\omega \frac{\omega^3}{e^{\frac{\omega}{\omega_{T_1}}}-1}\\
& \ \ \ \ \ \ \ \ \ \ \ \ \ \ \ \ \ \ \ \ \times \Im[\alpha_1(\omega)]\Im[\alpha_2(\omega)],
\label{eq:HPPVac_tau0}
\end{align}
\end{subequations}
which are both positive\footnote{$ \lim_{\tau \to 0^+}H_{1\textrm{,vac}}^{(2)}(\tau) $ in Eq.~\eqref{eq:HPPVac_tau0} as well as the first term of $ \lim_{\tau \to 0^+}\dot{U}_{1\textrm{E,vac}}^{(2)}(\tau) $ in Eq.~\eqref{eq:UdotPPVac_tau0} are manifestly positive. The second term of Eq.~\eqref{eq:UdotPPVac_tau0} is positive for $ \varepsilon_{\infty} < -2 $ or $ \varepsilon_{\infty} > 1/4 $. In this paper, we assume the latter condition to be true, in order for $ \protect\widetilde{\beta} $ and $ \beta $, defined in Sec.~\ref{subsec:DL}, to be real.}, and whose sum (giving the full flux at $ \tau \to 0^+ $) is thus also positive. Interestingly, only $ d^{-3} $ and $ d^{-5} $ laws, which are absent in the stationary heat transfer [see Eq.~\eqref{eq:FluxStPPvac}], survive in this limit, and $ \lim_{\tau \to 0^+}H_{1\textrm{,vac}}^{(2)}(\tau) \propto d^{-3} $ for any $ d $.

\begin{figure}[!t]
\begin{center}
\includegraphics[width=1.0\linewidth]{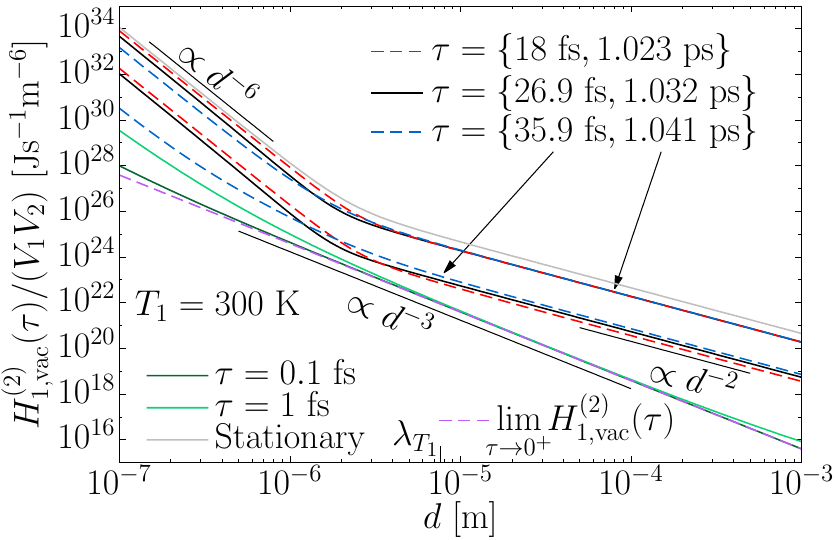}
\end{center}
\caption{\label{fig:T300ddep}Distance dependence of the nonstationary heat transfer for $ T_1 = 300 \ \textrm{K} $ and different values of $ \tau $. Red dashed, black solid, and blue dashed lines give the transfer for $ \tau $ corresponding to the maxima, averages, and minima, respectively, of the heat transfer in Fig.~\ref{fig:T300nf}.}
\end{figure}

How does the nonstationary heat transfer depend on the distance for other (not necessarily small) values of $ \tau $? Figure~\ref{fig:T300ddep} shows this dependence. For small $ \tau = 0.1 \ \textrm{fs} $, the heat transfer follows the limit in Eq.~\eqref{eq:HPPVac_tau0} for all the range plotted except the near-field distances. Increasing $ \tau $ by an order of magnitude leads to a stronger deviation from Eq.~\eqref{eq:HPPVac_tau0} in the near field and also for submillimeter distances, but the agreement still remains for intermediate $ d $. This indicates that $ d^{-3} $ law in Eq.~\eqref{eq:HPPVac_tau0} \enquote{bends} to $ d^{-6} $ and $ d^{-2} $ laws for small and large $ d $, respectively, as $ \tau $ increases. These laws are already clearly seen when $ \tau $ increases up to $ \pi/\omega_{0\alpha} = 18 \ \textrm{fs} $, which is roughly the first maximum of the near-field heat transfer [see Fig.~\ref{fig:T300nf}(b)]. Interestingly, for $ \tau = 2\pi/\omega_{0\alpha} = 35.9 \ \textrm{fs} $, corresponding to the subsequent near-field minimum, the transition from $ d^{-6} $ to $ d^{-2} $ law appears at a smaller $ d $, indicating a qualitative change of the distance dependence within one oscillation cycle. The corresponding average $ \tau = 3\pi/(2\omega_{0\alpha}) = 26.9 \ \textrm{fs} $ gives the distance dependence, which is qualitatively in between those for $ \tau = \pi/\omega_{0\alpha} $ and $ \tau = 2\pi/\omega_{0\alpha} $. The same picture applies for other oscillation cycles, e.g., $ \tau \in [57\pi/\omega_{\alpha} = 1.023 \ \textrm{ps}, 58\pi/\omega_{\alpha} = 1.041 \ \textrm{ps}] $, as demonstrated in Fig.~\ref{fig:T300ddep}, with the difference between the behaviors for different $ \tau $ within a cycle becoming smaller for a later cycle, finally converging to the stationary distance dependence.

\subsubsection{Nonstationary heat flux at small temperature}
\label{subsubsec:NonstatT30}
\begin{figure}[!t]
\begin{center}
\includegraphics[width=1.0\linewidth]{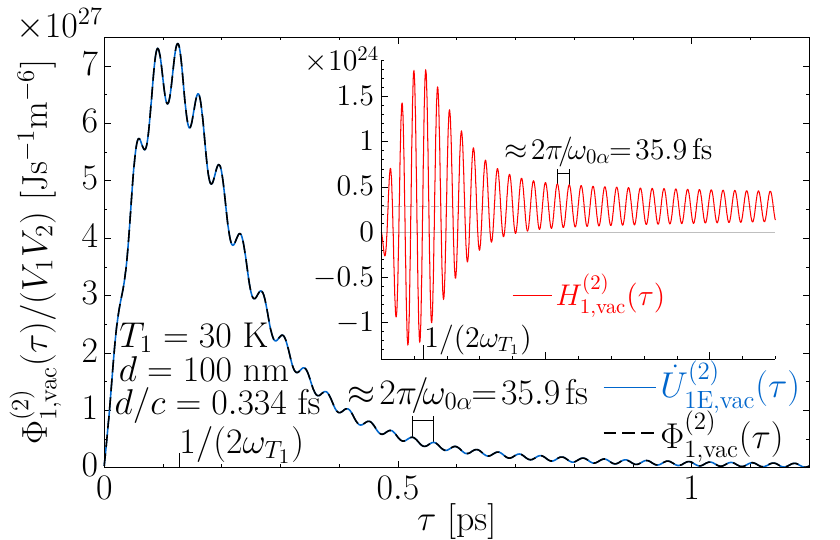}
\end{center}
\caption{\label{fig:T30nf}The same as Fig.~\ref{fig:T300nf}, but for $ T_1 = 30 \ \textrm{K} $. In contrast to the room-temperature flux in Fig.~\ref{fig:T300nf}, here, the relaxation time, being determined not by $ \gamma $ but the thermal frequency $ \omega_{T_1} $, is about nine times smaller. Compared to the nondissipative energy change (main plot), the heat transfer (inset) is negligible for the range of $ \tau $ plotted but it has stronger relative oscillations.}
\end{figure}

\begin{figure}[!t]
\begin{center}
\includegraphics[width=1.0\linewidth]{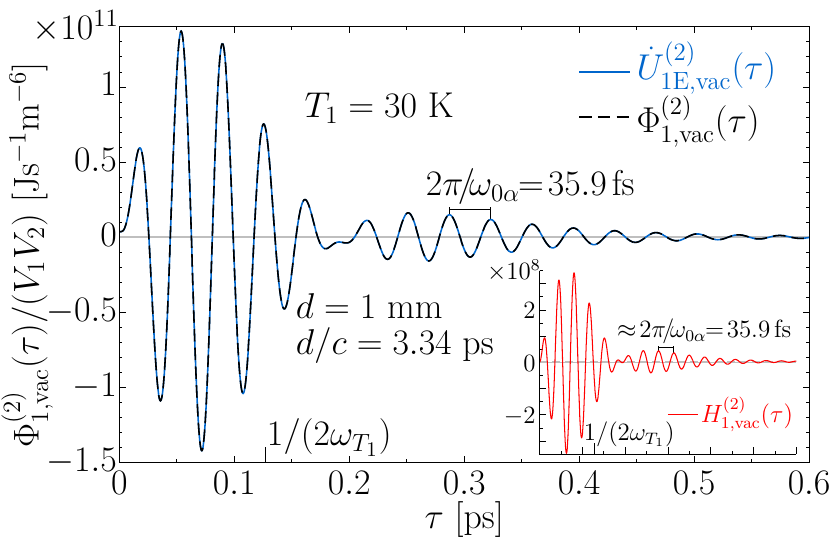}
\end{center}
\caption{\label{fig:T30ff}The same as Fig.~\ref{fig:T30nf}, but for $ d = 1 \ \textrm{mm} $ (far field). In contrast to the near-field flux in Fig.~\ref{fig:T30nf}, here, $ \dot{U}_{1\textrm{E,vac}}^{(2)}(\tau) $ (main plot) and $ H_{1\textrm{,vac}}^{(2)}(\tau) $ (inset) behave qualitatively similarly, but the latter quantity is still overall much smaller than the former one for the plotted range of $ \tau $.}
\end{figure}

If particle $ 1 $ starts radiating with temperature $ T_1 = 30 \ \textrm{K} $, the relevant frequencies of the heat flux are small compared to $ \omega_{0\alpha} $, such that the resonant structure of $ \alpha_i(\omega) $ plays a smaller role. The time dependence of the flux becomes qualitatively different compared to the room-temperature case, as shown in Figs.~\ref{fig:T30nf} and~\ref{fig:T30ff}.

Figure~\ref{fig:T30nf} demonstrates the near-field flux. For the plotted range of $ \tau $, the heat transfer (plotted in the inset) is overall three orders of magnitude smaller than the energy change (shown in the main plot), such that the full flux is determined by the latter quantity. This can be explained by the fact that the imaginary part of $ \alpha_2(\omega) $ is small compared to its real part or to $ V_2 $ for $ \omega \ll \omega_{0\alpha} $ [compare Eqs.~\eqref{eq:alphaDLRe} and~\eqref{eq:alphaDLIm}], which is the other way round for $ \omega \approx \omega_{0\alpha} $ (relevant for room temperature). Since $ H_{1\textrm{,vac}}^{(2)}(\tau) $ converges to $ H_{1\textrm{st,vac}}^{(2)} $, at large $ \tau $ (beyond the plotted range), the heat transfer becomes dominant over $ \dot{U}_{1\textrm{E,vac}}^{(2)}(\tau) $ (which converges to zero). 

The oscillations with a period of approximately $ 2\pi/\omega_{0\alpha} $ are still present in Fig.~\ref{fig:T30nf}, indicating that the resonance frequency still affects the time dependence at low temperature. Their relative amplitude is stronger for $ H_{1\textrm{,vac}}^{(2)}(\tau) $ than for $ \dot{U}_{1\textrm{E,vac}}^{(2)}(\tau) $. Both quantities have a global maximum at $ \tau \approx 1/(2\omega_{T_1}) = 0.13 \ \textrm{ps} $, which can also be considered as the relaxation time (it is roughly nine times smaller than the relaxation time at room temperature). Remarkably, and in contrast to Fig.~\ref{fig:T300nf}, here, the heat transfer takes also negative values, with a global minimum at $ \tau = 0.0815 \ \textrm{ps} $.

When the particles are separated by a large distance, $ \dot{U}_{1\textrm{E,vac}}^{(2)}(\tau) $ shows a relatively stronger oscillations, being symmetric around zero, as shown in the main plot of Fig.~\ref{fig:T30ff} for $ d = 1 \ \textrm{mm} $. The global maximum is shifted to a smaller time compared to Fig.~\ref{fig:T30nf}. Interestingly, one can distinguish two \enquote{wave packets} of different amplitude, reminiscent of the temporal dependence of the local fields at the nanostructures after the femtosecond pulse excitation~\cite{Stockman2004}. For the plotted range of $ \tau $, $ H_{1\textrm{,vac}}^{(2)}(\tau) $, plotted in the inset, has a similar behavior, but with a much smaller amplitude. The stationary heat transfer is $ H_{1\textrm{st,vac}}^{(2)}/(V_1V_2) = 2.44 \times 10^6 \ \textrm{J}\textrm{s}^{-1}\textrm{m}^{-6} $, which is $ 140 $ times smaller than the the maximum value of the transfer.

\section{Conclusion}
\label{sec:Conclusion}
In this paper, we developed a theoretical formalism for computing radiative heat flux sourced by an object starting from a certain time (say $ t = 0 $). The appearance of this flux brings the considered system to a transient state, where physical observables, including the flux itself, depend on time. In contrast to the well-studied stationary scenario, where radiated fields exist forever and where one can work entirely in frequency domain, the considered nonstationary problem requires to use time-dependent EM fields. Applying the fluctuational electrodynamics in time domain, we derived the formula for the time-dependent flux between two arbitrary objects in the presence of a collection of other arbitrary objects, also finding its limit in the case where the former two objects reduce to small spherical particles.

We applied the derived formula to study the transient flux between two isolated particles whose dielectric response is described by the Drude-Lorentz model. We found analytically the expression for the flux spectrum as a function of frequency, time, the interparticle distance $ d $, and the parameters of the dielectric function, and identified its dissipative (energy change) and nondissipative (heat transfer) parts. This spectrum accounts for EM retardation, being finite only after time $ t = d/c $ elapsed from the start of the radiation.

For $ t > d/c $, the frequency integral of the spectrum was performed numerically and studied for two SiC particles. When one particle starts emitting the room-temperature radiation ($ T_1 = 300 \ \textrm{K} $), the flux to the other one shows oscillatory exponential relaxation to the stationary value. The relaxation time $ 1/\gamma = 1.12 \ \textrm{ps} $ and the oscillation period $ 2\pi/\omega_{0\alpha} = 35.9 \ \textrm{fs} $ (for near-field $ d $) or $ \pi/\omega_{0\alpha} = 18 \ \textrm{fs} $ (for far-field $ d $) are determined by the damping rate $ \gamma $ and the resonance frequency $ \omega_{0\alpha} $ of the particle polarizability, respectively. The oscillations are stronger in the near field, where the flux can be significantly larger than its stationary limit. For small temperature, $ T_1 = 30 \ \textrm{K} $, the relaxation time, determined by the thermal frequency, is about nine times smaller. Here, the oscillations are relatively stronger in the far field than in the near field, and the maximum of the flux is more than four orders of magnitude larger than its stationary value. For small $ T_1 $ and intermediate times, the overall amplitude of the nondissipative part of the flux is several orders of magnitude larger than that of the dissipative part; this is very different compared to the room-temperature case, where the dissipative part dominates.

Femtosecond resolution required to experimentally measure the demonstrated time dependence was achieved in the experiments for related problems, e.g., electron energy transport~\cite{Grolleau2021}.

Future work may study the time dependence for more complex geometrical configurations (e.g., two particles above a plate), consider object's radiation during a finite time interval, as well as consider moving objects. The time dependence of the nonequilibrium Casimir forces~\cite{Kruger2012} is also a promising avenue for future research (see the recent study on the dynamical Casimir effect~\cite{Yu2024_2}).

\begin{acknowledgments}
I thank Matthias Kr\"uger for stimulating and enlightening discussions, and for a critical reading of the manuscript. I also thank Thorsten Emig, David Gelbwaser-Klimovsky, Noah Graham, Mehran Kardar, Riccardo Messina, and Stefan Scheel for discussions. Riccardo Messina is also acknowledged for a critical reading of the manuscript. This work was funded by the Deutsche Forschungsgemeinschaft (DFG, German Research Foundation) through the Walter Benjamin fellowship (Project No. 453458207).
\end{acknowledgments}

% MAIN_PART_end ------------------------------------------

\onecolumngrid
% APPENDIX_begin ------------------------------------------
\appendix
\section{Derivations of stationary heat flux}
\label{app:sec:FluxSt}
\subsection{Allowing the object to radiate far away in the past in the time-dependent formula}
\label{app:subsec:FluxSt_from_tdep}
In this appendix, we derive the stationary heat flux in two ways. The first way is about replacing the lower integration limits in Eq.~\eqref{eq:Flux}, as mentioned in Sec.~\ref{subsec:Formula}: Instead of starting the time integrals at zero, we start them at $ -\infty $, i.e., the object starts radiating far away in the past. In this case, the first term of Eq.~\eqref{eq:Flux} vanishes, because it is the time derivative of the stationary equal-time autocorrelation function [see the first term of Eq.~\eqref{eq:FluxPT}]. The remaining second term reads
\begin{align}
\notag \Phi_{1\textrm{st}}^{(2)} = \ & \frac{\hbar}{(2\pi)^3}\int_0^{\infty}\! d\omega \frac{\omega^2}{e^{\frac{\omega}{\omega_{T_1}}}-1}\int\! dt' \int\! dt''\cos[\omega(t'-t'')]\int\! d\tilde{t}\int_{V_1}\! d^3r_1\int_{V_2}\! d^3r_2\\
& \times \Tr\left\{\Im[\bbeps(\vct{r}_1,\omega)]\frac{\partial\mathbb{G}^{\textrm{E}}(\vct{r}_1,\vct{r}_2;t-\tilde{t}-t')}{\partial t}\bbeps(\vct{r}_2,\tilde{t})\mathbb{G}^{\textrm{E}T}(\vct{r}_1,\vct{r}_2;t-t'')\right\}.
\label{eq:FluxStA1}
\end{align}

Since, in frequency domain, the relations between the fields and the current via the GFs contain additional prefactors [see Eqs.~\eqref{eq:GEomegadef} and~\eqref{eq:GHomegadef}], the time-domain and frequency-domain GFs are not related via the Fourier transform, but via its modified versions:
\begin{subequations}
\begin{equation}
\mathbb{G}^{\textrm{E}}(\vct{r},\vct{r}';t-t') = \frac{2i}{c^2}\int\! d\omega \omega\mathbb{G}^{\textrm{E}}(\vct{r},\vct{r}';\omega)e^{-i\omega(t-t')},
\label{eq:GEtviaGEomega}
\end{equation}
\begin{equation}
\mathbb{G}^{\textrm{H}}(\vct{r},\vct{r}';t-t') = \frac{2}{c}\int\! d\omega \mathbb{G}^{\textrm{H}}(\vct{r},\vct{r}';\omega)e^{-i\omega(t-t')}.
\label{eq:GHtviaGHomega}
\end{equation}
\end{subequations}
Using these transforms and the Euler's formula for the cosine in Eq.~\eqref{eq:FluxStA1}, we obtain
\begin{align}
\notag \Phi_{1\textrm{st}}^{(2)} = \ & \frac{i\hbar}{8\pi^4c^4}\int_0^{\infty}\! d\omega \frac{\omega^2}{e^{\frac{\omega}{\omega_{T_1}}}-1}\int_{V_1}\! d^3r_1\int_{V_2}\! d^3r_2\int\! d\omega'\int\! d\omega''\int\! d\widetilde{\omega}\Tr\left\{\Im[\bbeps(\vct{r}_1,\omega)]\mathbb{G}^{\textrm{E}}(\vct{r}_1,\vct{r}_2;\omega')\bbeps(\vct{r}_2,\widetilde{\omega})\mathbb{G}^{\textrm{E}T}(\vct{r}_1,\vct{r}_2;\omega'')\right\}\\
& \times (\omega')^2\omega''e^{-i(\omega'+\omega'')t}\int\! dt'\int\! dt'' \int\! d\tilde{t}\left[e^{i(\omega'+\omega)t'}e^{i(\omega''-\omega)t''}+e^{i(\omega'-\omega)t'}e^{i(\omega''+\omega)t''}\right]e^{i(\omega'-\widetilde{\omega})\tilde{t}}.
\label{eq:FluxStA2}
\end{align}
The time integrals give delta functions,
\begin{align}
\notag \Phi_{1\textrm{st}}^{(2)} = \ & \frac{i\hbar}{\pi c^4}\int_0^{\infty} d\omega \frac{\omega^2}{e^{\frac{\omega}{\omega_{T_1}}}-1}\int_{V_1}d^3r_1\int_{V_2}d^3r_2\int d\omega'\int d\omega''\int d\widetilde{\omega}\Tr\left\{\Im[\bbeps(\vct{r}_1,\omega)]\mathbb{G}^{\textrm{E}}(\vct{r}_1,\vct{r}_2;\omega')\bbeps(\vct{r}_2,\widetilde{\omega})\mathbb{G}^{\textrm{E}T}(\vct{r}_1,\vct{r}_2;\omega'')\right\}\\
& \times (\omega')^2\omega''e^{-i(\omega'+\omega'')t}\left[\delta(\omega'+\omega)\delta(\omega''-\omega)+\delta(\omega'-\omega)\delta(\omega''+\omega)\right]\delta(\omega'-\widetilde{\omega}),
\label{eq:FluxStA3}
\end{align}
and the last three frequency integrals can be evaluated, canceling the dependence on $ t $,
\begin{align}
\notag \Phi_{1\textrm{st}}^{(2)} = \ & -\frac{i\hbar}{\pi c^4}\int_0^{\infty}\! d\omega \frac{\omega^5}{e^{\frac{\omega}{\omega_{T_1}}}-1}\int_{V_1}\! d^3r_1\int_{V_2}\! d^3r_2\\
& \times\Tr\left\{\Im[\bbeps(\vct{r}_1,\omega)]\left[\mathbb{G}^{\textrm{E}}(\vct{r}_1,\vct{r}_2;\omega)\bbeps(\vct{r}_2,\omega)\mathbb{G}^{\textrm{E}T}(\vct{r}_1,\vct{r}_2;-\omega)-\mathbb{G}^{\textrm{E}}(\vct{r}_1,\vct{r}_2;-\omega)\bbeps(\vct{r}_2,-\omega)\mathbb{G}^{\textrm{E}T}(\vct{r}_1,\vct{r}_2;\omega)\right]\right\}.
\label{eq:FluxStA4}
\end{align}
Because of the reality of $ \bbeps $ and $ \mathbb{G} $ in time domain, $ \bbeps(-\omega) = \bbeps^*(\omega) $ and $ \mathbb{G}^{\textrm{E}}(-\omega) = \mathbb{G}^{\textrm{E}*}(\omega) $~\cite{Eckhardt1984}, such that the second term inside the trace of Eq.~\eqref{eq:FluxStA4} is the complex conjugate of the first one,
\begin{align}
\notag & \Phi_{1\textrm{st}}^{(2)} = -\frac{i\hbar}{\pi c^4}\int_0^{\infty}\! d\omega \frac{\omega^5}{e^{\frac{\omega}{\omega_{T_1}}}-1}\int_{V_1}\! d^3r_1\int_{V_2}\! d^3r_2\\
& \times\left(\Tr\left\{\Im[\bbeps(\vct{r}_1,\omega)]\mathbb{G}^{\textrm{E}}(\vct{r}_1,\vct{r}_2;\omega)\bbeps(\vct{r}_2,\omega)\mathbb{G}^{\textrm{E}\dagger}(\vct{r}_1,\vct{r}_2;\omega)\right\}-\Tr\left\{\Im[\bbeps(\vct{r}_1,\omega)]\left[\mathbb{G}^{\textrm{E}}(\vct{r}_1,\vct{r}_2;\omega)\bbeps(\vct{r}_2,\omega)\mathbb{G}^{\textrm{E}\dagger}(\vct{r}_1,\vct{r}_2;\omega)\right]^*\right\}\right),
\label{eq:FluxStA5}
\end{align}
where we used $ \mathbb{G}^{\textrm{E}T}(-\omega) = \mathbb{G}^{\textrm{E}\dagger}(\omega) $. Using the properties of a trace and the reciprocity of $ \bbeps $, $ \bbeps = \bbeps^T $, we can rewrite the second trace as
\begin{align}
\notag \Tr &\left\{\Im[\bbeps(\vct{r}_1,\omega)]\left[\mathbb{G}^{\textrm{E}}(\vct{r}_1,\vct{r}_2;\omega)\bbeps(\vct{r}_2,\omega)\mathbb{G}^{\textrm{E}\dagger}(\vct{r}_1,\vct{r}_2;\omega)\right]^*\right\} = \Tr\left\{\left(\Im[\bbeps(\vct{r}_1,\omega)]\left[\mathbb{G}^{\textrm{E}}(\vct{r}_1,\vct{r}_2;\omega)\bbeps(\vct{r}_2,\omega)\mathbb{G}^{\textrm{E}\dagger}(\vct{r}_1,\vct{r}_2;\omega)\right]^*\right)^T\right\}\\
& = \Tr\left\{\left[\mathbb{G}^{\textrm{E}}(\vct{r}_1,\vct{r}_2;\omega)\bbeps(\vct{r}_2,\omega)\mathbb{G}^{\textrm{E}\dagger}(\vct{r}_1,\vct{r}_2;\omega)\right]^{\dagger}\Im[\bbeps(\vct{r}_1,\omega)]\right\} = \Tr\left\{\Im[\bbeps(\vct{r}_1,\omega)]\mathbb{G}^{\textrm{E}}(\vct{r}_1,\vct{r}_2;\omega)\bbeps^*(\vct{r}_2,\omega)\mathbb{G}^{\textrm{E}\dagger}(\vct{r}_1,\vct{r}_2;\omega)\right\},
\label{eq:TraceRewrite}
\end{align}
which differs from the first trace in Eq.~\eqref{eq:FluxStA5} by the conjugation of $ \bbeps $. Substituting Eq.~\eqref{eq:TraceRewrite} into Eq.~\eqref{eq:FluxStA5}, we get
\begin{equation}
\Phi_{1\textrm{st}}^{(2)} = -\frac{i\hbar}{\pi c^4}\int_0^{\infty}\! d\omega \frac{\omega^5}{e^{\frac{\omega}{\omega_{T_1}}}-1}\int_{V_1}\! d^3r_1\int_{V_2}\! d^3r_2\Tr\left\{\Im[\bbeps(\vct{r}_1,\omega)]\mathbb{G}^{\textrm{E}}(\vct{r}_1,\vct{r}_2;\omega)\left[\bbeps(\vct{r}_2,\omega)-\bbeps^*(\vct{r}_2,\omega)\right]\mathbb{G}^{\textrm{E}\dagger}(\vct{r}_1,\vct{r}_2;\omega)\right\},
\label{eq:FluxStA6}
\end{equation}
which leads to Eq.~\eqref{eq:FluxSt} after using $ \bbeps - \bbeps^* = 2i\Im[\bbeps] $.

\subsection{Derivation in frequency domain}
\label{app:subsec:FluxSt_omega}
Since we consider a stationary state, explicit details of the time dependence [e.g., the lower limit in Eqs.~\eqref{eq:E} and~\eqref{eq:H}] are not important. We can thus start from Maxwell equations in frequency domain~\cite{Rahi2009, Bimonte2017, Jackson1998},
\begin{subequations}
\begin{equation}
\nabla \times \vct{E} = i\frac{\omega}{c}\vct{B},
\label{eq:MEomega1}
\end{equation}
\begin{equation}
\nabla \times \vct{H} = \frac{4\pi}{c}\vct{j} - i\frac{\omega}{c}\vct{D}.
\label{eq:MEomega2}
\end{equation}
\end{subequations}
The constitutive relations read~\cite{Rahi2009, Bimonte2017, Jackson1998}
\begin{subequations}
\begin{equation}
\vct{D}(\vct{r},\omega) = \bbeps(\vct{r},\omega)\vct{E}(\vct{r},\omega),
\label{eq:ConstEomega}
\end{equation}
\begin{equation}
\vct{B}(\vct{r},\omega) = \vct{H}(\vct{r},\omega).
\label{eq:ConstHomega}
\end{equation}
\end{subequations}
The fields and the current are related via the frequency-dependent GFs (neglecting the homogeneous solutions)~\cite{Bimonte2017, Kruger2012, Zhu2018, Yang2022},
\begin{subequations}
\begin{equation}
\vct{E}(\vct{r},\omega) = \frac{4\pi i\omega}{c^2}\int_{V_1}\! d^3r' \mathbb{G}^{\textrm{E}}(\vct{r},\vct{r}';\omega)\vct{j}(\vct{r}',\omega),
\label{eq:GEomegadef}
\end{equation}
\begin{equation}
\vct{H}(\vct{r},\omega) = \frac{4\pi}{c}\int_{V_1}\! d^3r' \mathbb{G}^{\textrm{H}}(\vct{r},\vct{r}';\omega)\vct{j}(\vct{r}',\omega).
\label{eq:GHomegadef}
\end{equation}
\end{subequations}

Performing the Fourier transform of the fields in Eq.~\eqref{eq:nablaS}, we obtain
\begin{equation}
\nabla\cdot\vct{S}(\vct{r},t) = \frac{c}{16\pi^3}\int\! d\omega\int\! d\omega'\left\{\vct{H}^*(\vct{r},\omega')\cdot\left[\nabla\times\vct{E}(\vct{r},\omega)\right]-\vct{E}(\vct{r},\omega)\cdot\left[\nabla\times\vct{H}^*(\vct{r},\omega')\right]\right\}e^{-i(\omega-\omega')t}.
\label{eq:nablaSomega1}
\end{equation}
Using Eqs.~\eqref{eq:MEomega1}--\eqref{eq:ConstHomega}, Eq.~\eqref{eq:nablaSomega1} can be written as
\begin{align}
\notag -\nabla\cdot\vct{S}(\vct{r},t) = \ & -\frac{i}{16\pi^3}\int\! d\omega\int\! d\omega'\omega\vct{H}(\vct{r},\omega)\cdot\vct{H}^*(\vct{r},\omega')e^{-i(\omega-\omega')t}\\
& + \frac{i}{16\pi^3}\int\! d\omega\int\! d\omega'\omega'\vct{E}(\vct{r},\omega)\bbeps^*(\vct{r},\omega')\vct{E}^*(\vct{r},\omega')e^{-i(\omega-\omega')t} + \frac{1}{4\pi^2}\int\! d\omega\int\! d\omega'\vct{E}(\vct{r},\omega)\cdot\vct{j}^*(\vct{r},\omega')e^{-i(\omega-\omega')t},
\label{eq:nablaSomega2}
\end{align}
where each term equals the corresponding term in Eq.~\eqref{eq:PoyntingsTheorem}. Substituting Eq.~\eqref{eq:nablaSomega2} into Eq.~\eqref{eq:Flux_def}, we get
\begin{align}
\notag & \Phi_1^{(2)}(t) = -\frac{i}{16\pi^3}\int_{V_2}\! d^3r\int\! d\omega\int\! d\omega'\omega\langle\vct{H}(\vct{r},\omega)\cdot\vct{H}^*(\vct{r},\omega')\rangle e^{-i(\omega-\omega')t}\\
& + \frac{i}{16\pi^3}\int_{V_2}\! d^3r\int\! d\omega\int\! d\omega'\omega'\langle\vct{E}(\vct{r},\omega)\bbeps^*(\vct{r},\omega')\vct{E}^*(\vct{r},\omega')\rangle e^{-i(\omega-\omega')t} + \frac{1}{4\pi^2}\int_{V_2}\! d^3r\int\! d\omega\int\! d\omega'\langle\vct{E}(\vct{r},\omega)\cdot\vct{j}^*(\vct{r},\omega')\rangle e^{-i(\omega-\omega')t}.
\label{eq:Flux_omega1}
\end{align}
Using Eqs.~\eqref{eq:GEomegadef} and~\eqref{eq:GHomegadef} in Eq.~\eqref{eq:Flux_omega1}, we obtain
\begin{align}
\notag & \Phi_1^{(2)}(t) = -\frac{i}{\pi c^2}\int_{V_1}\! d^3r_1\int_{V_1}\! d^3r_1'\int_{V_2}\! d^3r_2\int\! d\omega\int\! d\omega'\omega\Tr\left\{\langle\vct{j}(\vct{r}_1,\omega)\otimes\vct{j}^*(\vct{r}_1',\omega')\rangle\mathbb{G}^{\textrm{H}\dagger}(\vct{r}_2,\vct{r}_1';\omega')\mathbb{G}^{\textrm{H}}(\vct{r}_2,\vct{r}_1;\omega)\right\} e^{-i(\omega-\omega')t}\\
\notag & +\frac{i}{\pi c^4}\int_{V_1}\! d^3r_1\int_{V_1}\! d^3r_1'\int_{V_2}\! d^3r_2\int\! d\omega\int\! d\omega'\omega(\omega')^2\Tr\left\{\langle\vct{j}(\vct{r}_1,\omega)\otimes\vct{j}^*(\vct{r}_1',\omega')\rangle\mathbb{G}^{\textrm{E}\dagger}(\vct{r}_2,\vct{r}_1';\omega')\bbeps^{\dagger}(\vct{r}_2,\omega')\mathbb{G}^{\textrm{E}}(\vct{r}_2,\vct{r}_1;\omega)\right\} e^{-i(\omega-\omega')t}\\
& + \frac{i}{\pi c^2}\int_{V_1}\! d^3r_1\int_{V_2}\! d^3r_2\int\! d\omega\int\! d\omega'\Tr\left\{\langle\vct{j}(\vct{r}_1,\omega)\otimes\vct{j}^*(\vct{r}_2,\omega')\rangle\mathbb{G}^{\textrm{E}}(\vct{r}_2,\vct{r}_1;\omega)\right\}e^{-i(\omega-\omega')t},
\label{eq:Flux_omega2}
\end{align}
which is an alternative form of Eq.~\eqref{eq:FluxEH} and Eq.~\eqref{eq:FluxEH} with $ 0 $ replaced by $ -\infty $ in the lower integration limits. It is evident from Eq.~\eqref{eq:Flux_omega2} that the time dependence cancels out if the current correlator is proportional to $ \delta(\omega-\omega') $, i.e., the current is in a stationary state for all times. Otherwise, the flux acquires the time dependence.

We continue with reciprocal objects and stationary correlator~\eqref{eq:FDTomega}. Substituting the latter into Eq.~\eqref{eq:Flux_omega2}, we get [since the currents inside different objects are uncorrelated, the last term of Eq.~\eqref{eq:Flux_omega2} vanishes]
\begin{align}
\notag \Phi_{1\textrm{st}}^{(2)} = \ & -\frac{i\hbar}{\pi c^2}\int\! d\omega\frac{\omega^3\sgn(\omega)}{e^{\frac{|\omega|}{\omega_{T_1}}}-1}\int_{V_1}\! d^3r_1\int_{V_2}\! d^3r_2\Tr\left\{\Im[\bbeps(\vct{r}_1,\omega)]\mathbb{G}^{\textrm{H}\dagger}(\vct{r}_2,\vct{r}_1;\omega)\mathbb{G}^{\textrm{H}}(\vct{r}_2,\vct{r}_1;\omega)\right\}\\
& +\frac{i\hbar}{\pi c^4}\int\! d\omega\frac{\omega^5\sgn(\omega)}{e^{\frac{|\omega|}{\omega_{T_1}}}-1}\int_{V_1}\! d^3r_1\int_{V_2}\! d^3r_2\Tr\left\{\Im[\bbeps(\vct{r}_1,\omega)]\mathbb{G}^{\textrm{E}\dagger}(\vct{r}_2,\vct{r}_1;\omega)\bbeps^{\dagger}(\vct{r}_2,\omega)\mathbb{G}^{\textrm{E}}(\vct{r}_2,\vct{r}_1;\omega)\right\}.
\label{eq:Flux_omega3}
\end{align}
Since the trace in the first term of Eq.~\eqref{eq:Flux_omega3} is real (because $ \Im[\bbeps] $ and $ \mathbb{G}^{\textrm{H}\dagger}\mathbb{G}^{\textrm{H}} $ are positive-semidefinite matrices\textsuperscript{\ref{fn:positive_matrix}}), that term must vanish. Indeed, since $ \Im[\bbeps(-\omega)] = -\Im[\bbeps(\omega)] $ and $ \mathbb{G}^{\textrm{H}}(-\omega) = \mathbb{G}^{\textrm{H}*}(\omega) $,
\begin{align}
\notag \Tr &\left\{\Im[\bbeps(\vct{r}_1,-\omega)]\mathbb{G}^{\textrm{H}\dagger}(\vct{r}_2,\vct{r}_1;-\omega)\mathbb{G}^{\textrm{H}}(\vct{r}_2,\vct{r}_1;-\omega)\right\} = -\Tr\left\{\Im[\bbeps(\vct{r}_1,\omega)]\mathbb{G}^{\textrm{H}T}(\vct{r}_2,\vct{r}_1;\omega)\mathbb{G}^{\textrm{H}*}(\vct{r}_2,\vct{r}_1;\omega)\right\}\\
& = -\Tr\left\{\mathbb{G}^{\textrm{H}\dagger}(\vct{r}_2,\vct{r}_1;\omega)\mathbb{G}^{\textrm{H}}(\vct{r}_2,\vct{r}_1;\omega)\Im[\bbeps(\vct{r}_1,\omega)]^T\right\} = -\Tr\left\{\Im[\bbeps(\vct{r}_1,\omega)]\mathbb{G}^{\textrm{H}\dagger}(\vct{r}_2,\vct{r}_1;\omega)\mathbb{G}^{\textrm{H}}(\vct{r}_2,\vct{r}_1;\omega)\right\},
\label{eq:TraceH}
\end{align} 
i.e., the trace is an odd function of $ \omega $, such that the integrand is also an odd function: the integral over $ \omega $ vanishes. The trace of the second term can be written as
\begin{align}
\notag &\Tr\left\{\Im[\bbeps(\vct{r}_1,\omega)]\mathbb{G}^{\textrm{E}\dagger}(\vct{r}_2,\vct{r}_1;\omega)\bbeps^{\dagger}(\vct{r}_2,\omega)\mathbb{G}^{\textrm{E}}(\vct{r}_2,\vct{r}_1;\omega)\right\} = \Tr\left\{\Im[\bbeps(\vct{r}_1,\omega)]\mathbb{G}^{\textrm{E}\dagger}(\vct{r}_2,\vct{r}_1;\omega)\bbeps^*(\vct{r}_2,\omega)\mathbb{G}^{\textrm{E}}(\vct{r}_2,\vct{r}_1;\omega)\right\}\\
& = \Tr\left\{\Im[\bbeps(\vct{r}_1,\omega)]\mathbb{G}^{\textrm{E}\dagger}(\vct{r}_2,\vct{r}_1;\omega)\Re[\bbeps(\vct{r}_2,\omega)]\mathbb{G}^{\textrm{E}}(\vct{r}_2,\vct{r}_1;\omega)\right\} -i\Tr\left\{\Im[\bbeps(\vct{r}_1,\omega)]\mathbb{G}^{\textrm{E}\dagger}(\vct{r}_2,\vct{r}_1;\omega)\Im[\bbeps(\vct{r}_2,\omega)]\mathbb{G}^{\textrm{E}}(\vct{r}_2,\vct{r}_1;\omega)\right\}.
\label{eq:TraceE}
\end{align} 
Similar to Eq.~\eqref{eq:TraceH}, it can be shown that the first term of Eq.~\eqref{eq:TraceE} is an odd function of $ \omega $, whereas the second term is an even function, such that only the latter contributes to the integral in the second term of Eq.~\eqref{eq:Flux_omega3},
\begin{equation}
\Phi_{1\textrm{st}}^{(2)} = \frac{2\hbar}{\pi c^4}\int_0^{\infty}\! d\omega\frac{\omega^5}{e^{\frac{\omega}{\omega_{T_1}}}-1}\int_{V_1}\! d^3r_1\int_{V_2}\! d^3r_2\Tr\left\{\Im[\bbeps(\vct{r}_1,\omega)]\mathbb{G}^{\textrm{E}\dagger}(\vct{r}_2,\vct{r}_1;\omega)\Im[\bbeps(\vct{r}_2,\omega)]\mathbb{G}^{\textrm{E}}(\vct{r}_2,\vct{r}_1;\omega)\right\},
\label{eq:Flux_omega4}
\end{equation}
which, after using the properties of a trace and the properties of the reciprocity, becomes identical to Eq.~\eqref{eq:FluxSt}.

\section{Derivation of the nonstationary heat flux in the point particle limit}
\label{app:sec:FluxPP}
In this appendix, we derive Eq.~\eqref{eq:FluxPP}, the point particle limit of Eq.~\eqref{eq:Flux}. This limit implies an expansion of the flux in terms of the scattering operators of objects $ 1 $ and $ 2 $, not in terms of their permittivity tensors~\cite{Asheichyk2017}. Therefore, we shall write Eq.~\eqref{eq:Flux} in terms of the scattering operators, which are typically used in frequency domain~\cite{Bimonte2017, Kruger2011, Messina2011, Kruger2012, Muller2017, Asheichyk2017}. We thus first make the Fourier transforms of the GFs [according to Eqs.~\eqref{eq:GEtviaGEomega} and~\eqref{eq:GHtviaGHomega}] and $ \bbeps_2 $ in Eq.~\eqref{eq:Flux}, such that the traces (together with the spatial integrals) of Eq.~\eqref{eq:Flux} become
\begin{subequations}
\begin{align}
\notag \int_{V_1}\! d^3r_1&\int_{V_2}\! d^3r_2\Tr\left\{\Im[\bbeps(\vct{r}_1,\omega)]\mathbb{G}^{\textrm{H}}(\vct{r}_1,\vct{r}_2;t-t') \mathbb{G}^{\textrm{H}T}(\vct{r}_1,\vct{r}_2;t-t'')\right\}\\
& = \frac{4}{c^2}\int\! d\omega'\int\! d\omega''e^{-i\omega'(t-t')}e^{i\omega''(t-t'')}\int_{V_1}\! d^3r_1\int_{V_2}\! d^3r_2\Tr\left\{\Im[\bbeps(\vct{r}_1,\omega)]\mathbb{G}^{\textrm{H}}(\vct{r}_1,\vct{r}_2;\omega')\mathbb{G}^{\textrm{H}\dagger}(\vct{r}_1,\vct{r}_2;\omega'')\right\},
\label{eq:PPtraceH1}\\
\notag \int_{V_1}\! d^3r_1&\int_{V_2}\! d^3r_2\Tr\left\{\Im[\bbeps(\vct{r}_1,\omega)]\frac{\partial\mathbb{G}^{\textrm{E}}(\vct{r}_1,\vct{r}_2;t-\tilde{t}-t')}{\partial t}\bbeps(\vct{r}_2,\tilde{t})\mathbb{G}^{\textrm{E}T}(\vct{r}_1,\vct{r}_2;t-t'')\right\}\\
\notag & = -\frac{2i}{\pi c^4}\int\! d\omega'\int\! d\omega'' \int\! d\widetilde{\omega}(\omega')^2\omega''e^{-i\omega'(t-\tilde{t}-t')}e^{i\omega''(t-t'')}e^{-i\widetilde{\omega}\tilde{t}}\\
& \ \ \ \times \int_{V_1}\! d^3r_1\int_{V_2}\! d^3r_2\Tr\left\{\Im[\bbeps(\vct{r}_1,\omega)]\mathbb{G}^{\textrm{E}}(\vct{r}_1,\vct{r}_2;\omega')\bbeps(\vct{r}_2,\widetilde{\omega})\mathbb{G}^{\textrm{E}\dagger}(\vct{r}_1,\vct{r}_2;\omega'')\right\}.
\label{eq:PPtraceE1}
\end{align}
\end{subequations}

For the aforementioned expansion, it is convenient to use the operator notation~\cite{Rahi2009, Kruger2012, Muller2017, Asheichyk2017}, where the operator multiplication involves both a matrix product and an integration over a common spatial argument, i.e., for operators $ \mathbb{A}(\vct{r},\widetilde{\vct{r}}) $ and $ \mathbb{B}(\widetilde{\vct{r}},\vct{r}') $, their operator product $ \mathbb{C} = \mathbb{A}\mathbb{B} $ is defined as $ \mathbb{C}(\vct{r},\vct{r}') = \int\! d^3\widetilde{r}\mathbb{A}(\vct{r},\widetilde{\vct{r}})\mathbb{B}(\widetilde{\vct{r}},\vct{r}') $, where the product inside the integral is a matrix product. Similarly, the operator trace is defined as $ \Tr{\mathbb{A}} = \int\! d^3r \Tr\mathbb{A}(\vct{r},\vct{r}) $, where the trace inside the integral is the matrix trace. In what follows in this appendix, the operator notation is understood if no spatial arguments are given explicitly. To proceed towards the operator notation, we rewrite the traces (together with the spatial integrals) of Eqs.~\eqref{eq:PPtraceH1} and~\eqref{eq:PPtraceE1} as
\begin{subequations}
\begin{align}
\notag \int_{V_1}\! &d^3r_1\int_{V_2}\! d^3r_2\Tr\left\{\Im[\bbeps(\vct{r}_1,\omega)]\mathbb{G}^{\textrm{H}}(\vct{r}_1,\vct{r}_2;\omega')\mathbb{G}^{\textrm{H}\dagger}(\vct{r}_1,\vct{r}_2;\omega'')\right\}\\
& = \int_{V_1}\! d^3r_1\int\! d^3r'\int_{V_2}\!d^3r_2\int\! d^3r''\Tr\left\{\Im[\bbeps(\vct{r}_1,\omega)]\delta^{(3)}(\vct{r}_1-\vct{r}')\mathbb{G}^{\textrm{H}}(\vct{r}',\vct{r}_2;\omega')\mathcal{I}\delta^{(3)}(\vct{r}_2-\vct{r}'')\mathbb{G}^{\textrm{H}*}(\vct{r}'',\vct{r}_1;\omega'')\right\},
\label{eq:PPtraceH2}\\
\notag \int_{V_1}\! &d^3r_1\int_{V_2}\! d^3r_2\Tr\left\{\Im[\bbeps(\vct{r}_1,\omega)]\mathbb{G}^{\textrm{E}}(\vct{r}_1,\vct{r}_2;\omega')\bbeps(\vct{r}_2,\widetilde{\omega})\mathbb{G}^{\textrm{E}\dagger}(\vct{r}_1,\vct{r}_2;\omega'')\right\}\\
& = \int_{V_1}\! d^3r_1\int\! d^3r'\int_{V_2}\! d^3r_2\int\! d^3r''\Tr\left\{\Im[\bbeps(\vct{r}_1,\omega)]\delta^{(3)}(\vct{r}_1-\vct{r}')\mathbb{G}^{\textrm{E}}(\vct{r}',\vct{r}_2;\omega')\bbeps(\vct{r}_2,\widetilde{\omega})\delta^{(3)}(\vct{r}_2-\vct{r}'')\mathbb{G}^{\textrm{E}*}(\vct{r}'',\vct{r}_1;\omega'')\right\},
\label{eq:PPtraceE2}
\end{align}
\end{subequations}
where $ \mathcal{I} $ is the identity matrix. Next, we introduce operators
\begin{equation}
\mathbb{I}_i(\vct{r},\vct{r}') =
\begin{cases}
\mathcal{I}\delta^{(3)}(\vct{r}-\vct{r}'), & \vct{r},\vct{r}' \in V_i,\\
0, & \textrm{otherwise}, 
\end{cases}\ \ \ \ \ \ \ \
\bbeps_i(\vct{r},\vct{r}') =
\begin{cases}
\bbeps(\vct{r})\delta^{(3)}(\vct{r}-\vct{r}'), & \vct{r},\vct{r}' \in V_i,\\
0, & \textrm{otherwise},
\end{cases}\ \ \ \ \ \ \ \
\mathbb{V}_i = \frac{\omega^2}{c^2}\left(\bbeps_i-\mathbb{I}_i\right),
\label{eq:Ii}
\end{equation}
where $ \mathbb{V}_i $ is identified with the potential of object $ i $~\cite{Rahi2009, Kruger2012, Muller2017, Asheichyk2017}, allowing us to rewrite Eqs.~\eqref{eq:PPtraceH2} and~\eqref{eq:PPtraceE2} in the operator form,
\begin{subequations}
\begin{align}
\int_{V_1}\! d^3r_1 & \int_{V_2}\! d^3r_2\Tr\left\{\Im[\bbeps(\vct{r}_1,\omega)]\mathbb{G}^{\textrm{H}}(\vct{r}_1,\vct{r}_2;\omega')\mathbb{G}^{\textrm{H}\dagger}(\vct{r}_1,\vct{r}_2;\omega'')\right\} = \frac{c^2}{\omega^2}\Tr\left\{\Im[\mathbb{V}_1(\omega)]\mathbb{G}^{\textrm{H}}(\omega')\mathbb{I}_2\mathbb{G}^{\textrm{H}*}(\omega'')\right\},
\label{eq:PPtraceH3}\\
\notag \int_{V_1}\! d^3r_1 & \int_{V_2}\! d^3r_2\Tr\left\{\Im[\bbeps(\vct{r}_1,\omega)]\mathbb{G}^{\textrm{E}}(\vct{r}_1,\vct{r}_2;\omega')\bbeps(\vct{r}_2,\widetilde{\omega})\mathbb{G}^{\textrm{E}\dagger}(\vct{r}_1,\vct{r}_2;\omega'')\right\}\\
& = \frac{c^2}{\omega^2}\Tr\left\{\Im[\mathbb{V}_1(\omega)]\mathbb{G}^{\textrm{E}}(\omega')\mathbb{I}_2\mathbb{G}^{\textrm{E}*}(\omega'')\right\} + \frac{c^4}{\omega^2\widetilde{\omega}^2}\Tr\left\{\Im[\mathbb{V}_1(\omega)]\mathbb{G}^{\textrm{E}}(\omega')\mathbb{V}_2(\widetilde{\omega})\mathbb{G}^{\textrm{E}*}(\omega'')\right\}.
\label{eq:PPtraceE3}
\end{align}
\end{subequations}

The scattering operator of object $ i $ is defined as~\cite{Rahi2009, Kruger2012, Muller2017, Asheichyk2017}
\begin{equation}
\mathbb{T}_i = \mathbb{V}_i\left(\mathbb{I}-\mathbb{G}^{\textrm{E}}_0\mathbb{V}_i\right)^{-1},
\label{eq:Toperator}
\end{equation}
where $ \mathbb{G}^{\textrm{E}}_0 $ is the free-space electric GF and $ \mathbb{I}(\vct{r},\vct{r}') = \mathcal{I}\delta^{(3)}(\vct{r}-\vct{r}') $ is the identity operator. From Eq.~\eqref{eq:Toperator}, it follows that
\begin{equation}
\mathbb{V}_i = \mathbb{T}_i\left(\mathbb{I}+\mathbb{G}^{\textrm{E}}_0\mathbb{T}_i\right)^{-1}.
\label{eq:VviaT}
\end{equation}
The GF of the entire system can also be written in terms of the scattering operator~\cite{Rahi2009, Kruger2012, Muller2017, Asheichyk2017},
\begin{equation}
\mathbb{G}^{\textrm{E}} = \mathbb{G}^{\textrm{E}}_0 + \mathbb{G}^{\textrm{E}}_0\mathbb{T}\mathbb{G}^{\textrm{E}}_0,
\label{eq:GEviaT}
\end{equation}
where $ \mathbb{T} $ is the scattering operator of all objects, which can be expressed via $ \mathbb{G}^{\textrm{E}}_0 $ and the scattering operators of individual objects (see Appendix~G in Ref.~\cite{Muller2017}). Since~\cite{Tai1994}
\begin{equation}
\mathbb{G}^{\textrm{H}} = \nabla\times\mathbb{G}^{\textrm{E}} \equiv 
\begin{pmatrix}
0 & -\frac{\partial}{\partial z} & \frac{\partial}{\partial y}\\
\frac{\partial}{\partial z} & 0 & -\frac{\partial}{\partial x}\\
-\frac{\partial}{\partial y} & \frac{\partial}{\partial x} & 0
\end{pmatrix}
\mathbb{G}^{\textrm{E}},
\label{eq:GHviaGE}
\end{equation}
the relation between the magnetic GF and the scattering operator follows directly from Eq.~\eqref{eq:GEviaT},
\begin{equation}
\mathbb{G}^{\textrm{H}} = \mathbb{G}^{\textrm{H}}_0 + \mathbb{G}^{\textrm{H}}_0\mathbb{T}\mathbb{G}^{\textrm{E}}_0.
\label{eq:GHviaT}
\end{equation}
Substituting Eqs.~\eqref{eq:VviaT},~\eqref{eq:GEviaT}, and~\eqref{eq:GHviaT} into Eqs.~\eqref{eq:PPtraceH3} and~\eqref{eq:PPtraceE3}, we get
\begin{subequations}
\begin{align}
& \frac{c^2}{\omega^2}\Tr\left\{\Im[\mathbb{V}_1(\omega)]\mathbb{G}^{\textrm{H}}(\omega')\mathbb{I}_2\mathbb{G}^{\textrm{H}*}(\omega'')\right\} = \frac{c^2}{\omega^2}\Tr\left\{\Im[\mathbb{T}_1\left(\mathbb{I}+\mathbb{G}^{\textrm{E}}_0\mathbb{T}_1\right)^{-1}(\omega)][\mathbb{G}^{\textrm{H}}_0 + \mathbb{G}^{\textrm{H}}_0\mathbb{T}\mathbb{G}^{\textrm{E}}_0](\omega')\mathbb{I}_2[\mathbb{G}^{\textrm{H}*}_0 + \mathbb{G}^{\textrm{H}*}_0\mathbb{T}^*\mathbb{G}^{\textrm{E}*}_0](\omega'')\right\},
\label{eq:PPtraceH4}\\
\notag & \frac{c^2}{\omega^2}\Tr\left\{\Im[\mathbb{V}_1(\omega)]\mathbb{G}^{\textrm{E}}(\omega')\mathbb{I}_2\mathbb{G}^{\textrm{E}*}(\omega'')\right\} + \frac{c^4}{\omega^2\widetilde{\omega}^2}\Tr\left\{\Im[\mathbb{V}_1(\omega)]\mathbb{G}^{\textrm{E}}(\omega')\mathbb{V}_2(\widetilde{\omega})\mathbb{G}^{\textrm{E}*}(\omega'')\right\}\\
\notag & \ \ \ \ = \frac{c^2}{\omega^2}\Tr\left\{\Im[\mathbb{T}_1\left(\mathbb{I}+\mathbb{G}^{\textrm{E}}_0\mathbb{T}_1\right)^{-1}(\omega)][\mathbb{G}^{\textrm{E}}_0 + \mathbb{G}^{\textrm{E}}_0\mathbb{T}\mathbb{G}^{\textrm{E}}_0](\omega')\mathbb{I}_2[\mathbb{G}^{\textrm{E}*}_0 + \mathbb{G}^{\textrm{E}*}_0\mathbb{T}^*\mathbb{G}^{\textrm{E}*}_0](\omega'')\right\}\\
& \ \ \ \ \ \ \ + \frac{c^4}{\omega^2\widetilde{\omega}^2}\Tr\left\{\Im[\mathbb{T}_1\left(\mathbb{I}+\mathbb{G}^{\textrm{E}}_0\mathbb{T}_1\right)^{-1}(\omega)][\mathbb{G}^{\textrm{E}}_0 + \mathbb{G}^{\textrm{E}}_0\mathbb{T}\mathbb{G}^{\textrm{E}}_0](\omega')\mathbb{T}_2\left(\mathbb{I}+\mathbb{G}^{\textrm{E}}_0\mathbb{T}_2\right)^{-1}(\widetilde{\omega})[\mathbb{G}^{\textrm{E}*}_0 + \mathbb{G}^{\textrm{E}*}_0\mathbb{T}^*\mathbb{G}^{\textrm{E}*}_0](\omega'')\right\}.
\label{eq:PPtraceE4}
\end{align}
\end{subequations}

In the point particle limit, we leave only terms linear in $ \mathbb{T}_1 $ and $ \mathbb{T}_2 $. Note that the terms $ \sim \mathbb{I}_2\mathbb{T}_2 $ are also neglected, because they imply two integrals over a small volume $ V_2 $. Since, up to the linear order, $ \mathbb{T}_i\left(\mathbb{I}+\mathbb{G}^{\textrm{E}}_0\mathbb{T}_i\right)^{-1} = \mathbb{T}_i $, only the part of $ \mathbb{T} $ which contains neither $ \mathbb{T}_1 $ nor $ \mathbb{T}_2 $ should be left in Eqs.~\eqref{eq:PPtraceH4} and~\eqref{eq:PPtraceE4}. This part is $ \mathbb{T}_{\overline{12}} $, the scattering operator of the system not including particles $ 1 $ and $ 2 $~\cite{Muller2017}. $ \mathbb{G}^{\textrm{H}}_0 + \mathbb{G}^{\textrm{H}}_0\mathbb{T}_{\overline{12}}\mathbb{G}^{\textrm{E}}_0 $ and $ \mathbb{G}^{\textrm{E}}_0 + \mathbb{G}^{\textrm{E}}_0\mathbb{T}_{\overline{12}}\mathbb{G}^{\textrm{E}}_0 $ are then identified with the corresponding GFs, and Eqs.~\eqref{eq:PPtraceH4} and~\eqref{eq:PPtraceE4} simplify to
\begin{subequations}
\begin{align}
& \frac{c^2}{\omega^2}\Tr\left\{\Im[\mathbb{V}_1(\omega)]\mathbb{G}^{\textrm{H}}(\omega')\mathbb{I}_2\mathbb{G}^{\textrm{H}*}(\omega'')\right\} = \frac{c^2}{\omega^2}\Tr\left\{\Im[\mathbb{T}_1(\omega)]\mathbb{G}_{\overline{12}}^{\textrm{H}}(\omega')\mathbb{I}_2\mathbb{G}_{\overline{12}}^{\textrm{H}*}(\omega'')\right\},
\label{eq:PPtraceH5}\\
\notag & \frac{c^2}{\omega^2}\Tr\left\{\Im[\mathbb{V}_1(\omega)]\mathbb{G}^{\textrm{E}}(\omega')\mathbb{I}_2\mathbb{G}^{\textrm{E}*}(\omega'')\right\} + \frac{c^4}{\omega^2\widetilde{\omega}^2}\Tr\left\{\Im[\mathbb{V}_1(\omega)]\mathbb{G}^{\textrm{E}}(\omega')\mathbb{V}_2(\widetilde{\omega})\mathbb{G}^{\textrm{E}*}(\omega'')\right\}\\
& \ \ \ \ = \frac{c^2}{\omega^2}\Tr\left\{\Im[\mathbb{T}_1(\omega)]\mathbb{G}_{\overline{12}}^{\textrm{E}}(\omega')\mathbb{I}_2\mathbb{G}^{\textrm{E}*}_{\overline{12}}(\omega'')\right\} + \frac{c^4}{\omega^2\widetilde{\omega}^2}\Tr\left\{\Im[\mathbb{T}_1(\omega)]\mathbb{G}^{\textrm{E}}_{\overline{12}}(\omega')\mathbb{T}_2(\widetilde{\omega})\mathbb{G}^{\textrm{E}*}_{\overline{12}}(\omega'')\right\}.
\label{eq:PPtraceE5}
\end{align}
\end{subequations}
For $ \mathbb{T}_1 $ and $ \mathbb{T}_2 $, we use the scattering operator of a homogeneous isotropic nonmagnetic small sphere~\citep{Asheichyk2017},
\begin{equation}
\mathbb{T}_i(\vct{r},\vct{r}';\omega) = 3\frac{\omega^2}{c^2}\frac{\varepsilon_i(\omega)-1}{\varepsilon_i(\omega)+2}\mathbb{I}_i(\vct{r},\vct{r}'),
\label{eq:Tss}
\end{equation}
such that Eqs.~\eqref{eq:PPtraceH5} and~\eqref{eq:PPtraceE5} become
\begin{subequations}
\begin{align}
& \frac{c^2}{\omega^2}\Tr\left\{\Im[\mathbb{T}_1(\omega)]\mathbb{G}_{\overline{12}}^{\textrm{H}}(\omega')\mathbb{I}_2\mathbb{G}_{\overline{12}}^{\textrm{H}*}(\omega'')\right\} = 3\Im\left[\frac{\varepsilon_1(\omega)-1}{\varepsilon_1(\omega)+2}\right]\int_{V_1}\! d^3r_1\int_{V_2}\! d^3r_2\Tr\left\{\mathbb{G}_{\overline{12}}^{\textrm{H}}(\vct{r}_1,\vct{r}_2;\omega')\mathbb{G}_{\overline{12}}^{\textrm{H}\dagger}(\vct{r}_1,\vct{r}_2;\omega'')\right\},
\label{eq:PPtraceH6}\\
\notag & \frac{c^2}{\omega^2}\Tr\left\{\Im[\mathbb{T}_1(\omega)]\mathbb{G}_{\overline{12}}^{\textrm{E}}(\omega')\mathbb{I}_2\mathbb{G}^{\textrm{E}*}_{\overline{12}}(\omega'')\right\} + \frac{c^4}{\omega^2\widetilde{\omega}^2}\Tr\left\{\Im[\mathbb{T}_1(\omega)]\mathbb{G}^{\textrm{E}}_{\overline{12}}(\omega')\mathbb{T}_2(\widetilde{\omega})\mathbb{G}^{\textrm{E}*}_{\overline{12}}(\omega'')\right\}\\
\notag & \ \ \ \ = 3\Im\left[\frac{\varepsilon_1(\omega)-1}{\varepsilon_1(\omega)+2}\right]\int_{V_1}\! d^3r_1\int_{V_2}\! d^3r_2\Tr\left\{\mathbb{G}_{\overline{12}}^{\textrm{E}}(\vct{r}_1,\vct{r}_2;\omega')\mathbb{G}_{\overline{12}}^{\textrm{E}\dagger}(\vct{r}_1,\vct{r}_2;\omega'')\right\}\\
& \ \ \ \ \ \ \ \ + 9\Im\left[\frac{\varepsilon_1(\omega)-1}{\varepsilon_1(\omega)+2}\right]\frac{\varepsilon_2(\widetilde{\omega})-1}{\varepsilon_2(\widetilde{\omega})+2}\int_{V_1}\! d^3r_1\int_{V_2}\! d^3r_2\Tr\left\{\mathbb{G}_{\overline{12}}^{\textrm{E}}(\vct{r}_1,\vct{r}_2;\omega')\mathbb{G}_{\overline{12}}^{\textrm{E}\dagger}(\vct{r}_1,\vct{r}_2;\omega'')\right\}.
\label{eq:PPtraceE6}
\end{align}
\end{subequations}
Furthermore, since the GFs hardly vary for different points inside each particle, we can fix $ \vct{r}_1 $ and $ \vct{r}_2 $ to be positions of the particles' centers, such that the integrals yield the volumes,
\begin{subequations}
\begin{align}
\notag & 3\Im\left[\frac{\varepsilon_1(\omega)-1}{\varepsilon_1(\omega)+2}\right]\int_{V_1}\! d^3r_1\int_{V_2}\! d^3r_2\Tr\left\{\mathbb{G}_{\overline{12}}^{\textrm{H}}(\vct{r}_1,\vct{r}_2;\omega')\mathbb{G}_{\overline{12}}^{\textrm{H}\dagger}(\vct{r}_1,\vct{r}_2;\omega'')\right\}\\
& \ \ \ \ = 4\pi\Im[\alpha_1(\omega)]V_2\Tr\left\{\mathbb{G}_{\overline{12}}^{\textrm{H}}(\vct{r}_1,\vct{r}_2;\omega')\mathbb{G}_{\overline{12}}^{\textrm{H}\dagger}(\vct{r}_1,\vct{r}_2;\omega'')\right\},
\label{eq:PPtraceHfinal}\\
\notag & 3\Im\left[\frac{\varepsilon_1(\omega)-1}{\varepsilon_1(\omega)+2}\right]\int_{V_1}\! d^3r_1\int_{V_2}\! d^3r_2\Tr\left\{\mathbb{G}_{\overline{12}}^{\textrm{E}}(\vct{r}_1,\vct{r}_2;\omega')\mathbb{G}_{\overline{12}}^{\textrm{E}\dagger}(\vct{r}_1,\vct{r}_2;\omega'')\right\}\\
\notag & \ \ \ \ \ \ \ \ \ \ \ \ + 9\Im\left[\frac{\varepsilon_1(\omega)-1}{\varepsilon_1(\omega)+2}\right]\frac{\varepsilon_2(\widetilde{\omega})-1}{\varepsilon_2(\widetilde{\omega})+2}\int_{V_1}\! d^3r_1\int_{V_2}\! d^3r_2\Tr\left\{\mathbb{G}_{\overline{12}}^{\textrm{E}}(\vct{r}_1,\vct{r}_2;\omega')\mathbb{G}_{\overline{12}}^{\textrm{E}\dagger}(\vct{r}_1,\vct{r}_2;\omega'')\right\}\\
& \ \ \ \ = 4\pi\Im[\alpha_1(\omega)]V_2\Tr\left\{\mathbb{G}_{\overline{12}}^{\textrm{E}}(\vct{r}_1,\vct{r}_2;\omega')\mathbb{G}_{\overline{12}}^{\textrm{E}\dagger}(\vct{r}_1,\vct{r}_2;\omega'')\right\} + 16\pi^2\Im[\alpha_1(\omega)]\alpha_2(\widetilde{\omega})\Tr\left\{\mathbb{G}_{\overline{12}}^{\textrm{E}}(\vct{r}_1,\vct{r}_2;\omega')\mathbb{G}_{\overline{12}}^{\textrm{E}\dagger}(\vct{r}_1,\vct{r}_2;\omega'')\right\},
\label{eq:PPtraceEfinal}
\end{align}
\end{subequations}
where $ \alpha_i $ is the particle polarizability defined in Eq.~\eqref{eq:polarizability}.

Using Eqs.~\eqref{eq:PPtraceHfinal} and~\eqref{eq:PPtraceEfinal} for the traces (together with the spatial integrals) in Eqs.~\eqref{eq:PPtraceH1} and~\eqref{eq:PPtraceE1}, respectively, performing the inverse Fourier transforms, substituting them into Eq.~\eqref{eq:Flux}, and using
\begin{align}
\notag \int_0^{\infty}\! dt' & \int_0^{\infty}\! dt''\cos[\omega(t'-t'')]\Tr\left\{\frac{\partial\mathbb{G}_{\overline{12}}^{\textrm{E}}(\vct{r}_1,\vct{r}_2;t-t')}{\partial t}\mathbb{G}_{\overline{12}}^{\textrm{E}T}(\vct{r}_1,\vct{r}_2;t-t'')\right\}\\
& = \frac{1}{2}\frac{\partial}{\partial t}\int_0^{\infty}\! dt' \int_0^{\infty}\! dt''\cos[\omega(t'-t'')]\Tr\left\{\mathbb{G}_{\overline{12}}^{\textrm{E}}(\vct{r}_1,\vct{r}_2;t-t')\mathbb{G}_{\overline{12}}^{\textrm{E}T}(\vct{r}_1,\vct{r}_2;t-t'')\right\},
\label{eq:derivtive_identity}
\end{align}
we arrive at Eq.~\eqref{eq:FluxPP}.

\section{Derivation of the stationary magnetic energy density in the point particle limit}
\label{app:sec:uHst}
Here we derive Eq.~\eqref{eq:uHst} starting from Eq.~\eqref{eq:uH}. As in Appendix~\ref{app:subsec:FluxSt_from_tdep}, the stationary limit of Eq.~\eqref{eq:uH} can be obtained by changing the lower integration limits from $ 0 $ to $ -\infty $. Performing the Fourier transform of the GF according to Eq.~\eqref{eq:GHtviaGHomega}, and using the Euler's formula for the cosine, we get
\begin{align}
\notag u_{1\textrm{Hst}}^{(2)} = \ & \frac{\hbar}{2\pi^2c^2}\int_0^{\infty}\! d\omega \frac{\omega^2}{e^{\frac{\omega}{\omega_{T_1}}}-1}\Im[\alpha_1(\omega)]\int\! d\omega' \int\! d\omega'' \Tr\left\{\mathbb{G}_{\overline{12}}^{\textrm{H}}(\vct{r}_1,\vct{r}_2;\omega')\mathbb{G}_{\overline{12}}^{\textrm{H}T}(\vct{r}_1,\vct{r}_2;\omega'')\right\}e^{-i(\omega'+\omega'')t}\\
& \times \int\! dt' \int\! dt'' \left[e^{i(\omega'+\omega)t'}e^{i(\omega''-\omega)t''}+e^{i(\omega'-\omega)t'}e^{i(\omega''+\omega)t''}\right].
\label{eq:uHst1}
\end{align}
The time integrals give delta functions,
\begin{align}
\notag u_{1\textrm{Hst}}^{(2)} = \ & \frac{2\hbar}{c^2}\int_0^{\infty}\! d\omega \frac{\omega^2}{e^{\frac{\omega}{\omega_{T_1}}}-1}\Im[\alpha_1(\omega)]\int\! d\omega' \int\! d\omega'' \Tr\left\{\mathbb{G}_{\overline{12}}^{\textrm{H}}(\vct{r}_1,\vct{r}_2;\omega')\mathbb{G}_{\overline{12}}^{\textrm{H}T}(\vct{r}_1,\vct{r}_2;\omega'')\right\}e^{-i(\omega'+\omega'')t}\\
& \times \left[\delta(\omega'+\omega)\delta(\omega''-\omega)+\delta(\omega'-\omega)\delta(\omega''+\omega)\right],
\label{eq:uHst2}
\end{align}
and the last two frequency integrals can be evaluated, canceling the dependence on $ t $,
\begin{equation}
u_{1\textrm{Hst}}^{(2)} = \frac{2\hbar}{c^2}\int_0^{\infty}\! d\omega \frac{\omega^2}{e^{\frac{\omega}{\omega_{T_1}}}-1}\Im[\alpha_1(\omega)] \Tr\left\{\mathbb{G}_{\overline{12}}^{\textrm{H}}(\vct{r}_1,\vct{r}_2;\omega)\mathbb{G}_{\overline{12}}^{\textrm{H}T}(\vct{r}_1,\vct{r}_2;-\omega)+\mathbb{G}_{\overline{12}}^{\textrm{H}}(\vct{r}_1,\vct{r}_2;-\omega)\mathbb{G}_{\overline{12}}^{\textrm{H}T}(\vct{r}_1,\vct{r}_2;\omega)\right\}.
\label{eq:uHst3}
\end{equation}
Since $ \mathbb{G}^{\textrm{H}}(-\omega) = \mathbb{G}^{\textrm{H}*}(\omega) $ (owing to the reality of the GF in time domain), the second term inside the trace is the complex conjugate of the first one, such that
\begin{equation}
u_{1\textrm{Hst}}^{(2)} = \frac{4\hbar}{c^2}\int_0^{\infty}\! d\omega \frac{\omega^2}{e^{\frac{\omega}{\omega_{T_1}}}-1}\Im[\alpha_1(\omega)] \Re\Tr\left\{\mathbb{G}_{\overline{12}}^{\textrm{H}}(\vct{r}_1,\vct{r}_2;\omega)\mathbb{G}_{\overline{12}}^{\textrm{H}\dagger}(\vct{r}_1,\vct{r}_2;\omega)\right\},
\label{eq:uHst4}
\end{equation}
where we used $ \mathbb{G}^{\textrm{H}T}(-\omega) = \mathbb{G}^{\textrm{H}\dagger}(\omega) $. Since the matrix inside the trace of Eq.~\eqref{eq:uHst4} is a positive semi-definite one~\cite{Horn2012}, its trace is a real nonnegative number~\cite{Horn2012}, such that Eq.~\eqref{eq:uHst} follows from Eq.~\eqref{eq:uHst4}.

The derivation of Eq.~\eqref{eq:uE0st} from Eq.~\eqref{eq:uE0} is analogous. However, one should take into account the difference between the Fourier transforms for the electric and magnetic GFs [compare Eqs.~\eqref{eq:GEtviaGEomega} and~\eqref{eq:GHtviaGHomega}].

\section{Evaluation of the interaction term for the nonstationary heat flux between two isolated particles}
\label{app:sec:FluxPPVacInt}
\subsection{Far-field term}
\label{app:subsec:d2}
In this appendix, we evaluate the time integrals of the third term of Eq.~\eqref{eq:FluxPP} for two particles in vacuum. The trace of the GFs is given by Eq.~\eqref{eq:TrGEprimeGETt}, and polarizability $ \alpha_2(t) $ is given by Eq.~\eqref{eq:alphat} (here, we consider only the contribution of its second term).

We start with the far-field term,
\begin{equation}
I_{d^{-2}} \equiv \frac{2}{c^4d^2}\frac{3\omega_{\textrm{p}}^2R_2^3}{(\varepsilon_{\infty}+2)^2\beta}\int_0^{\infty}\! dt' \int_0^{\infty}\! dt''\cos[\omega(t'-t'')]\int\! d\tilde{t}f(\tilde{t})\delta''(\tau-\tilde{t}-t')\delta'(\tau-t''),
\label{eq:Id2}
\end{equation}
where 
\begin{equation}
f(\tilde{t}) \equiv e^{-\frac{\gamma}{2}\tilde{t}}\sin(\beta \tilde{t})\theta(\tilde{t}),
\label{eq:f}
\end{equation}
and the derivatives of the delta functions are with respect to their full arguments. In the integral over $ \tilde{t} $, the double derivative acting on the delta function can be moved to $ f $ by applying partial integration, such that the integral equals the double derivative of $ f $ evaluated at $ \tau - t' $,
\begin{equation}
\int\! d\tilde{t}f(\tilde{t})\delta''(\tau-\tilde{t}-t') = \int\! d\tilde{t}\frac{d^2f(\tilde{t})}{d\tilde{t}^2}\delta(\tau-\tilde{t}-t') = \frac{d^2f(\tilde{t})}{d\tilde{t}^2}\Bigg|_{\tilde{t}=\tau-t'}.
\label{eq:Id2_1}
\end{equation}
The integral over $ t'' $ can be performed in a similar way, but zero in the lower limit leads to a more complicated structure,
\begin{align}
\notag \int_0^{\infty}\! dt''\cos[\omega(t'-t'')]\delta'(\tau-t'') & = -\cos[\omega(t'-t'')]\delta(\tau-t'')\big|_{t''=0}^{\infty} + \int_0^{\infty}\! dt''\frac{d\cos[\omega(t'-t'')]}{dt''}\delta(\tau-t'')\\
& = \cos(\omega t')\delta(\tau) + \omega\int_0^{\infty}\! dt''\sin[\omega(t'-t'')]\delta(\tau-t'').
\label{eq:Id2_2}
\end{align}
Further evaluation of Eq.~\eqref{eq:Id2_2} depends on the sign of $ \tau $. If $ \tau < 0 $, both terms in Eq.~\eqref{eq:Id2_2} are zero, and hence $ I_{d^{-2}} = 0 $. For $ \tau = 0 $, $ \delta(\tau) $ remains; we do not consider this case in the subsequent computations. If $ \tau > 0 $, the first term vanishes, whereas the second one evaluates to $ \omega\sin[\omega(t'-\tau)] $. Using Eqs.~\eqref{eq:Id2_1} and~\eqref{eq:Id2_2} in Eq.~\eqref{eq:Id2}, we get for $ \tau > 0 $,
\begin{equation}
I_{d^{-2}} = \frac{2}{c^4d^2}\frac{3\omega_{\textrm{p}}^2R_2^3}{(\varepsilon_{\infty}+2)^2\beta}\omega\int_0^{\infty}\! dt' \sin[\omega(t'-\tau)]\frac{d^2f(\tilde{t})}{d\tilde{t}^2}\Bigg|_{\tilde{t}=\tau-t'}.
\label{eq:Id2_3}
\end{equation}

Making the substitution $ \tau - t' \equiv s $ and using partial integration, the integral in Eq.~\eqref{eq:Id2_3} can be rewritten as
\begin{align}
\notag \int_0^{\infty}\! dt' \sin[\omega(t'-\tau)]\frac{d^2f(\tilde{t})}{d\tilde{t}^2}\Bigg|_{\tilde{t}=\tau-t'} & = -\int_{-\infty}^{\tau}\! dt' \sin(\omega s)f''(s) = -\sin(\omega s)f'(s)\big|_{s=-\infty}^{\tau} + \omega\int_{-\infty}^{\tau}\! ds \cos(\omega s)f'(s)\\
& = -\sin(\omega s)f'(s)\big|_{s=-\infty}^{\tau} + \omega\cos(\omega s)f(s)\big|_{s=-\infty}^{\tau} + \omega^2\int_{-\infty}^{\tau}\! ds\sin(\omega s)f(s).
\label{eq:Id2_4}
\end{align}
The lower limits of the first two terms in the last line of Eq.~\eqref{eq:Id2_4} are zero, because $ f(s) \propto \theta(s) $. For the same reason, the integral in the third term can start from zero. Substituting $ f $ from Eq.~\eqref{eq:f}, we get
\begin{align}
\notag \int_0^{\infty}\! dt' \sin[\omega(t'-\tau)]\frac{d^2f(\tilde{t})}{d\tilde{t}^2}\Bigg|_{\tilde{t}=\tau-t'} & = -\sin(\omega\tau)\left[e^{-\frac{\gamma}{2}\tau}\sin(\beta\tau)\right]' + e^{-\frac{\gamma}{2}\tau}\omega\sin(\beta\tau)\cos(\omega\tau) + \omega^2\int_0^{\tau}\! dse^{-\frac{\gamma}{2}s}\sin(\beta s)\sin(\omega s)\\
\notag & = e^{-\frac{\gamma}{2}\tau}\left[\frac{\gamma}{2}\sin(\beta\tau)\sin(\omega\tau) - \beta\cos(\beta\tau)\sin(\omega\tau) + \omega\sin(\beta\tau)\cos(\omega\tau)\right]\\
& \ \ \ \ \ \ \ \ \ \ \ \ \ \ \ \ + \frac{\omega^2}{2}\Bigg\{\int_0^{\tau}\! dse^{-\frac{\gamma}{2}s}\cos[(\beta-\omega)s]-\int_0^{\tau}\! dse^{-\frac{\gamma}{2}s}\cos[(\beta+\omega)s]\Bigg\}.
\label{eq:Id2_5}
\end{align}
The integrals in the last line of Eq.~\eqref{eq:Id2_5} can be evaluated using the Euler's formula for the cosine,
\begin{align}
\notag & \int_0^{\tau}\! dse^{-\frac{\gamma}{2}s}\cos[(\beta-\omega)s]-\int_0^{\tau}\! dse^{-\frac{\gamma}{2}s}\cos[(\beta+\omega)s] = \frac{1}{2\left[\frac{\gamma^2}{4}+(\omega-\beta)^2\right]}\Big\{\gamma-e^{-\frac{\gamma}{2}\tau}\big[\gamma\cos[(\beta-\omega)\tau]\\
\notag & \ \ \ \ \ \ \ \ -2(\beta-\omega)\sin[(\beta-\omega)\tau]\big]\Big\} - \frac{1}{2\left[\frac{\gamma^2}{4}+(\omega+\beta)^2\right]}\Big\{\gamma-e^{-\frac{\gamma}{2}\tau}\big[\gamma\cos[(\beta+\omega)\tau]-2(\beta+\omega)\sin[(\beta+\omega)\tau]\big]\Big\}\\
\notag & = \frac{1}{\left(\omega^2-\omega_{0\alpha}^2\right)^2+\gamma^2\omega^2}\Bigg\{2\gamma\beta\omega - e^{-\frac{\gamma}{2}\tau}\Bigg[2\gamma\beta\omega\cos(\beta\tau)\cos(\omega\tau) + \gamma\left(\omega^2+\omega_{0\alpha}^2\right)\sin(\beta\tau)\sin(\omega\tau)\\
& \ \ \ \ \ \ \ \ \ \ \ \ \ \ \ \ \ \ \ \ \ \ \ \ \ \ \ \ \ \ \ \ -2\beta\left(\omega^2-\omega_{0\alpha}^2\right)\cos(\beta\tau)\sin(\omega\tau) + 2\omega\left(\omega^2-\omega_{0\alpha}^2+\frac{\gamma^2}{2}\right)\sin(\beta\tau)\cos(\omega\tau)\Bigg]\Bigg\},
\label{eq:Id2_6}
\end{align}
where we used the definition of $ \beta $ in Sec.~\ref{subsec:DL}. Substituting the result of Eq.~\eqref{eq:Id2_6} into Eq.~\eqref{eq:Id2_5}, combining the terms with different powers of $ \gamma $, and substituting the result into Eq.~\eqref{eq:Id2_3}, we obtain
\begin{align}
\notag & \frac{c^4d^2}{2\omega^3}I_{d^{-2}} = \frac{3\omega_{\textrm{p}}^2R_2^3}{(\varepsilon_{\infty}+2)^2}\frac{\gamma\omega}{\left(\omega^2-\omega_{0\alpha}^2\right)^2+\gamma^2\omega^2} + \frac{3\omega_{\textrm{p}}^2R_2^3}{(\varepsilon_{\infty}+2)^2}\frac{\gamma\omega}{\left(\omega^2-\omega_{0\alpha}^2\right)^2+\gamma^2\omega^2}e^{-\frac{\gamma}{2}\tau}\Bigg\{-\cos(\beta\tau)\cos(\omega\tau)\\
\notag & -\frac{\omega_{0\alpha}^2\left(3\omega^2-\omega_{0\alpha}^2\right)}{2\beta\omega^3}\sin(\beta\tau)\sin(\omega\tau)-\frac{\gamma}{\omega}\left(\cos(\beta\tau)\sin(\omega\tau)-\frac{\omega}{2\beta}\sin(\beta\tau)\cos(\omega\tau)\right)+\frac{\gamma^2}{2\beta\omega}\sin(\beta\tau)\sin(\omega\tau)\Bigg\}\\
& - \frac{3\omega_{\textrm{p}}^2R_2^3}{(\varepsilon_{\infty}+2)^2}\frac{\omega^2-\omega_{0\alpha}^2}{\left(\omega^2-\omega_{0\alpha}^2\right)^2+\gamma^2\omega^2}\frac{\omega_{0\alpha}^2}{\omega^2} e^{-\frac{\gamma}{2}\tau}\Bigg\{-\cos(\beta\tau)\sin(\omega\tau)+\frac{\omega}{\beta}\sin(\beta\tau)\cos(\omega\tau)\Bigg\}.
\label{eq:Id2_final}
\end{align}
Comparing Eq.~\eqref{eq:Id2_final} with Eqs.~\eqref{eq:alphaDLRe} and~\eqref{eq:alphaDLIm}, we can see that the first two terms of Eq.~\eqref{eq:Id2_final} are proportional to $ \Im[\alpha_2(\omega)] $ and they vanish if $ \gamma \to 0 $, whereas the third term is proportional to $ \Re[\alpha_2(\omega)] - \alpha_{\infty} $ and it remains finite as $ \gamma \to 0 $. We hence conclude that the first two terms correspond to the heat transfer, while the third term is the contribution to the nondissipative energy change. The first term, being time independent, corresponds to the stationary heat transfer, while the last two terms vanish as $ \tau \to \infty $.

\subsection{The term proportional to the inverse distance to the third power}
\label{app:subsec:d3}
For the $ d^{-3} $ term, we have
\begin{equation}
I_{d^{-3}} \equiv \frac{2}{c^3d^3}\frac{3\omega_{\textrm{p}}^2R_2^3}{(\varepsilon_{\infty}+2)^2\beta}\int_0^{\infty}\! dt' \int_0^{\infty}\! dt''\cos[\omega(t'-t'')]\int\! d\tilde{t}f(\tilde{t})\left[\delta'(\tau-\tilde{t}-t')\delta'(\tau-t'')+\delta''(\tau-\tilde{t}-t')\delta(\tau-t'')\right].
\label{eq:Id3}
\end{equation}
For the first term in the brackets, the integral over $ t'' $ is given by Eq.~\eqref{eq:Id2_2}: it is zero for $ \tau < 0 $ and evaluates to $ \omega\sin[\omega(t'-\tau)] $ for $ \tau > 0 $. For the second term, the integral over $ t'' $ is zero for $ \tau < 0 $ and evaluates to $ \cos[\omega(t'-\tau)] $ for $ \tau > 0 $. Therefore, for $ \tau > 0 $, we have
\begin{equation}
I_{d^{-3}} = \frac{2}{c^3d^3}\frac{3\omega_{\textrm{p}}^2R_2^3}{(\varepsilon_{\infty}+2)^2\beta}\left[\omega\int_0^{\infty}\! dt'\int\! d\tilde{t}f(\tilde{t})\delta'(\tau-\tilde{t}-t')\sin[\omega(t'-\tau)] + \int_0^{\infty}\! dt'\int\! d\tilde{t}f(\tilde{t})\delta''(\tau-\tilde{t}-t')\cos[\omega(t'-\tau)]\right].
\label{eq:Id3_1}
\end{equation}
For $ \tau < 0 $, $ I_{d^{-3}}  = 0 $. Performing the integral over $ t' $ in the second term of Eq.~\eqref{eq:Id3_1},
\begin{equation}
\int_0^{\infty}\! dt'\delta''(\tau-\tilde{t}-t')\cos[\omega(t'-\tau)] = \cos(\omega\tau)\delta'(\tau-\tilde{t}) -\omega\int_0^{\infty}\! dt'\delta'(\tau-\tilde{t}-t')\sin[\omega(t'-\tau)], 
\label{eq:Id3_2}
\end{equation}
we can see that the second term of Eq.~\eqref{eq:Id3_2}, being substituted into the second term of Eq.~\eqref{eq:Id3_1}, cancels the first term of Eq.~\eqref{eq:Id3_1}. The remaining part reads as
\begin{equation}
I_{d^{-3}} = \frac{2}{c^3d^3}\frac{3\omega_{\textrm{p}}^2R_2^3}{(\varepsilon_{\infty}+2)^2\beta}\cos(\omega\tau)\int\! d\tilde{t}f(\tilde{t})\delta'(\tau-\tilde{t}) = \frac{2}{c^3d^3}\frac{3\omega_{\textrm{p}}^2R_2^3}{(\varepsilon_{\infty}+2)^2\beta}\cos(\omega\tau)f'(\tau).
\label{eq:Id3_3}
\end{equation}

Using Eq.~\eqref{eq:f}, we find
\begin{equation}
I_{d^{-3}} = \frac{2}{c^3d^3}\frac{3\omega_{\textrm{p}}^2R_2^3}{(\varepsilon_{\infty}+2)^2}e^{-\frac{\gamma}{2}\tau}\left[-\frac{\gamma}{2\beta}\sin(\beta\tau)\cos(\omega\tau)+\cos(\beta\tau)\cos(\omega\tau)\right].
\label{eq:Id3_4}
\end{equation}

In contrast to the far-field contribution in Eq.~\eqref{eq:Id2_final}, Eq.~\eqref{eq:Id3_4} is more simple but shows no direct proportionality to real or imaginary part of $ \alpha_2(\omega) $. These parts can be identified by multiplying and dividing the first term of Eq.~\eqref{eq:Id3_4} by $ \omega\left[\left(\omega^2-\omega_{0\alpha}^2\right)^2+\gamma^2\omega^2\right] $ and by using $ 1 = \left(\omega^2-\omega_{0\alpha}^2\right)^2\Big/\left[\left(\omega^2-\omega_{0\alpha}^2\right)^2+\gamma^2\omega^2\right] + \gamma^2\omega^2\Big/\left[\left(\omega^2-\omega_{0\alpha}^2\right)^2+\gamma^2\omega^2\right] $ in the second term,
\begin{align}
\notag I_{d^{-3}} = & \ \frac{2}{c^3d^3}\frac{3\omega_{\textrm{p}}^2R_2^3}{(\varepsilon_{\infty}+2)^2}e^{-\frac{\gamma}{2}\tau}\Bigg\{-\frac{\gamma\omega}{\left(\omega^2-\omega_{0\alpha}^2\right)^2+\gamma^2\omega^2}\frac{\left(\omega^2-\omega_{0\alpha}^2\right)^2+\gamma^2\omega^2}{2\beta\omega}\sin(\beta\tau)\cos(\omega\tau)\\
& +\frac{\gamma\omega}{\left(\omega^2-\omega_{0\alpha}^2\right)^2+\gamma^2\omega^2}\gamma\omega\cos(\beta\tau)\cos(\omega\tau)+\frac{\omega^2-\omega_{0\alpha}^2}{\left(\omega^2-\omega_{0\alpha}^2\right)^2+\gamma^2\omega^2}\left(\omega^2-\omega_{0\alpha}^2\right)\cos(\beta\tau)\cos(\omega\tau)\Bigg\},
\label{eq:Id3_5}
\end{align}
which can be rewritten as
\begin{align}
\notag \frac{c^3d^3}{2\omega^2}I_{d^{-3}} = & \ \frac{3\omega_{\textrm{p}}^2R_2^3}{(\varepsilon_{\infty}+2)^2}\frac{\gamma\omega}{\left(\omega^2-\omega_{0\alpha}^2\right)^2+\gamma^2\omega^2}e^{-\frac{\gamma}{2}\tau}\Bigg\{-\frac{\left(\omega^2-\omega_{0\alpha}^2\right)^2}{2\beta\omega^3}\sin(\beta\tau)\cos(\omega\tau)+\frac{\gamma}{\omega}\cos(\beta\tau)\cos(\omega\tau)\\
& -\frac{\gamma^2}{2\beta\omega}\sin(\beta\tau)\cos(\omega\tau)\Bigg\} + \frac{3\omega_{\textrm{p}}^2R_2^3}{(\varepsilon_{\infty}+2)^2}\frac{\omega^2-\omega_{0\alpha}^2}{\left(\omega^2-\omega_{0\alpha}^2\right)^2+\gamma^2\omega^2}e^{-\frac{\gamma}{2}\tau}\frac{\omega^2-\omega_{0\alpha}^2}{\omega^2}\cos(\beta\tau)\cos(\omega\tau),
\label{eq:Id3_final}
\end{align}
where the first term is proportional $ \Im[\alpha_2(\omega)] $ and corresponds to the heat transfer, while the second term is proportional to $ \Re[\alpha_2(\omega)] - \alpha_{\infty} $ and corresponds to the energy change. As expected, $ I_{d^{-3}} $ contains no steady-state part and vanishes as $ \tau \to \infty $.

\subsection{The term proportional to the inverse distance to the fourth power}
\label{app:subsec:d4}  
The $ d^{-4} $ term contains the following integrals:
\begin{align}
\notag & I_{d^{-4}} \equiv \frac{2}{c^2d^4}\frac{3\omega_{\textrm{p}}^2R_2^3}{(\varepsilon_{\infty}+2)^2\beta}\\
& \times\int_0^{\infty}\! dt' \int_0^{\infty}\! dt''\cos[\omega(t'-t'')]\int\! d\tilde{t}f(\tilde{t})\left[\delta(\tau-\tilde{t}-t')\delta'(\tau-t'')+3\delta'(\tau-\tilde{t}-t')\delta(\tau-t'')+\delta''(\tau-\tilde{t}-t')\theta(\tau-t'')\right].
\label{eq:Id4}
\end{align}
The integrals of the first term can be evaluated using Eq.~\eqref{eq:Id2_2},
\begin{align}
\notag & \int_0^{\infty}\! dt' \int_0^{\infty}\! dt''\cos[\omega(t'-t'')]\int\! d\tilde{t}f(\tilde{t})\delta(\tau-\tilde{t}-t')\delta'(\tau-t'') \overset{\tau > 0}{=} \omega\int_0^{\infty}\! dt'\sin[\omega(t'-\tau)]\int\! d\tilde{t}f(\tilde{t})\delta(\tau-\tilde{t}-t')\\
& = -\omega\int_0^{\infty}\! dt'\sin[\omega(\tau-t')]f(\tau-t') \overset{\tau-t' \equiv s}{=} -\omega\int_{-\infty}^{\tau}\! ds\sin(\omega s)f(s).
\label{eq:Id4_1}
\end{align}
For the second term, using partial integration, we obtain
\begin{align}
\notag & 3\int_0^{\infty}\! dt' \int_0^{\infty}\! dt''\cos[\omega(t'-t'')]\int\! d\tilde{t}f(\tilde{t})\delta'(\tau-\tilde{t}-t')\delta(\tau-t'') \overset{\tau > 0}{=} 3\int\! d\tilde{t}f(\tilde{t})\int_0^{\infty}\! dt'\cos[\omega(t'-\tau)]\delta'(\tau-\tilde{t}-t')\\
\notag & = 3\cos(\omega\tau)\int\! d\tilde{t}f(\tilde{t})\delta(\tau-\tilde{t}) - 3\omega\int\! d\tilde{t} f(\tilde{t})\int_0^{\infty}\! dt'\sin[\omega(t'-\tau)]\delta(\tau-\tilde{t}-t') = 3\cos(\omega\tau)f(\tau)\\
& + 3\omega\int_0^{\infty}\! dt'\sin[\omega(\tau-t')]f(\tau-t') \overset{\tau-t' \equiv s}{=} 3\cos(\omega\tau)f(\tau) + 3\omega\int_{-\infty}^{\tau}\! ds\sin(\omega s)f(s).
\label{eq:Id4_2}
\end{align}
The third term reads as
\begin{align}
\notag & \int_0^{\infty}\! dt' \int_0^{\infty}\! dt''\cos[\omega(t'-t'')]\int\! d\tilde{t}f(\tilde{t})\delta''(\tau-\tilde{t}-t')\theta(\tau-t'') \overset{\tau > 0}{=} \frac{1}{\omega}\int_0^{\infty}\! dt'\sin[\omega(\tau-t')]\int\! d\tilde{t}f(\tilde{t})\delta''(\tau-\tilde{t}-t')\\
\notag & + \frac{1}{\omega}\int_0^{\infty}\! dt'\sin(\omega t')\int\! d\tilde{t}f(\tilde{t})\delta''(\tau-\tilde{t}-t') = \frac{1}{\omega}\int_0^{\infty}\! dt' \sin[\omega(\tau-t')]\frac{d^2f(\tilde{t})}{d\tilde{t}^2}\Bigg|_{\tilde{t}=\tau-t'} + \frac{1}{\omega}\int_0^{\infty}\! dt' \sin(\omega t')\frac{d^2f(\tilde{t})}{d\tilde{t}^2}\Bigg|_{\tilde{t}=\tau-t'}\\
\notag & \overset{\tau-t' \equiv s}{=} \frac{\sin(\omega\tau)}{\omega}f'(\tau) - \cos(\omega\tau)f(\tau) -\omega\int_{-\infty}^{\tau}\! ds\sin(\omega s)f(s) + f(\tau) -\omega\int_{-\infty}^{\tau}\! ds\sin[\omega(\tau-s)]f(s)\\
& = \frac{\sin(\omega\tau)}{\omega}f'(\tau) + \left[1-\cos(\omega\tau)\right]f(\tau) -\omega\left[1-\cos(\omega\tau)\right]\int_{-\infty}^{\tau}ds\sin(\omega s)f(s) -\omega\sin(\omega\tau)\int_{-\infty}^{\tau}ds\cos(\omega s)f(s),
\label{eq:Id4_3}
\end{align}
where we used partial integration, and Eqs.~\eqref{eq:Id2_1} and \eqref{eq:Id2_4}. Substituting Eqs.~\eqref{eq:Id4_1}--\eqref{eq:Id4_3} into Eq.~\eqref{eq:Id4}, we get for $ \tau > 0 $,
\begin{align}
\notag & I_{d^{-4}} = \frac{2}{c^2d^4}\frac{3\omega_{\textrm{p}}^2R_2^3}{(\varepsilon_{\infty}+2)^2\beta}\\
& \times \left\{\frac{\sin(\omega\tau)}{\omega}f'(\tau) + \left[1+2\cos(\omega\tau)\right]f(\tau) + \omega\left[1+\cos(\omega\tau)\right]\int_{-\infty}^{\tau}\! ds\sin(\omega s)f(s) - \omega\sin(\omega\tau)\int_{-\infty}^{\tau}\! ds\cos(\omega s)f(s)\right\}.
\label{eq:Id4_4}
\end{align}
For $ \tau < 0 $, $ I_{d^{-4}} = 0 $.

The integral with the sine in Eq.~\eqref{eq:Id4_4} is one half of Eq.~\eqref{eq:Id2_6}. The integral with the cosine reads as
\begin{align}
\notag & \int_{-\infty}^{\tau}\! ds\cos(\omega s)f(s) = \int_0^{\tau}\! dse^{-\frac{\gamma}{2}s}\sin(\beta s)\cos(\omega s) = \frac{1}{2}\int_0^{\tau}\! dse^{-\frac{\gamma}{2}s}\sin[(\beta-\omega)s] + \frac{1}{2}\int_0^{\tau}\! dse^{-\frac{\gamma}{2}s}\sin[(\beta+\omega)s]\\
\notag & = \frac{1}{2\left[\frac{\gamma^2}{4}+(\omega-\beta)^2\right]}\Bigg\{\beta-\omega-e^{-\frac{\gamma}{2}\tau}\left[\frac{\gamma}{2}\sin[(\beta-\omega)\tau] + (\beta-\omega)\cos[(\beta-\omega)\tau]\right]\Bigg\} + \frac{1}{2\left[\frac{\gamma^2}{4}+(\omega+\beta)^2\right]}\Bigg\{\beta+\omega\\
\notag & \ \ \ -e^{-\frac{\gamma}{2}\tau}\left[\frac{\gamma}{2}\sin[(\beta+\omega)\tau] + (\beta+\omega)\cos[(\beta+\omega)\tau]\right]\Bigg\} = \frac{1}{\left(\omega^2-\omega_{0\alpha}^2\right)^2+\gamma^2\omega^2}\Bigg\{-\beta\left(\omega^2-\omega_{0\alpha}^2\right)\\
\notag & \ \ \ -e^{-\frac{\gamma}{2}\tau}\Bigg[-\beta\left(\omega^2-\omega_{0\alpha}^2\right)\cos(\beta\tau)\cos(\omega\tau) - \omega\left(\omega^2-\omega_{0\alpha}^2+\frac{\gamma^2}{2}\right)\sin(\beta\tau)\sin(\omega\tau) -\gamma\beta\omega\cos(\beta\tau)\sin(\omega\tau)\\
& \ \ \ \ \ \ \ \ \ \ \ \ \ \ \ +\frac{\gamma}{2}\left(\omega^2+\omega_{0\alpha}^2\right)\sin(\beta\tau)\cos(\omega\tau)\Bigg]\Bigg\}.
\label{eq:Id4_5}
\end{align}
Summing up all terms of Eq.~\eqref{eq:Id4_4} and combining the terms with different powers of $ \gamma $, we obtain
\begin{align}
\notag & \frac{c^2 d^4}{2\omega}I_{d^{-4}} = \frac{3\omega_{\textrm{p}}^2R_2^3}{(\varepsilon_{\infty}+2)^2}\frac{\gamma\omega}{\left(\omega^2-\omega_{0\alpha}^2\right)^2+\gamma^2\omega^2} + \frac{3\omega_{\textrm{p}}^2R_2^3}{(\varepsilon_{\infty}+2)^2}\frac{\gamma\omega}{\left(\omega^2-\omega_{0\alpha}^2\right)^2+\gamma^2\omega^2}\Bigg\{\cos(\omega\tau) +e^{-\frac{\gamma}{2}\tau}\Big[-\cos(\beta\tau)\left[1+\cos(\omega\tau)\right]\\
\notag & -\frac{\omega^2+\omega_{0\alpha}^2}{2\beta\omega}\sin(\beta\tau)\sin(\omega\tau) - \frac{\gamma}{2\beta}\sin(\beta\tau)\left[1+\cos(\omega\tau)\right]\Big]\Bigg\} - \frac{3\omega_{\textrm{p}}^2R_2^3}{(\varepsilon_{\infty}+2)^2}\frac{\omega^2-\omega_{0\alpha}^2}{\left(\omega^2-\omega_{0\alpha}^2\right)^2+\gamma^2\omega^2}\Bigg\{-\sin(\omega\tau)\\
\notag & +e^{-\frac{\gamma}{2}\tau}\left[-\cos(\beta\tau)\sin(\omega\tau)+\frac{\omega}{\beta}\sin(\beta\tau)\left[1+\cos(\omega\tau)\right]\right]\Bigg\}\\
& + \frac{3\omega_{\textrm{p}}^2R_2^3}{(\varepsilon_{\infty}+2)^2}\frac{1}{\omega^2}e^{-\frac{\gamma}{2}\tau}\left[-\frac{\gamma}{2\beta}\sin(\beta\tau)\sin(\omega\tau)+\cos(\beta\tau)\sin(\omega\tau)+\frac{\omega}{\beta}\sin(\beta\tau)\left[1+2\cos(\omega\tau)\right]\right],
\label{eq:Id4_6}
\end{align}
where, similarly to Eq.~\eqref{eq:Id2_final}, the first two terms, $ \propto \Im[\alpha_2(\omega)] $, correspond to the heat transfer, whereas the third term, $ \propto \Re[\alpha_2(\omega)] - \alpha_{\infty} $, is related to the energy change. The last term, showing no direct proportionality to $ \Im[\alpha_2(\omega)] $ or $ \Re[\alpha_2(\omega)] $, is similar to Eq.~\eqref{eq:Id3_4}. We rewrite this term similarly to what we did in Sec.~\ref{app:subsec:d3}. Combining the result with the first three terms of Eq.~\eqref{eq:Id4_6}, we get
\begin{align}
\notag & \frac{c^2 d^4}{2\omega}I_{d^{-4}} = \frac{3\omega_{\textrm{p}}^2R_2^3}{(\varepsilon_{\infty}+2)^2}\frac{\gamma\omega}{\left(\omega^2-\omega_{0\alpha}^2\right)^2+\gamma^2\omega^2} + \frac{3\omega_{\textrm{p}}^2R_2^3}{(\varepsilon_{\infty}+2)^2}\frac{\gamma\omega}{\left(\omega^2-\omega_{0\alpha}^2\right)^2+\gamma^2\omega^2}\Bigg\{\cos(\omega\tau) +e^{-\frac{\gamma}{2}\tau}\Bigg[-\cos(\beta\tau)\left[1+\cos(\omega\tau)\right]\\
\notag & -\frac{2\omega^4-\omega_{0\alpha}^2\omega^2+\omega_{0\alpha}^4}{2\beta\omega^3}\sin(\beta\tau)\sin(\omega\tau) +\frac{\gamma}{\omega}\left(\cos(\beta\tau)\sin(\omega\tau) + \frac{\omega}{2\beta}\sin(\beta\tau)\left[1+3\cos(\omega\tau)\right]\right) -\frac{\gamma^2}{2\beta\omega}\sin(\beta\tau)\sin(\omega\tau)\Bigg]\Bigg\}\\
\notag & - \frac{3\omega_{\textrm{p}}^2R_2^3}{(\varepsilon_{\infty}+2)^2}\frac{\omega^2-\omega_{0\alpha}^2}{\left(\omega^2-\omega_{0\alpha}^2\right)^2+\gamma^2\omega^2}\Bigg\{-\sin(\omega\tau) + e^{-\frac{\gamma}{2}\tau}\Bigg[-\frac{2\omega^2-\omega_{0\alpha}^2}{\omega^2}\cos(\beta\tau)\sin(\omega\tau)\\
& \ \ \ \ \ \ \ \ \ \ \ \ \ \ \ \ \ \ \ \ \ \ \ \ \ \ \ \ \ \ \ \ \ \ \ \ \ \ \ \ \ \ \ \ \ \ +\frac{\omega}{\beta}\sin(\beta\tau)\left(\frac{\omega_{0\alpha}^2}{\omega^2}-\frac{\omega^2-2\omega_{0\alpha}^2}{\omega^2}\cos(\omega\tau)\right)\Bigg]\Bigg\},
\label{eq:Id4_final}
\end{align}
where the first two terms are proportional to $ \Im[\alpha_2(\omega)] $ and the last term is proportional to $ \Re[\alpha_2(\omega)] - \alpha_{\infty} $. The first (time-independent) term of Eq.~\eqref{eq:Id4_final} corresponds to the stationary heat transfer. Note that, in contrast to Eqs.~\eqref{eq:Id2_final} and~\eqref{eq:Id3_final}, the time-dependent part of Eq.~\eqref{eq:Id4_final} is not entirely proportional to $ e^{-\frac{\gamma}{2}\tau} $, but contains the terms $ \propto \cos(\omega\tau) $ and $ \propto \sin(\omega\tau) $. Nevertheless, after integration over $ \omega $ in Eq.~\eqref{eq:FluxPP}, these terms also vanish as $ \tau \to \infty $ (we show this numerically in Sec.~\ref{subsec:FluxPPSiC}).

\subsection{The term proportional to the inverse distance to the fifth power}
\label{app:subsec:d5}
For the $ d^{-5} $ term, we have 
\begin{equation}
I_{d^{-5}} \equiv \frac{6}{cd^5}\frac{3\omega_{\textrm{p}}^2R_2^3}{(\varepsilon_{\infty}+2)^2\beta}\int_0^{\infty}\! dt' \int_0^{\infty}\! dt''\cos[\omega(t'-t'')]\int\! d\tilde{t}f(\tilde{t})\left[\delta(\tau-\tilde{t}-t')\delta(\tau-t'')+\delta'(\tau-\tilde{t}-t')\theta(\tau-t'')\right].
\label{eq:Id5}
\end{equation}
For $ \tau < 0 $, $ I_{d^{-5}} = 0 $. For $ \tau > 0 $, the integrals for the first term are given by Eq.~\eqref{eq:Id4_5}. The second term reads as (using partial integration)
\begin{align}
\notag & \int_0^{\infty}\! dt' \int\! d\tilde{t}f(\tilde{t})\delta'(\tau-\tilde{t}-t')\int_0^{\infty}\! dt''\cos[\omega(t'-t'')]\theta(\tau-t'') = \frac{1}{\omega}\int_0^{\infty}\! dt' \int\! d\tilde{t}f(\tilde{t})\delta'(\tau-\tilde{t}-t')\sin(\omega t')\\
\notag & +\frac{1}{\omega}\int_0^{\infty}\! dt' \int\! d\tilde{t}f(\tilde{t})\delta'(\tau-\tilde{t}-t')\sin[\omega(\tau-t')] = \int_0^{\infty}\! dt'\cos(\omega t')\int\! d\tilde{t}f(\tilde{t})\delta(\tau-\tilde{t}-t') + \frac{\sin(\omega\tau)}{\omega}f(\tau)\\
\notag & - \int_0^{\infty}\! dt'\cos[\omega(\tau-t')]\int\! d\tilde{t}f(\tilde{t})\delta(\tau-\tilde{t}-t') = \int_0^{\infty}\! dt'\cos(\omega t')f(\tau-t') + \frac{\sin(\omega\tau)}{\omega}f(\tau)\\
& - \int_0^{\infty}\! dt'\cos[\omega(\tau-t')]f(\tau-t') \overset{\tau-t' \equiv s}{=} \int_{-\infty}^{\tau}\! ds\cos[\omega(\tau-s)]f(s) + \frac{\sin(\omega\tau)}{\omega}f(\tau) - \int_{-\infty}^{\tau}\! ds\cos(\omega s)f(s),
\label{eq:Id5_1}
\end{align}
where the third term in the last line is minus Eq.~\eqref{eq:Id4_5}, and it hence cancels the first term of Eq.~\eqref{eq:Id5}. We get
\begin{equation}
I_{d^{-5}} \equiv \frac{6}{cd^5}\frac{3\omega_{\textrm{p}}^2R_2^3}{(\varepsilon_{\infty}+2)^2\beta}\left\{\frac{\sin(\omega\tau)}{\omega}f(\tau) + \cos(\omega\tau)\int_{-\infty}^{\tau}\! ds\cos(\omega s)f(s) + \sin(\omega\tau)\int_{-\infty}^{\tau}\! ds\sin(\omega s)f(s)\right\}.
\label{eq:Id5_2}
\end{equation}

The integral with the cosine in Eq.~\eqref{eq:Id5_2} is given by Eq.~\eqref{eq:Id4_5}; the integral with the sine is one half of Eq.~\eqref{eq:Id2_6}; $ f $ is given by Eq.~\eqref{eq:f}. We obtain
\begin{align}
\notag \frac{c d^5}{6}I_{d^{-5}} = \ & \frac{3\omega_{\textrm{p}}^2R_2^3}{(\varepsilon_{\infty}+2)^2}\frac{\gamma\omega}{\left(\omega^2-\omega_{0\alpha}^2\right)^2+\gamma^2\omega^2}\Bigg\{\sin(\omega\tau) -e^{-\frac{\gamma}{2}\tau}\frac{\omega^2+\omega_{0\alpha}^2}{2\beta\omega}\sin(\beta\tau)\Bigg\}\\
& - \frac{3\omega_{\textrm{p}}^2R_2^3}{(\varepsilon_{\infty}+2)^2}\frac{\omega^2-\omega_{0\alpha}^2}{\left(\omega^2-\omega_{0\alpha}^2\right)^2+\gamma^2\omega^2}\Bigg\{\cos(\omega\tau)-e^{-\frac{\gamma}{2}\tau}\cos(\beta\tau)\Bigg\} + \frac{3\omega_{\textrm{p}}^2R_2^3}{(\varepsilon_{\infty}+2)^2}\frac{1}{\beta\omega}e^{-\frac{\gamma}{2}\tau}\sin(\beta\tau)\sin(\omega\tau),
\label{eq:Id5_3}
\end{align}
which can be rewritten as
\begin{align}
\notag \frac{c d^5}{6}I_{d^{-5}} = \ & \frac{3\omega_{\textrm{p}}^2R_2^3}{(\varepsilon_{\infty}+2)^2}\frac{\gamma\omega}{\left(\omega^2-\omega_{0\alpha}^2\right)^2+\gamma^2\omega^2}\Bigg\{\sin(\omega\tau) -e^{-\frac{\gamma}{2}\tau}\left[\frac{\omega^2+\omega_{0\alpha}^2}{2\beta\omega}\sin(\beta\tau) -\frac{\gamma}{\beta}\sin(\beta\tau)\sin(\omega\tau)\right]\Bigg\}\\
& - \frac{3\omega_{\textrm{p}}^2R_2^3}{(\varepsilon_{\infty}+2)^2}\frac{\omega^2-\omega_{0\alpha}^2}{\left(\omega^2-\omega_{0\alpha}^2\right)^2+\gamma^2\omega^2}\Bigg\{\cos(\omega\tau)-e^{-\frac{\gamma}{2}\tau}\left[\cos(\beta\tau) + \frac{\omega^2-\omega_{0\alpha}^2}{\beta\omega}\sin(\beta\tau)\sin(\omega\tau)\right]\Bigg\},
\label{eq:Id5_final}
\end{align}
where the first and second terms correspond to the heat transfer and energy change, respectively. As expected, no stationary part is present in Eq.~\eqref{eq:Id5_final}: The terms $ \propto e^{-\frac{\gamma}{2}\tau} $ vanish as $ \tau \to \infty $ for any $ \omega $, whereas the other terms vanish after integration over $ \omega $ in Eq.~\eqref{eq:FluxPP}.

\subsection{Near-field term}
\label{app:subsec:d6}
The integrals for the near-field term read as
\begin{equation}
I_{d^{-6}} \equiv \frac{6}{d^6}\frac{3\omega_{\textrm{p}}^2R_2^3}{(\varepsilon_{\infty}+2)^2\beta}\int_0^{\infty}\! dt' \int_0^{\infty}\! dt''\cos[\omega(t'-t'')]\int\! d\tilde{t}f(\tilde{t})\delta(\tau-\tilde{t}-t')\theta(\tau-t'').
\label{eq:Id6}
\end{equation}
Here, the integral over $ \tilde{t} $ can be computed directly, leading to
\begin{align}
\notag & \int_0^{\infty}\! dt' \int_0^{\infty}\! dt''\cos[\omega(t'-t'')]\int\! d\tilde{t}f(\tilde{t})\delta(\tau-\tilde{t}-t')\theta(\tau-t'') = \int_0^{\infty}\! dt' \int_0^{\infty}\! dt''\cos[\omega(t'-t'')]f(\tau-t')\theta(\tau-t'')\\
& = \int_0^{\infty}\! dt' \int_0^{\tau}\! dt''\cos[\omega(t'-t'')]f(\tau-t') = \frac{1}{\omega}\int_0^{\infty}\! dt'\sin[\omega(\tau-t')]f(\tau-t') + \frac{1}{\omega}\int_0^{\infty}\! dt'\sin(\omega t')f(\tau-t')
\label{eq:Id6_1}
\end{align}
for $ \tau > 0 $, and to zero for $ \tau < 0 $ (such that $ I_{d^{-6}}  = 0 $ in this case).

Making the substitution $ \tau - t' \equiv s $ in Eq.~\eqref{eq:Id6_1}, we find
\begin{align}
\notag \int_0^{\infty}\! dt' \int_0^{\tau}\! dt''\cos[\omega(t'-t'')]f(\tau-t') & = \frac{1}{\omega}\int_0^{\infty}\! dt'\sin[\omega(\tau-t')]f(\tau-t') + \frac{1}{\omega}\int_0^{\infty}\! dt'\sin(\omega t')f(\tau-t')\\
& = \frac{1-\cos(\omega\tau)}{\omega}\int_{-\infty}^{\tau}\! ds\sin(\omega s)f(s) + \frac{\sin(\omega\tau)}{\omega}\int_{-\infty}^{\tau}\! ds\cos(\omega s)f(s),
\label{eq:Id6_2}
\end{align}
where the integral in first term is the same as one half of Eq.~\eqref{eq:Id2_6}. The integral in the second term is given by Eq.~\eqref{eq:Id4_5}. We obtain
\begin{align}
\notag & \omega\int_0^{\infty}\! dt' \int_0^{\tau}\! dt''\cos[\omega(t'-t'')]f(\tau-t') = \frac{1}{\left(\omega^2-\omega_{0\alpha}^2\right)^2+\gamma^2\omega^2}\Bigg\{\gamma\beta\omega\left[1-\cos(\omega\tau)\right] -\beta\left(\omega^2-\omega_{0\alpha}^2\right)\sin(\omega\tau)\\
\notag & -e^{-\frac{\gamma}{2}\tau}\Bigg[-\gamma\beta\omega\cos(\beta\tau)-\omega\left(\omega^2-\omega_{0\alpha}^2+\frac{\gamma^2}{2}\right)\sin(\beta\tau) + \gamma\beta\omega\cos(\beta\tau)\cos(\omega\tau) + \frac{\gamma}{2}\left(\omega^2+\omega_{0\alpha}^2\right)\sin(\beta\tau)\sin(\omega\tau)\\
& \ \ \ \ \ \ \ \ \ \ \ \ -\beta\left(\omega^2-\omega_{0\alpha}^2\right)\cos(\beta\tau)\sin(\omega\tau) + \omega\left(\omega^2-\omega_{0\alpha}^2+\frac{\gamma^2}{2}\right)\sin(\beta\tau)\cos(\omega\tau)\Bigg]\Bigg\}.
\label{eq:Id6_3}
\end{align}
Combining the terms with different powers of $ \gamma $ and substituting the result into Eq.~\eqref{eq:Id6}, we get
\begin{align}
\notag & \frac{\omega d^6}{6}I_{d^{-6}} = \frac{3\omega_{\textrm{p}}^2R_2^3}{(\varepsilon_{\infty}+2)^2}\frac{\gamma\omega}{\left(\omega^2-\omega_{0\alpha}^2\right)^2+\gamma^2\omega^2} + \frac{3\omega_{\textrm{p}}^2R_2^3}{(\varepsilon_{\infty}+2)^2}\frac{\gamma\omega}{\left(\omega^2-\omega_{0\alpha}^2\right)^2+\gamma^2\omega^2}\Bigg\{-\cos(\omega\tau) +e^{-\frac{\gamma}{2}\tau}\Bigg[\cos(\beta\tau)\left[1-\cos(\omega\tau)\right]\\
\notag & -\frac{\omega^2+\omega_{0\alpha}^2}{2\beta\omega}\sin(\beta\tau)\sin(\omega\tau) + \frac{\gamma}{2\beta}\sin(\beta\tau)\left[1-\cos(\omega\tau)\right]\Bigg]\Bigg\} - \frac{3\omega_{\textrm{p}}^2R_2^3}{(\varepsilon_{\infty}+2)^2}\frac{\omega^2-\omega_{0\alpha}^2}{\left(\omega^2-\omega_{0\alpha}^2\right)^2+\gamma^2\omega^2}\Bigg\{\sin(\omega\tau)\\
& +e^{-\frac{\gamma}{2}\tau}\left[-\cos(\beta\tau)\sin(\omega\tau)-\frac{\omega}{\beta}\sin(\beta\tau)\left[1-\cos(\omega\tau)\right]\right]\Bigg\}.
\label{eq:Id6_final}
\end{align}
Regarding the physics of different terms of Eq.~\eqref{eq:Id6_final}, the same discussion as after Eq.~\eqref{eq:Id4_final} applies here.

\section{Estimation of the formula for the nonstationary near-field heat flux at room temperature between two isolated particles}
\label{app:sec:FluxPPVacFormula}
\subsection{Estimating the curve for the heat flux maxima}
\label{app:subsec:FluxPPVacFormulaMax}
In this appendix, we describe how we came up with Eq.~\eqref{eq:FluxAnNF}. First, let us find a curve which approximates the maxima of the flux for $ d = 100 \ \textrm{nm} $ (black filled circles in Fig.~\ref{fig:T300nf}). We observed that, for small $ \tau $, the maxima scale approximately as $ \tau^{\frac{3}{4}} $, whereas for large $ \tau $, they approach the stationary value as  $ H_{1\textrm{st,vac}}^{(2)} + b\tau^{\frac{3}{4}}e^{-\frac{\gamma}{2}\tau} $, with $ b $ being a $ \tau $-independent parameter (but it depends on the interparticle distance). Furthermore, since the maxima form a global maximum (i.e., the maximum of the desired curve), we include another exponent $ e^{-\gamma\tau} $ [as appears in Eq.~\eqref{eq:FluxAv}], and parameter $ a $ to accurately model this maximum. Based on these considerations, we write the analytical approximation as ($ \tau > 0 $)
\begin{equation}
\Phi_{1\textrm{,vac}}^{(2)\textrm{,max}}(\tau) = H_{1\textrm{st,vac}}^{(2)}\left(1-e^{-\gamma\tau}\right) + \tau^{\frac{3}{4}}\left(ae^{-\gamma\tau}+be^{-\frac{\gamma}{2}\tau}\right),
\label{eq:FluxMax}
\end{equation}
where the first term is the average flux given by Eq.~\eqref{eq:FluxAv}.

We determine parameters $ a $ and $ b $ in Eq.~\eqref{eq:FluxMax} by the position and value of the global maximum, $ \tau_{\textrm{max}} $ and $ \Phi_{1\textrm{,vac}}^{(2)\textrm{,max}}(\tau_{\textrm{max}}) $, respectively, which are known numerically, i.e., we require the maximum of Eq.~\eqref{eq:FluxMax} to coincide with the numerical maximum. Using the condition for the maximum,
\begin{equation}
\frac{d}{d\tau}\Phi_{1\textrm{,vac}}^{(2)\textrm{,max}}(\tau)\Big|_{\tau=\tau_{\textrm{max}}} = 0,
\label{eq:max_condition}
\end{equation} 
we can express $ b $ via $ a $,
\begin{equation}
b = \frac{\left[\gamma \tau_{\textrm{max}}^{\frac{1}{4}}H_{1\textrm{st,vac}}^{(2)} - a\left(\gamma\tau_{\textrm{max}}-\frac{3}{4}\right)\right]e^{-\frac{\gamma}{2}\tau_{\textrm{max}}}}{\frac{\gamma}{2}\tau_{\textrm{max}}-\frac{3}{4}}.
\label{eq:bviaa}
\end{equation}
Substituting Eq.~\eqref{eq:bviaa} into Eq.~\eqref{eq:FluxMax} and considering that the right-hand side of the latter evaluated at $ \tau_{\textrm{max}} $ equals $ \Phi_{1\textrm{,vac}}^{(2)\textrm{,max}}(\tau_{\textrm{max}}) $ (known numerically), we find $ a $,
\begin{equation}
a = \frac{\left(\frac{\gamma}{2}\tau_{\textrm{max}}+\frac{3}{4}\right)H_{1\textrm{st,vac}}^{(2)} + \left(\frac{\gamma}{2}\tau_{\textrm{max}}-\frac{3}{4}\right)\left[H_{1\textrm{st,vac}}^{(2)}-\Phi_{1\textrm{,vac}}^{(2)\textrm{,max}}(\tau_{\textrm{max}})\right]e^{\gamma\tau_{\textrm{max}}}}{\frac{\gamma}{2}\tau_{\textrm{max}}^{\frac{7}{4}}}.
\label{eq:a}
\end{equation}
Note that both $ a $ and $ b $ depend on the interparticle distance. For $ d = 100 \ \textrm{nm} $, $ \tau_{\textrm{max}} = 2.854 \ \textrm{ps} $, $ \Phi_{1\textrm{,vac}}^{(2)\textrm{,max}}(\tau_{\textrm{max}}) = 1.88 \times 10^{34} \ \textrm{J}\textrm{s}^{-1}\textrm{m}^{-6} $, $ H_{1\textrm{st,vac}}^{(2)} = 1.15 \times 10^{34} \ \textrm{J}\textrm{s}^{-1}\textrm{m}^{-6} $, which are used to calculate $ a $ and $ b $. 

\begin{figure}[!t]
\begin{center}
\includegraphics[width=0.6\linewidth]{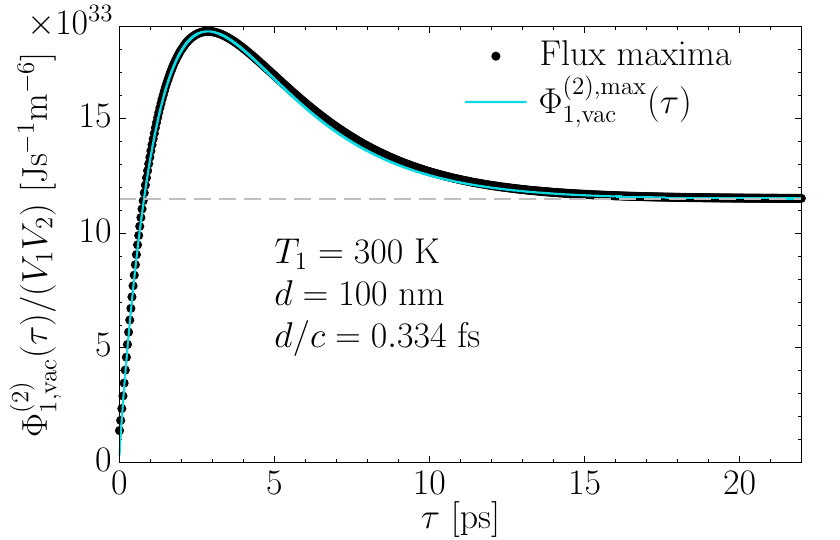}
\end{center}
\caption{\label{fig:FluxMax}Comparison of the maxima of the nonstationary near-field heat flux between two isolated particles with the analytical approximation $ \Phi_{1\textrm{,vac}}^{(2)\textrm{,max}}(\tau) $ given by Eq.~\eqref{eq:FluxMax}. The results are given as functions of time $ \tau \equiv t - d/c > 0 $. The temperature of particle $ 1 $ is $ T_1 = 300 \ \textrm{K} $; the interparticle distance is $ d = 100 \ \textrm{nm} $. Dashed line corresponds to the stationary heat flux (transfer).}
\end{figure}

Figure~\ref{fig:FluxMax} shows that formula~\eqref{eq:FluxMax} fits well the numerical data, with the relative discrepancy less than $ 2\% $ for $ \tau > 0.5 \ \textrm{ps} $.

\subsection{Formula for the nonstationary heat flux}
\label{app:subsec:FluxPPVacFormulaSelf}
Taking into account that the heat flux oscillates along the exponential relaxation in $ \overline{\Phi}_{1\textrm{,vac}}^{(2)}(\tau) $, with the frequency $ \omega_{0\alpha} $ and maxima appearing at $ \tau \approx \pi (2n-1)/\omega_{0\alpha} $ (for significantly large $ \tau $ and hence large natural $ n $), we assume that it can be approximated by Eq.~\eqref{eq:FluxAnNF}, where $ \eta(\tau) $, the amplitude of the oscillations, is determined by the fact that, at the maxima [where $ \cos(\omega_{0\alpha}\tau) = -1 $] Eq.~\eqref{eq:FluxAnNF} should be equal to $ \Phi_{1\textrm{,vac}}^{(2)\textrm{,max}}(\tau) $ in Eq.~\eqref{eq:FluxMax}. Therefore, $ \eta(\tau) $ is the last term of the latter equation,
\begin{equation}
\eta(\tau) = \tau^{\frac{3}{4}}\left(ae^{-\gamma\tau}+be^{-\frac{\gamma}{2}\tau}\right).
\label{eq:eta}
\end{equation}

% APPENDIX_end ------------------------------------------
\twocolumngrid

% BIBLIOGRAPHY_begin ------------------------------------------


% BIBLIOGRAPHY_end ------------------------------------------

\end{document}